\documentclass[aps,prx,nofootinbib,superscriptaddress,preprintnumbers,amsmath,amssymb,latexsym,array,enumerate,letter,twocolumn]{revtex4}
\pdfoutput=1

\usepackage{latexsym}
\usepackage{graphicx, graphics, hyperref, amsmath, amssymb, slashed, xcolor, bbm,bm,amsthm, array,multirow}
\usepackage{enumitem}
 \usepackage{subfigure}
 \usepackage{listings} 
 \usepackage{trimclip}
\usepackage{siunitx}
\usepackage[utf8]{inputenc}
\usepackage{rotating}
\usepackage{afterpage}
\usepackage{cancel}

\newcommand{\nc}{\newcommand}

\nc{\beq}{\begin{equation}}
\nc{\eeq}{\end{equation}}
\nc{\barray}{\begin{eqnarray}}
\nc{\earray}{\end{eqnarray}}
\nc{\barrayn}{\begin{eqnarray*}}
\nc{\earrayn}{\end{eqnarray*}}
\nc{\bcenter}{\begin{center}}
\nc{\ecenter}{\end{center}}
\nc{\mc}{\mathcal}
\nc{\er}[1]{(\ref{eq:#1})}
\nc{\onehalf}{\frac{1}{2}} 
\nc{\partialbar}{\bar{\partial}}
\nc{\psit}{\widetilde{\psi}}
\nc{\Tr}{\mbox{Tr}}
\nc{\hc}{\mbox{H.c.}}
\nc{\ev}{\;\mathrm{eV}}
\nc{\mev}{\;\mathrm{MeV}}
\nc{\gev}{\;\mathrm{GeV}}
\nc{\cm}{\;\mathrm{cm}}
\nc{\kev}{\;\mathrm{keV}}
\nc{\tev}{\;\mathrm{TeV}}
\nc{\mum}{\;\mu\mathrm{m}}
\nc{\m}{\;\mathrm{m}}
\nc{\mm}{\;\mathrm{mm}}
\nc{\um}{\;\mu\mathrm{m}}
\nc{\s}{\mathrm{s}}
\nc{\ms}{\;\mathrm{ms}}
\nc{\us}{\;\mu\mathrm{s}}
\nc{\mywidth}{0.3\textwidth}

\def\beq{\begin{equation}}
\def\eeq{\end{equation}}
\def\bea{\begin{eqnarray}}
\def\eea{\end{eqnarray}}

\def\eV{\textrm{eV}}
\def\eh{$e^-h^+$}

\def\bal{\begin{align*}}
\def\eal{\end{align*}}
\def\bea{\begin{eqnarray}}
\def\eea{\end{eqnarray}}

\def\om{\omega}
\def\epom{\epsilon(\omega)}

\newcommand{\Rs}{$R_{1e^-}$}

\graphicspath{{plots/}}

\begin{document}

\widetext

\title{Sources of Low-Energy Events in Low-Threshold Dark Matter and Neutrino Detectors}

\author{Peizhi Du}
\affiliation{C.N. Yang Institute for Theoretical Physics, Stony Brook University, Stony Brook, NY, 11794, USA}

\author{ Daniel Egana-Ugrinovic}
\affiliation{Perimeter Institute for Theoretical Physics, Waterloo, ON N2L 2Y5}

\author{Rouven Essig}
\affiliation{C.N. Yang Institute for Theoretical Physics, Stony Brook University, Stony Brook, NY, 11794, USA}

\author{Mukul Sholapurkar}
\affiliation{C.N. Yang Institute for Theoretical Physics, Stony Brook University, Stony Brook, NY, 11794, USA}

\preprint{YITP-SB-2020-37}
 \begin{abstract}
We discuss several low-energy backgrounds to sub-GeV dark matter searches, which arise from high-energy particles of cosmic or radioactive origin that interact with detector materials. 
We focus in particular on Cherenkov radiation, transition radiation, and luminescence or phonons from electron-hole pair recombination,
and show that these processes are an important source of backgrounds at both current and planned detectors.
We perform detailed analyses of these backgrounds at several existing and proposed experiments based on a wide variety of detection strategies and levels of shielding. 
We find that a large fraction of the observed single-electron events in the SENSEI 2020 run originate from Cherenkov photons generated by high-energy events in the Skipper Charge Coupled Device, and from recombination photons generated in a phosphorus-doped layer of the same instrument.
In a SuperCDMS HVeV 2020 run, Cherenkov photons produced in printed-circuit-boards located near the sensor likely explain the origin of most of the events containing 2 to 6 electrons. 
At SuperCDMS SNOLAB, 
radioactive contaminants inside the Cirlex located inside or on the copper side walls of their detectors will produce many Cherenkov photons, which could dominate the low-energy backgrounds. 
For the EDELWEISS experiment, Cherenkov or luminescence backgrounds are subdominant to their observed event rate, 
but could still limit the sensitivity of their future searches.  
We also point out that Cherenkov radiation, transition radiation, and recombination
could be a significant source of backgrounds at future experiments aiming to detect dark-matter via scintillation or phonon signals. 
We also discuss the implications of our results for the development of superconducting qubits and low-threshold searches for coherent neutrino scattering.
Fortunately, several design strategies to mitigate these backgrounds can be implemented, such as minimizing non-conductive materials near the target, implementing active and passive shielding, and using multiple nearby detectors.   
 \end{abstract}
 
  \maketitle


\tableofcontents

\begin{figure*}[t!]
\centering
\includegraphics[width=0.45\textwidth]{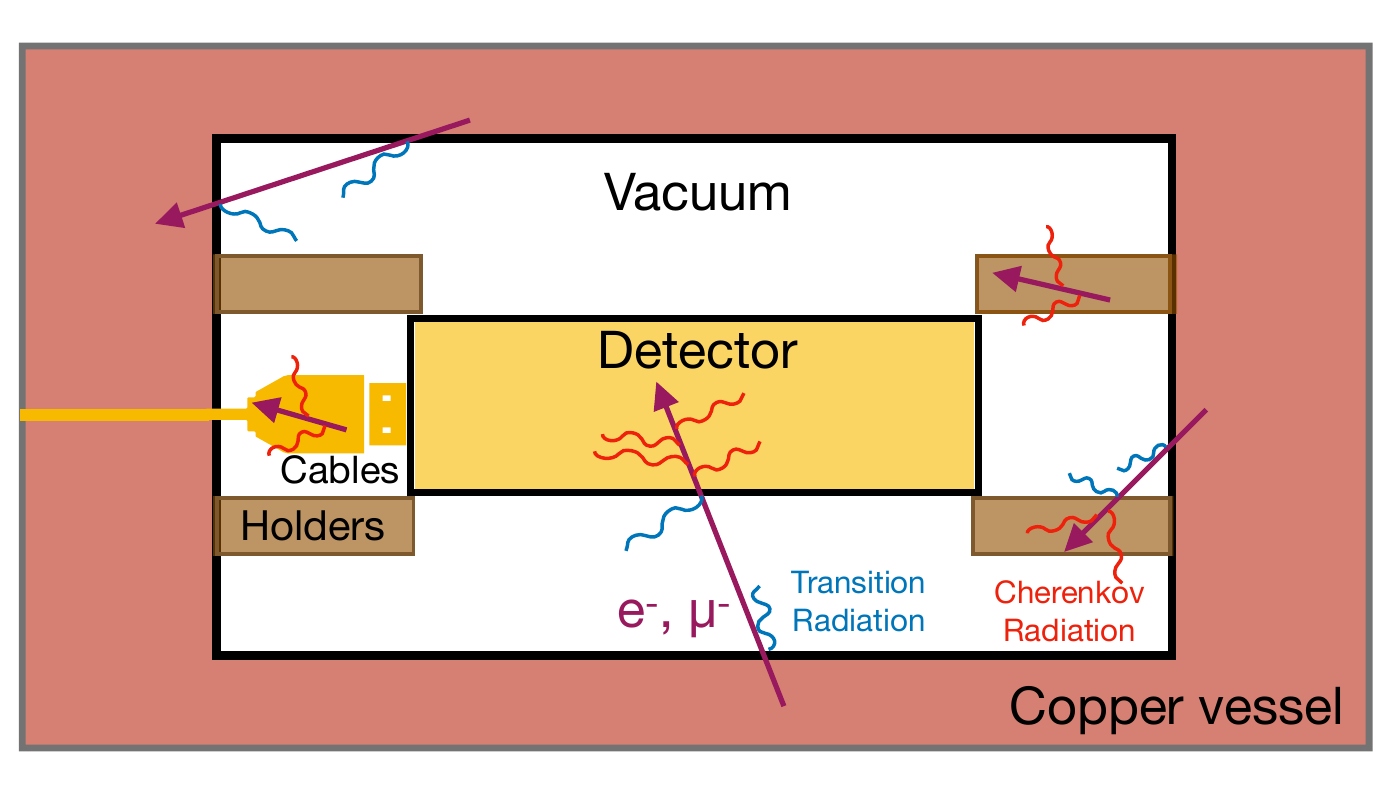}
\quad
\includegraphics[width=0.45\textwidth]{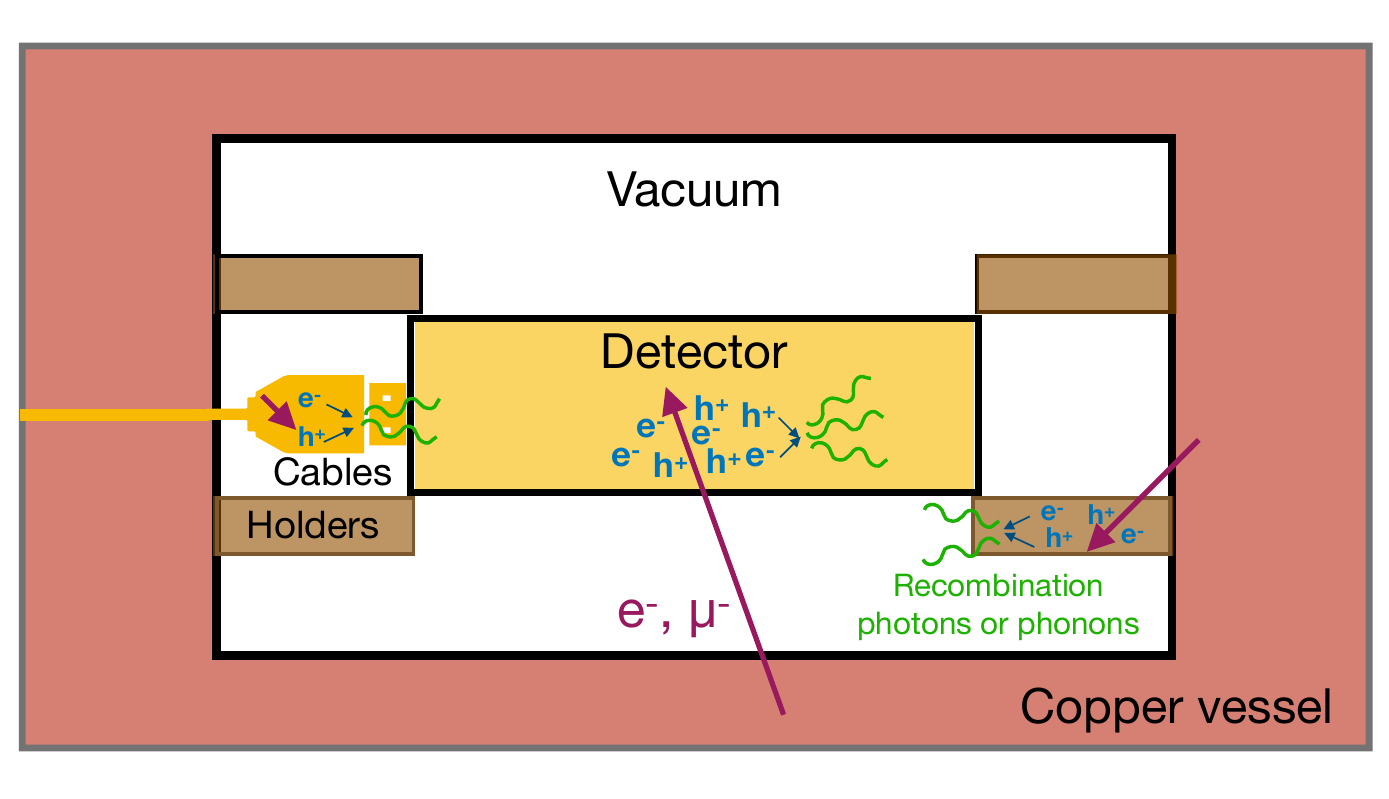}
\caption{
Schematic for the production of radiative backgrounds: Cherenkov and transition radiation (\textbf{left}) and recombination photons and phonons (\textbf{right}).  High-energy electrons (from ambient radioactivity or radioactive contaminants) and cosmic-ray muons (purple) can (i) interact in non-conductive materials, including the detector target and the surrounding materials, such as holders and cables, to produce Cherenkov radiation (red); (ii) cross surfaces to produce transition radiation (blue); and (iii) deposit energy to create luminescence or phonons via, \textit{e.g.}, the recombination of electron-hole pairs (green). 
The low energy photons or phonons obtained by these processes can then be absorbed in the detector target, producing a measurable signal that depends on the target properties, but that can take the form of, \textit{e.g.}, one or more electrons, photons, or phonons, which can mimic the low-energy signals produced by low-mass dark matter interacting with the detector target. 
\label{fig:intro}
} 
\end{figure*}

\section{Introduction}

The search for the particle nature of dark matter is currently among the most important topics in particle physics.  Direct-detection experiments provide an essential approach in our endeavor to detect dark-matter particles in the laboratory and identify their interactions.  
In the past decade, 
searches for dark-matter particles with masses in the previously unexplored range between $\sim$1~meV to 1~GeV  
have gained significant attention due to novel theoretical and experimental developments~\cite{Essig:2011nj,Essig:2012yx,Angle:2011th,Graham:2012su,An:2014twa,Aprile:2014eoa,Essig:2015cda,Lee:2015qva,Hochberg:2015pha,Hochberg:2015fth,Essig:2016crl,Derenzo:2016fse,Aguilar-Arevalo:2016zop,Bloch:2016sjj,Cavoto:2016lqo,Hochberg:2016ntt,Hochberg:2016ajh,Hochberg:2016sqx,Kouvaris:2016afs,Aprile:2016wwo,Agnese:2016cpb,Tiffenberg:2017aac,Romani:2017iwi,Essig:2017kqs,Budnik:2017sbu,Bunting:2017net,Cavoto:2017otc,Fichet:2017bng,Knapen:2017ekk,Hochberg:2017wce,Dolan:2017xbu,Ibe:2017yqa,Angloher:2017sxg,Essig:2018tss,Crisler:2018gci, Agnese:2018col, Agnes:2018oej,Settimo:2018qcm,Bringmann:2018cvk,Ema:2018bih,Akerib:2018hck,Hertel:2018aal,Szydagis:2018wjp,Essig:2019kfe,Essig:2019xkx,Abramoff:2019dfb, Aguilar-Arevalo:2019wdi,Armengaud:2019kfj,Liu:2019kzq,Aprile:2019jmx,Emken:2019tni,Bell:2019egg,Cappiello:2019qsw,Trickle:2019ovy,Griffin:2019mvc,Trickle:2019nya,Baxter:2019pnz,Catena:2019gfa,Blanco:2019lrf,Berlin:2019uco,Abdelhameed:2019hmk,Abdelhameed:2019mac,Aprile:2019xxb,Kurinsky:2019pgb,Kurinsky:2020dpb,Robinson:2020lqx,Kurinsky:2020fut,Barak:2020fql,Aprile:2020tmw,Arnaud:2020svb,Amaral:2020ryn,Alkhatib:2020slm,Griffin:2020lgd,Alexander:2016aln,Battaglieri:2017aum,BRN}. 
Early experimental successes in probing dark matter interacting with electrons or nuclei in parts of this mass range have come from 
XENON10~\cite{Angle:2011th,Essig:2012yx}, XENON100~\cite{Aprile:2016wwo,Essig:2017kqs}, SENSEI~\cite{Crisler:2018gci,Abramoff:2019dfb,Barak:2020fql}, DarkSide-50~\cite{Agnes:2018oej}, DAMIC~\cite{Aguilar-Arevalo:2019wdi}, XENON1T~\cite{Aprile:2019xxb,Aprile:2020tmw}, SuperCDMS~\cite{Agnese:2018col,Amaral:2020ryn,Alkhatib:2020slm}, EDELWEISS~\cite{Armengaud:2019kfj,Arnaud:2020svb}, and CRESST-III~\cite{Angloher:2017sxg,Abdelhameed:2019hmk,Abdelhameed:2019mac}, 
and rapid experimental progress is expected in this decade.
However,  
a common feature among \textit{all} these experiments is that they observe sizable rates of low-energy background events of unknown origin that affect the reach of dark-matter searches. 
These backgrounds will need to be identified, characterized, and mitigated to ensure the success of the low-mass dark matter search program in the coming decade as experiments improve their sensitivity. 

The research towards identifying the origins of several of these backgrounds is already underway, even if a satisfactory understanding of the observed events in each case is still lacking.  
For example, for the single-electron events in noble liquid detectors (\textit{e.g.}, XENON10/100/1T and DarkSide-50) several hypothesis have been proposed, such as the delayed release of electrons from the liquid-gas interface, impurities capturing and releasing electrons, the photoelectric effect from photons hitting metallic surfaces, and others~\cite{Santos:2011ju,Aprile_2014,Sorensen:2017ymt,Sorensen_2018,Tomas:2018pny,Akimov:2019ogx,LUX:2020vbj}.  
The low-energy excess in CRESST-III is thought to arise from cracking or micro-fracturing of the crystal or support holders 
that occur when the detectors are being tightly clamped~\cite{Astrom:2005zk}, but it remains unclear whether this explains the excess. 
Some single-electron events currently seen in Skipper-Charge Coupled Device (Skipper-CCD) detectors are known to originate from spurious charge generation during readout and can be reduced by optimizing the voltages that move the charge pixel-to-pixel, but a large single-electron event rate remains even after subtracting the spurious-charge contribution~\cite{Barak:2020fql}.  
A leakage current likely sources the single-electron events in the SuperCDMS HVeV detectors and EDELWEISS sub-GeV dark matter search~\cite{Agnese:2018col,Amaral:2020ryn,Arnaud:2020svb}, 
but its microphysical origin is not satisfactorily understood, 
nor is the microphysical origin understood of the events containing more than one electron.  
Other backgrounds that have been considered to produce low-energy events are the low-energy tail of the Compton and beta-decay spectra,  as well as coherent solar neutrino-nucleus scattering, see \textit{e.g.}~\cite{Billard:2013qya,Essig:2018tss,Hochberg:2015fth}, 
but in practice these backgrounds are subdominant at many experiments, and in fact these tails cannot explain 
the aforementioned excesses of low-energy events. 

In this paper, we take a significant step towards identifying several processes that produce backgrounds in low-mass dark-matter detectors.
In particular, we discuss and demonstrate the importance of three largely unexplored processes in the context of low-energy threshold detectors, 
which arise when high-energy particles such as radiogenic electrons or gamma-rays and cosmic muons interact with detector materials:
Cherenkov radiation,\footnote{Cherenkov radiation, in particular, has been explored as a background in the context of WIMP detection in DarkSide~\cite{xiang2018cherenkov}.} transition radiation, and luminescence or phonons from recombination. 
We provide a schematic depiction on how these backgrounds are obtained in detectors in Fig.~\ref{fig:intro}.
The Cherenkov process is realized when charged particles go through dielectric materials at velocities exceeding the speed of light in the medium, 
while transition radiation is obtained if charged tracks encounter interfaces separating media with different dielectric properties.
As a result of these two processes,
one or more photons with energies below a few-eV can be emitted by the high-energy particles.
Recombination photons or phonons, on the other hand, 
arise since high-energy particles efficiently deposit energy in materials by exciting a large number of \eh-pairs.
These pairs subsequently recombine radiatively or non-radiatively releasing an energy of order $\sim$eV per pair.
The low energy photons or phonons obtained by these processes
can then be absorbed in the detector target, 
producing a measurable signal that depends on the target properties, but can be in the form of one or more electrons, photons, phonons, or magnons, which potentially mimics the low-energy signals produced by low-mass dark matter. 

To demonstrate the importance of Cherenkov radiation, transition radiation, and recombination of \eh-pairs as sources of backgrounds for dark-matter detectors, 
the details of different experiments and their data analyses pipelines must be taken into account.
With this objective, 
we perform background rate estimates at several current and proposed experiments, including SENSEI, SuperCDMS HVeV/CPD/SNOLAB, EDELWEISS, and CRESST-III.
Different experiments employ different strategies to mitigate backgrounds arising from high-energy tracks.
Detectors such as the Skipper-CCDs used by SENSEI~\cite{Tiffenberg:2017aac,Crisler:2018gci,Abramoff:2019dfb,Barak:2020fql}, DAMIC-M~\cite{Castello-Mor:2020jhd}, and Oscura~\cite{BRN-announcement2},
which aim to read out electron-hole pairs created by dark-matter scattering, 
have very little timing information but excellent position resolution.
In this case,
events are vetoed if their position in the Si CCD is close to an observed high-energy track.
Despite these vetoes, single-electron-event backgrounds may still arise from track-induced radiation (which creates \eh-pairs upon absorption in the CCD), 
if the photons travel far away from their originating tracks.
In the SENSEI CCD, this happens when the photons have energies close or below the Si bandgap.
We will show that such photons can be created by the Cherenkov effect as tracks pass through the CCD or other surrounding dielectric materials,
and that they are also abundantly obtained from the radiative recombination of \eh-pairs created by tracks that pass 
through a layer on the CCD's backside with high phosphorus doping (in other CCD regions, the radiative recombination rate is comparatively smaller). 
By performing a quantitative analysis of the number of events obtained by these processes,
we find that Cherenkov and recombination radiation are responsible for at least a large fraction of the single-electron event rate reported by SENSEI.  

Most other solid-state detectors, 
including SuperCDMS, EDELWEISS, and CRESST-III and proposed detectors based on scintillation~\cite{Derenzo:2016fse} or single-phonon~\cite{Knapen:2017ekk} signals such as SPICE~\cite{SPICE}, have excellent timing resolution. 	
In this case  low-energy events coming from high-energy tracks that pass through the detector target are easily vetoed due to their time-coincidence with the observed track.\footnote{One exception to this is radiative backgrounds from slow luminescence (phosphorescence), in which case the emission could occur well after a track has passed. This ``afterglow'' will be challenging to veto, and could be important for some dark matter searches (see discussion in, \textit{e.g.},~\cite{Derenzo:2016fse,Derenzo:2018plr}).  A comprehensive study of luminescence for all materials typically present in dark matter detectors is still lacking.}
However, high-energy events interacting in uninstrumented materials that surround the target usually cannot be vetoed, and hence low-energy photons produced in those interactions (either from Cherenkov or transition radiation, or from luminescence) would provide a low-energy signal.
In this context our estimates indicate that at SuperCDMS-HVeV, 
Cherenkov radiation from tracks passing through PCBs located around the detector is very possibly the dominant source of the low-energy events containing two to six electrons seen in the data~\cite{Agnese:2018col,Amaral:2020ryn}.
We also show that transition radiation provides another, smaller contribution. 
At EDELWEISS~\cite{Arnaud:2020svb}, we find that given its well-shielded and underground location, track-induced backgrounds from any of the processes discussed here are currently expected to be subdominant, 
but may be relevant to assess the future sensitivity of this experiment. 
Track-related backgrounds are also expected to be subdominant at CRESST-III~\cite{Abdelhameed:2019hmk,Abdelhameed:2019mac}, 
since both the detector and surrounding materials have been designed to scintillate to veto events associated with tracks.
At SuperCDMS CPD~\cite{Alkhatib:2020slm}, we find that Cherenkov or transition radiation, or prompt radiation from radiative recombination due to the passage of high-energy tracks, make up less than $\sim$10\% of their large observed low-energy event rate. 
The fact that we find track-related radiative backgrounds to be subdominant to the observed background rates at EDELWEISS, CRESST-III, and SuperCDMS-CPD is consistent with the characteristics of the observed backgrounds: most of them seem to be ``low-yield'', i.e., they produce little or no ionization (and hence cannot originate from the absorption of an above-bandgap photon) and instead produce a heat or phonon signal~\cite{Arnaud:2020svb,Abdelhameed:2019hmk,Abdelhameed:2019mac,Alkhatib:2020slm,EXCESSworkshop,MattPyle-discussion}.  Nevertheless, our analysis indicates that track-related backgrounds are an important background for future versions of these detectors, at least once the low-yield backgrounds are understood. 
Regarding proposed detectors such as SPICE, 
we point out that Cherenkov or recombination radiation generated in materials surrounding the detector, 
or recombination phonons generated in the detector holders,
could potentially be a significant source of backgrounds that had not been previously considered in the literature. 
 
Going beyond dark matter detectors, the above backgrounds may be relevant for other devices that have eV or sub-eV sensitivity. For example, neutrino detectors based on Skipper-CCDs such as CONNIE~\cite{Aguilar-Arevalo:2019jlr} and $\nu$IOLETA~\cite{Fernandez-Moroni:2020yyl} may suffer from these backgrounds. More interestingly, these backgrounds maybe important for quantum computing. Superconducting quantum bit (qubit) typically has an energy gap of $O(\mathrm{meV})$. Photons above this energy can break Cooper pairs in superconductors and create quasiparticles. This may be a source of the observed loss of coherence in superconducting qubits~\cite{Vepsalainen:2020trd,Cardani_2021,Wilen:2020lgg,Mcewen:2021ood}. 

While the arguments and estimates that we perform clearly demonstrate that Cherenkov radiation, transition radiation, and recombination constitute important backgrounds, 
we also show that it is often surprisingly challenging to evaluate their magnitude precisely.  
In particular, this requires knowing or measuring the precise optical properties of all materials present in the detector and performing detailed simulations of the high-energy backgrounds, detector setup, and data analysis chain.  
With that being said, the required measurements and simulations are feasible, 
so the backgrounds discussed here can and need to be characterized precisely in the future.

Finally, and very importantly, having identified these backgrounds 
and their exact origin at detectors, we present specific and concrete strategies that can be implemented at future experiments in order to mitigate their impact. 

The remaining paper is organized as follows.  In Sec.~\ref{sec:cherenkov} and Sec.~\ref{sec:transition}, we discuss in detail Cherenkov radiation and transition radiation produced directly by charged particles interacting in a variety of materials typically found in detectors, respectively.  
In Sec.~\ref{sec:recombination}, we discuss luminescence and phonons that are produced from the recombination of \eh-pairs after the passage of a high-energy particle in different materials.  In Sec.~\ref{sec:experiments-current}, we show how these radiative backgrounds contribute to the events observed in recent results from SENSEI~\cite{Barak:2020fql}, SuperCDMS HVeV~\cite{Amaral:2020ryn}, EDELWEISS~\cite{Arnaud:2020svb}, CRESST-III~\cite{Abdelhameed:2019hmk,Abdelhameed:2019mac}, SuperCDMS CPD~\cite{Alkhatib:2020slm}, and EDELWEISS-Surf~\cite{Armengaud:2019kfj}.  Sec.~\ref{sec:experiments-future} discusses these backgrounds in the upcomingSuperCDMS SNOLAB experiment~\cite{Agnese:2016cpb}, future Skipper-CCD detectors (SENSEI at SNOLAB, DAMIC-M, Oscura), and future detectors searching for photons or phonons (including SPICE~\cite{SPICE}).  
Apart from dark matter detections, we point out in Sec.~\ref{subsec:other} that these radiative backgrounds may also be relevant for neutrino experiments and superconducting qubits. 
In Sec.~\ref{sec:mitigation}, we discuss how these radiative backgrounds can be mitigated.  
We summarize our findings in Sec.~\ref{sec:conclusions}.  
Several appendices include additional details on semiconductor properties (Appendix~\ref{app:semiconductors}), 
as well as our background estimates for SENSEI (Appendix~\ref{app:SENSEI}), 
SuperCDMS HVeV (Appendix~\ref{app:SuperCDMS-HVeV}), 
and SuperCDMS at SNOLAB (Appendix~\ref{app:SuperCDMS-SNOLAB}).


\section{Cherenkov Radiation}
\label{sec:cherenkov}

In this section, we discuss the emission of Cherenkov radiation from high-energy charged particles traversing a material.  This can produce low-energy photons, which can be absorbed in the target to produce low-energy events that mimic a dark matter signal.

\subsection{The Cherenkov Effect: Theory}\label{subsec:cherenkov-theory}
The Cherenkov effect is the spontaneous emission of radiation by a charged particle passing through a dielectric material. 
The emission of Cherenkov radiation depends on the properties of the dielectric,
which are set by its complex refraction index $\tilde{n}$, 
\begin{equation}
\tilde{n}\equiv n + i \kappa \quad , \quad n,\kappa \in \mathbb{R}  \quad .
\label{eq:refraction}
\end{equation}
The complex refraction index determines how light propagates in the medium, $e^{i(\omega \tilde{n} x -\omega t)}=e^{-\kappa \omega x} e^{i \omega(nx -t)}$. 
The real part of the refraction index, $n$, sets the phase-space velocity of light in the material, 
while the imaginary part $\kappa$ is the extinction coefficient. 
Alternatively, one can describe the material with the complex dielectric function, 
defined as the square of the complex refraction index, $\epom \equiv \tilde{n}^2$, 
or
\begin{equation}
\textrm{Re}\, \epom \equiv n^2-\kappa^2 \quad , \quad  \textrm{Im}\, \epom \equiv 2n\kappa \quad .
\label{eq:epom}
\end{equation}
Emission of Cherenkov photons with frequency $\omega$ happens when the product of the square of the particle velocity times the real part of the material's dielectric function is greater than one~\cite{PhysRev.89.1147,PhysRev.91.256}
\bea
\label{eq:cherenkovcon}
v^2 \textrm{Re}\,\epsilon(\omega)>1 \quad .
\eea
Given that $v\leq 1$, 
a necessary condition for Cherenkov radiation to occur is $\textrm{Re}\,\epsilon(\omega)>1
$. 
Note that for non-absorptive materials, $\kappa=0$, 
and the Cherenkov condition reduces to the requirement that the charged particle velocity exceeds the speed of light in the material, $v  > 1/n$. 

Cherenkov radiation has particular characteristics that differentiate it from other types of radiation that arise due to the passage of charged particles through matter.
While these features are most accurately described quantum-mechanically, 
they are already apparent in classical electrodynamics.  We first discuss the classical results and postpone commenting on quantum-mechanical corrections to the end of this section.

We begin by discussing the Cherenkov emission rate. 
The differential emission rate of Cherenkov photons by a unit-charge particle passing through a medium is given by~\cite{PhysRev.89.1147}
\beq
\label{eq:cherenkov}
\frac{d^2N_\gamma}{d\om dx}=\alpha \left(1-\frac{\textrm{Re} \,\epsilon(\omega) }{v^2 |{\epsilon(\omega)|^2}}\right) \quad , \quad v^2 \textrm{Re}\,\epsilon(\omega)>1 
\eeq
where $N_\gamma$ is the number of emitted photons, $\omega$ is the photon frequency, and $dx$ is the charged particle's path-length differential.
Given that the Cherenkov process arises at leading order in electrodynamics, 
the rate Eq.~\eqref{eq:cherenkov} is proportional to one power only of the fine-structure constant $\alpha$.\footnote{The power counting of the Cherenkov rate requires some explanation. For Cherenkov radiation to happen, the dielectric function needs to be larger than one. Given that $\epsilon(\omega)$ is set by electromagnetic interactions, it would seem that $1-\epsilon(\omega) \propto \alpha$, which would make Cherenkov radiation an $\mathcal{O}(\alpha^2)$ effect. However, it turns out that in dielectrics $\epsilon(\omega)$ deviates from one at zero-th order in $\alpha$. The reason is that while at small $\alpha$ photons couple less strongly to the electrons in the material, at the same time electrons become less bound to the atoms, making the material more polarizable. These competing effects cancel out, leading to $1-\epsilon(\omega) \sim \mathcal{O}(1)$ \cite{Jackson:1998nia}. Thus,  Cherenkov radiation is in fact $\mathcal{O}(\alpha)$ process in dielectric materials.
}
This must be compared, for example, to Bremsstrahlung, 
which originates from the collisions of the charged particles with the fixed ions and therefore arises at $\mathcal{O} (\alpha^3)$~\cite{Jackson:1998nia}.
As a consequence, 
for photon frequencies where the condition Eq.~\eqref{eq:cherenkovcon} is fulfilled, 
the Cherenkov effect is expected to be one of the leading means by which charged particles radiate in the 
material.
Note that the differential rate depends on the properties of the material through its dielectric function.
However,
for materials that are strongly dielectric $|\epsilon(\omega)|\gg 1$ (within some range of photon frequencies),
and if the charged particle is fast enough, 
the condition $v^2 |\epsilon(\omega)|\gg 1$ is satisfied.
In this case,
the differential rate Eq.~\eqref{eq:cherenkov} becomes insensitive to the details of the material,
and reduces to a constant $\frac{d^2N}{d\om dx} \simeq \alpha$ that is independent of the photon frequency.
In this limit, the total number of photons emitted within a frequency interval is simply proportional to the path length travelled by the charged particle times the frequency interval. 

The angular distribution of Cherenkov photons is sharply peaked at an angle $\theta_{\textrm{Ch}}$ with respect to the charged particle's momentum vector, 
which is\footnote{Eqns.~\ref{eq:cherenkovcon}, \ref{eq:angle}, and \ref{eq:cherenkov} do not take into account corrections due to angular aberration, which are of order  $\mathcal{O}\big(\textrm{Im}\, \epom/\textrm{Re}\, \epom\big)$~\cite{grichine2002energy}.
For all practical purposes, in this work such corrections can be safely neglected,
as when estimating backgrounds at dark-matter detectors we will only be interested in Cherenkov radiation in the transparent or semi-transparent regimes where $\textrm{Re}\, \epom \gg \textrm{Im}\, \epom$.}
\bea
\label{eq:angle}
\cos \theta_{\textrm{Ch}}= \frac{\sqrt{\textrm{Re}\, \epom}}{v |\epsilon(\omega)|} \quad .
\eea 
Finally, Cherenkov radiation is linearly polarized, 
with the polarization vector lying in a direction parallel to the plane containing the charged particle and photon momentum vectors.

We conclude this section by commenting on the quantum corrections to the classical description of Cherenkov radiation discussed up to now.
Quantum corrections must be considered when the time over which radiation is emitted is smaller than the inverse photon frequency, $\omega t < 1$~\cite{PhysRev.92.1362}. 
In this case, 
quantum effects introduce a spread in the Cherenkov angle \eqref{eq:angle} and modifications to the rate Eq.~\eqref{eq:cherenkov} (the polarization remains as in the classical description).
In order to estimate the size of these corrections, consider a high-energy electron track going through $100\, \mu \textrm{m}$ of Si at the speed of light, over a time $t \sim  3 \times 10^{-13} \textrm{s} \sim 500 \, \eV^{-1}$. 
Considering emission of Cherenkov photons in the 1 eV-range, relevant for current light-dark matter detectors, this leads to $\omega t \geq 500$, 
so the classical condition $\omega t \gg 1$ is satisfied
and quantum corrections are expected to be small, of order $1/(\omega t)\sim \mathcal{O}(0.2 \%)$.
However, 
note that when taking shorter charged tracks, or Cherenkov photons with much smaller energies, quantum corrections need to be taken into account.
For instance, for photon frequencies of order $10\, \textrm{meV}$, which are relevant for future single-phonon detectors (see Sec.~\ref{subsec:future-collective}), 
quantum corrections to the classical Cherenkov formulas can be of order  $\sim \mathcal{O}(20 \%)$.

\begin{figure*}[t!]
\centering
\includegraphics[width=0.27\textwidth]{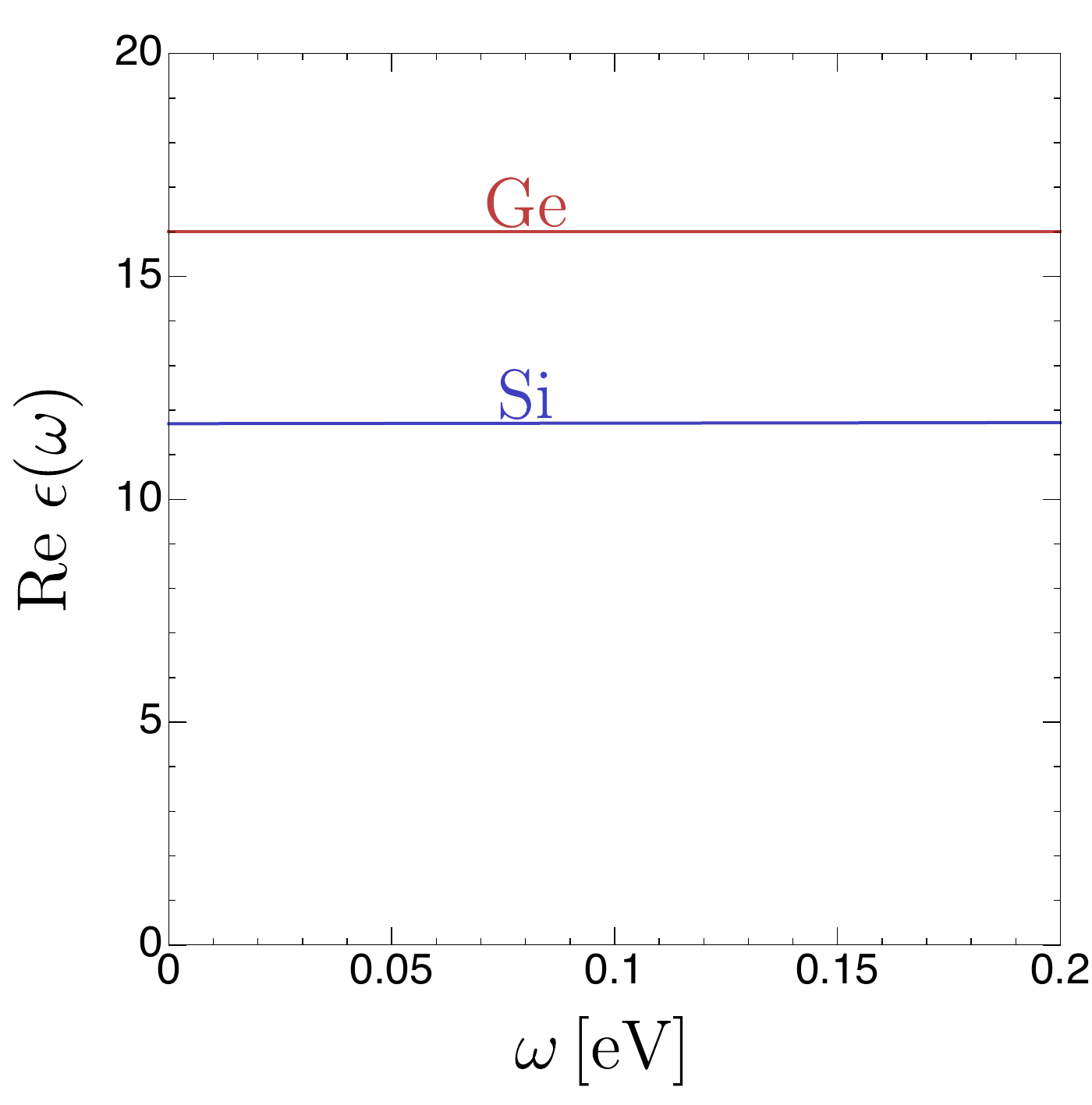}
\quad
\includegraphics[width=0.28\textwidth]{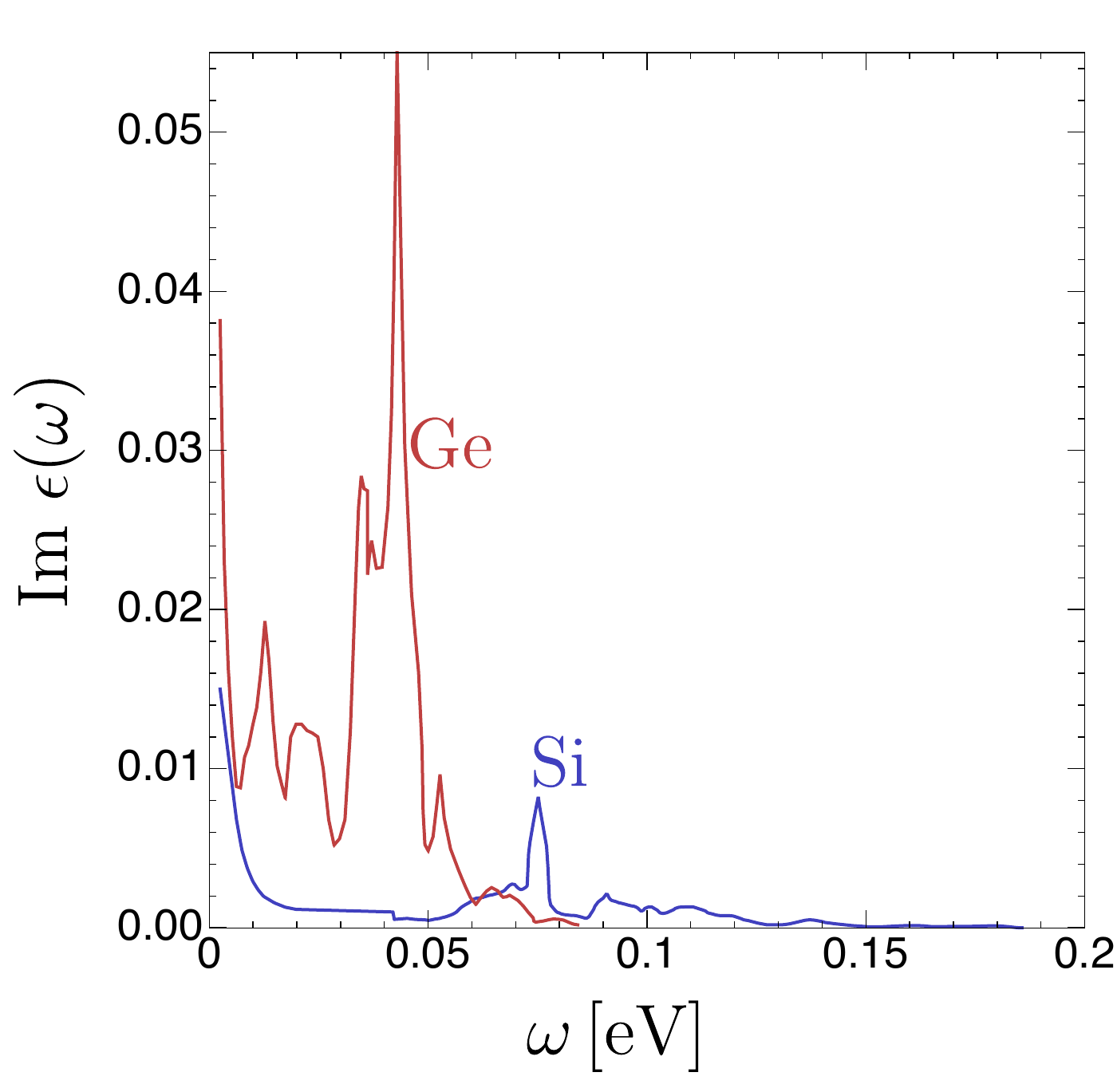}\\
\includegraphics[width=0.28\textwidth]{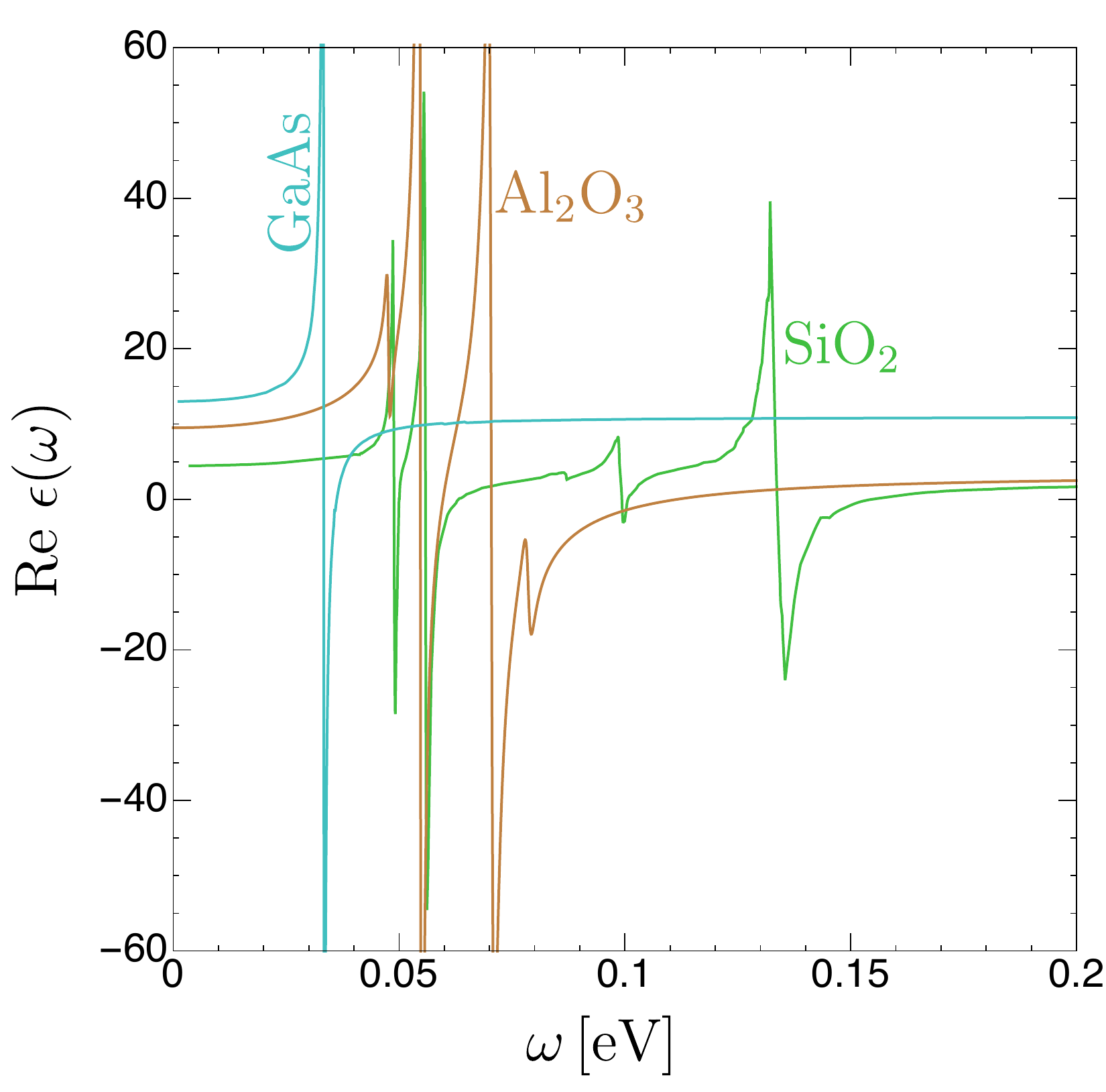}
\quad
\includegraphics[width=0.27\textwidth]{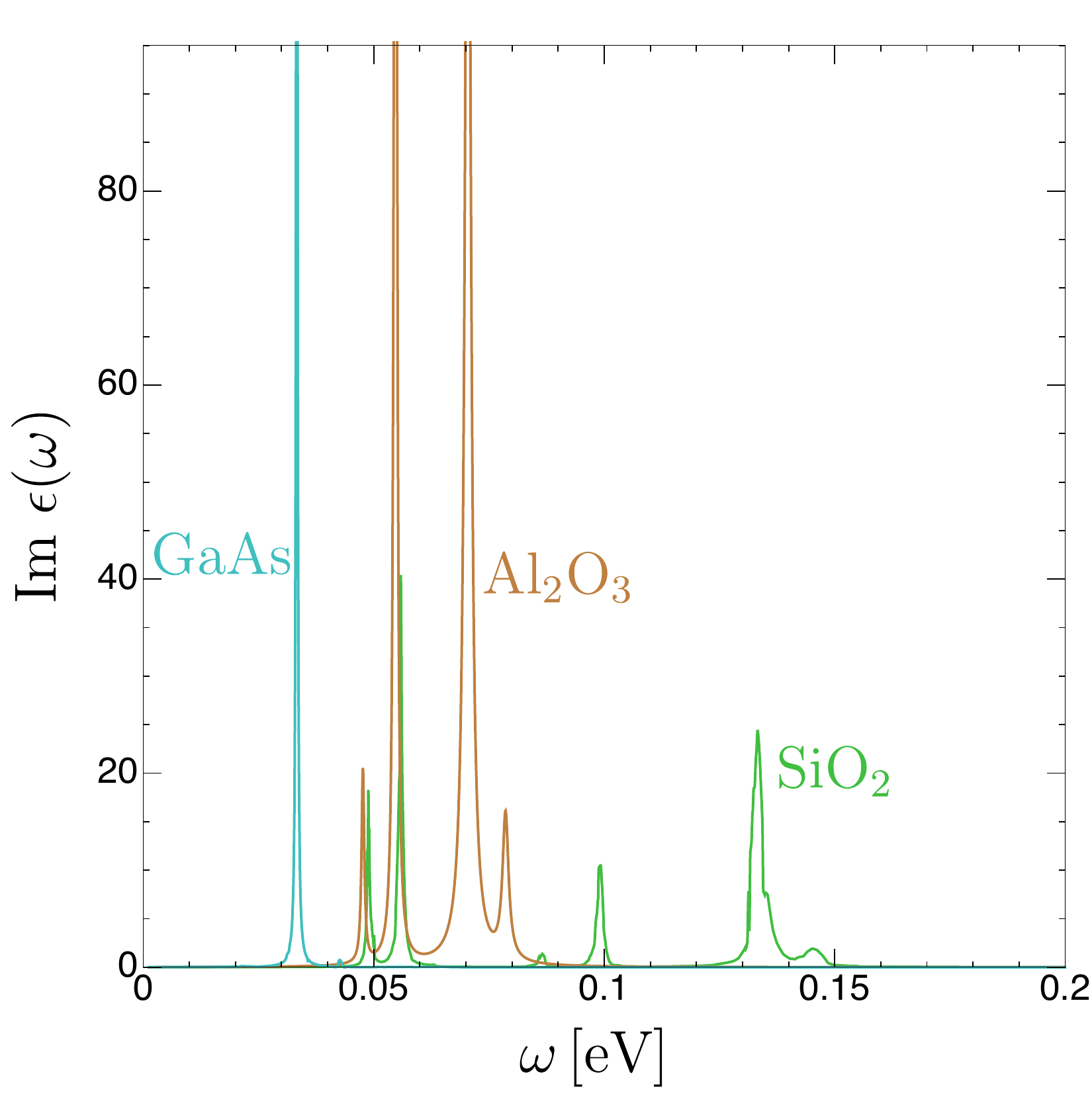}
\caption{
Real and imaginary part of the dielectric functions of Ge and Si (\textbf{top}), 
and of the polar materials $\alpha$-quartz, sapphire, and GaAs (\textbf{bottom}), 
in the range of photon energies $0\leq \omega \leq 0.2 \,\eV$.
For the case of sapphire and $\alpha$-quartz, 
which are optically anisotropic materials, we only show the dielectric function for the ordinary ray, with polarization perpendicular to the optical axis. 
All data is presented at room temperature.  
Data for Si is obtained from~\cite{1985hocs.book.....P} and~\cite{Johnson_1959}, 
for Ge from~\cite{1985hocs.book.....P} and~\cite{1954PhRv...93..674C},
for $\alpha$-quartz from~\cite{1985hocs.book.....P} and~\cite{1961PhRv..121.1324S}, 
for sapphire from~\cite{PhysRev.132.1474}, 
and for GaAs from~\cite{1985hocs.book.....P}. 
\label{fig:dielectricIR}
}
\centering
\end{figure*}


\textbf{Dielectric Properties of Detector Materials.}
In order to discuss the relevance of the Cherenkov effect for dark-matter detectors in more detail, 
the specifics of the dielectric materials in the detectors need to be studied.
While there is a variety of relevant materials, 
for concreteness let us now discuss the case of three semiconductors, germanium (Ge), silicon (Si) and gallium arsenide (GaAs), 
and two bandgap insulators, silicon dioxide $\textrm{SiO}_2$  ($\alpha$-quartz) and aluminum oxide $\textrm{Al}_2 \textrm{O}_3$ (sapphire). 
Spanning more than an order of magnitude in their room-temperature bandgaps, with $E_g^{\textrm{Ge}}=0.66\,\eV$, $E_g^{\textrm{Si}}=1.11\,\eV$, 
 $E_g^{\textrm{GaAs}}=1.43\,\eV$,
$E_g^{\textrm{SiO}_2}=8.9\,\eV$, and $E_g^{\textrm{A}_2\textrm{O}_3}=8.8\,\eV$~\cite{VARSHNI1967149,doi:10.1111/j.1151-2916.1990.tb06541.x,DISTEFANO19712259},
these materials are representative of a wide range of non-conducting media, 
and allow us to discuss Cherenkov radiation and photon absorption with enough generality. 
These materials can be found in a variety of current and planned experiments, 
such as SENSEI~\cite{Tiffenberg:2017aac}, SuperCDMS~\cite{Agnese:2014aze}, EDELWEISS~\cite{Arnaud:2020svb}, DAMIC~\cite{Barreto:2011zu}, and proposed athermal phonon detectors~\cite{Griffin:2019mvc}. 

\begin{figure*}[t!]
\centering
\includegraphics[width=\mywidth]{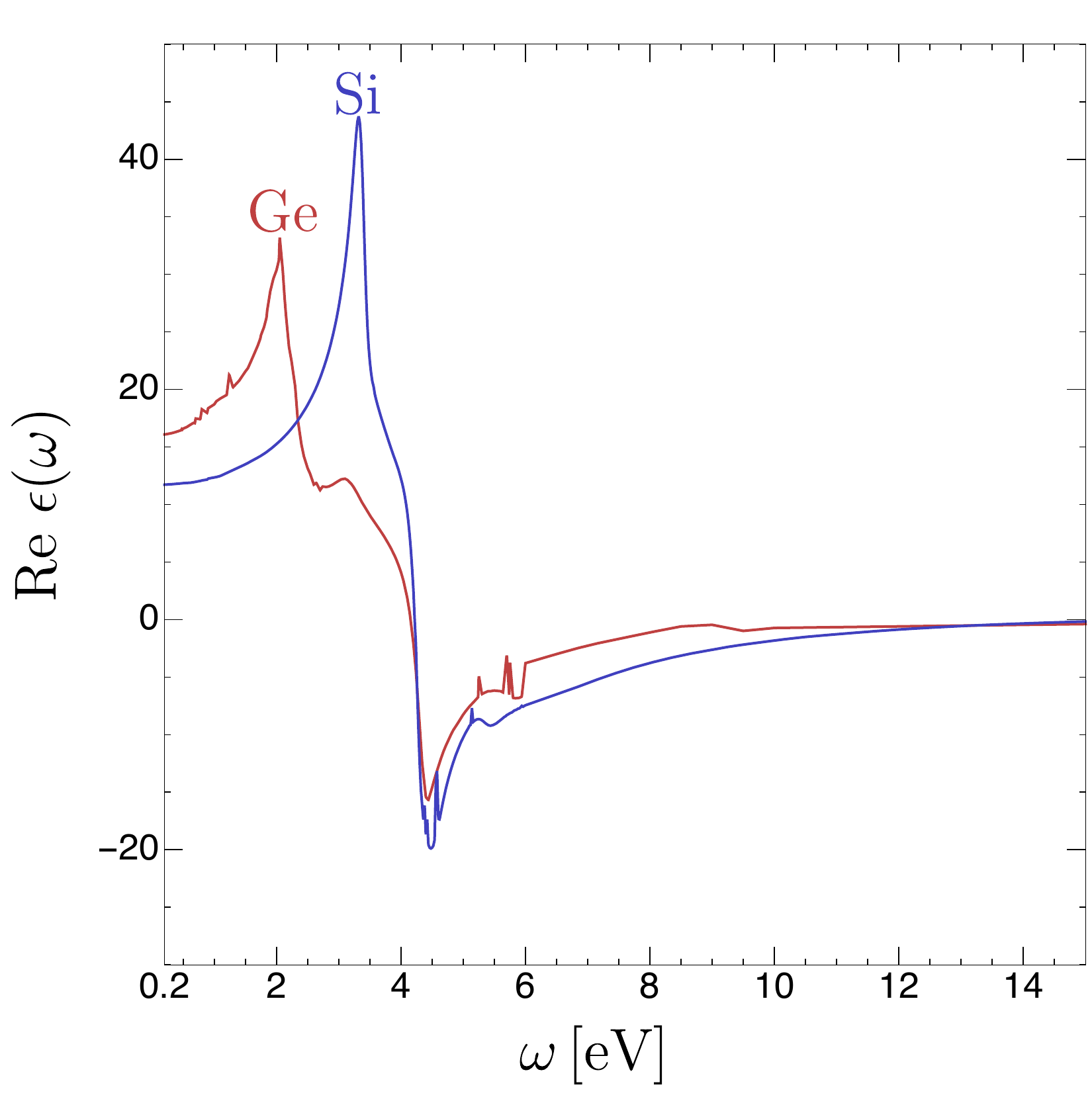}
\quad
\includegraphics[width=\mywidth]{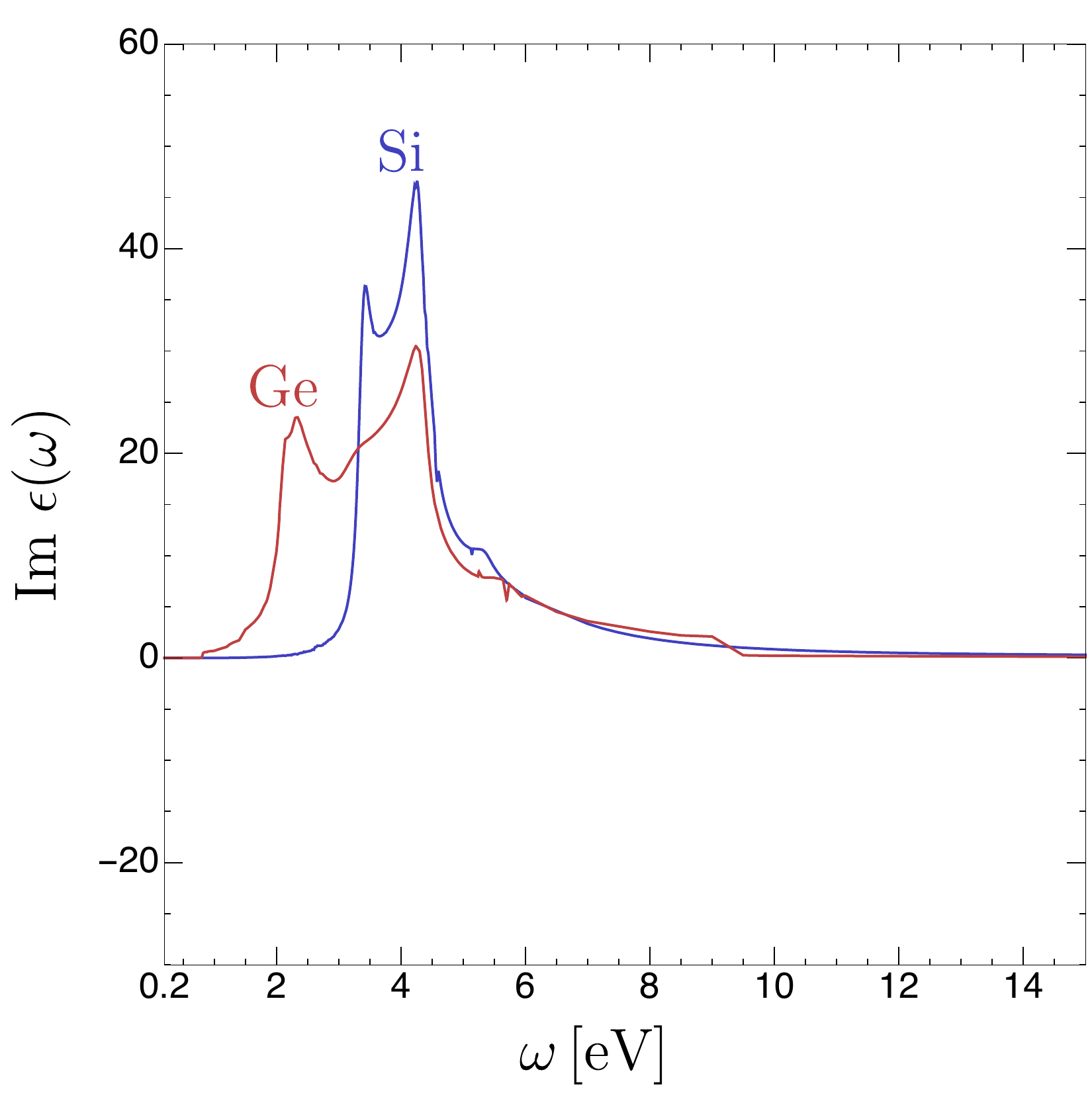}\\
\includegraphics[width=\mywidth]{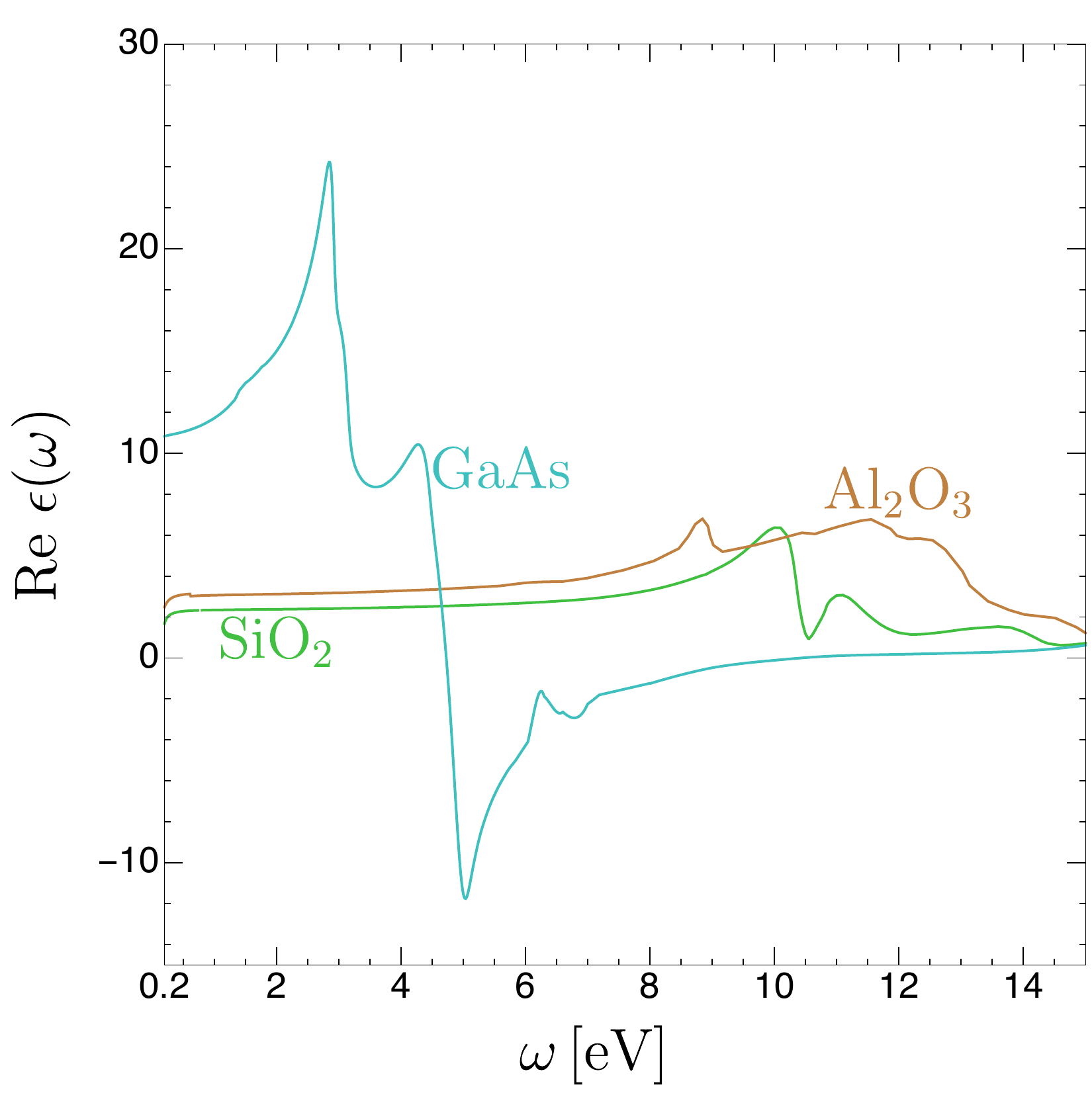}
\quad
\includegraphics[width=\mywidth]{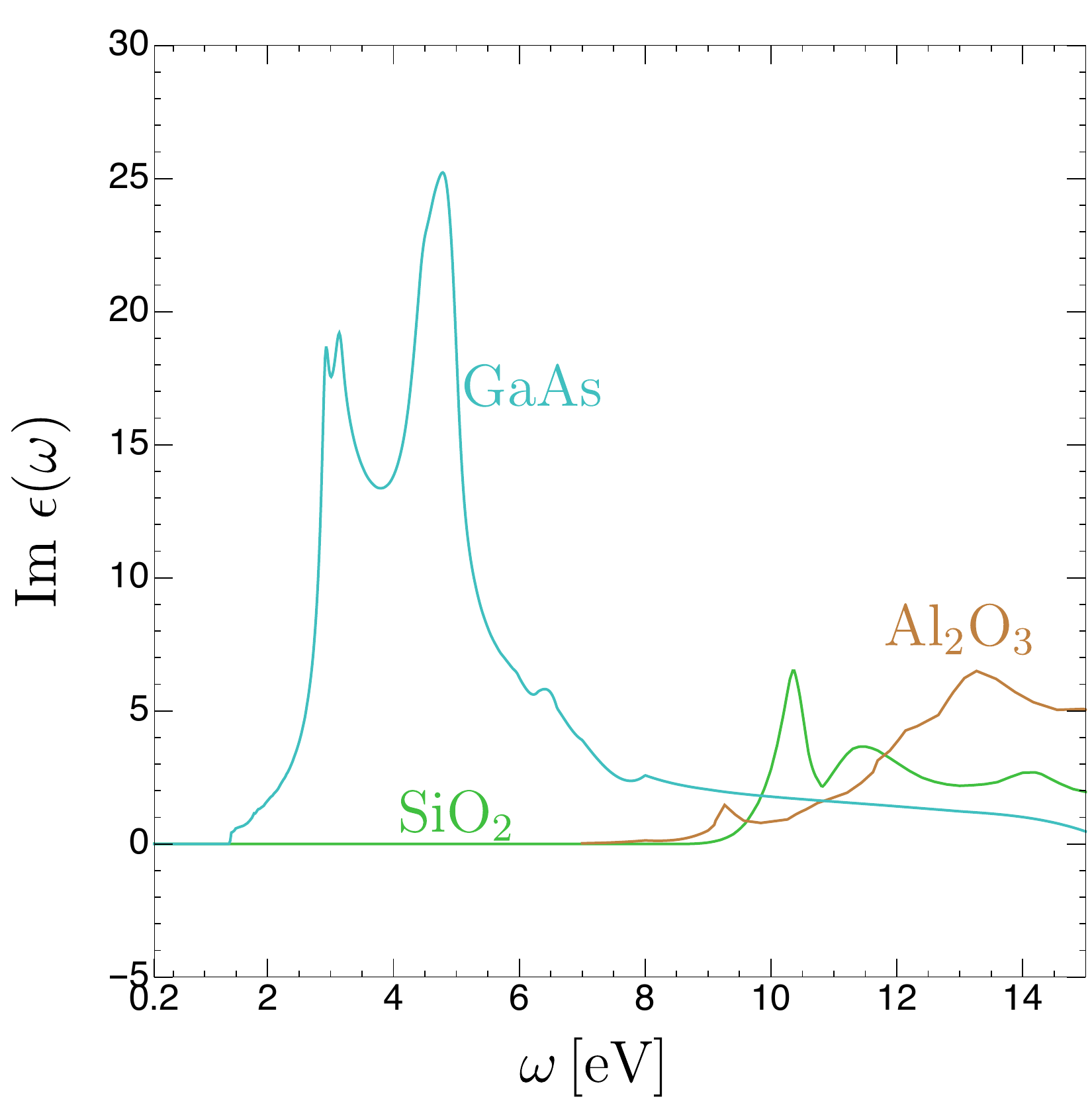}
\caption{
Real and imaginary part of the dielectric functions of Ge and Si (\textbf{top}), 
and of the polar materials $\alpha$-quartz, sapphire, and GaAs (\textbf{bottom}), 
in the range of photon energies $0.2\leq \omega \leq 15 \,\eV$.
For the case of sapphire and $\alpha$-quartz we show the dielectric function for the ordinary ray.
All data is at room temperature.  
Data for Si, Ge, $\alpha$-quartz, and GaAs is obtained from~\cite{1985hocs.book.....P}. 
Data for sapphire is obtained from~\cite{1985hocs.book.....P} and~\cite{doi:10.1143/JPSJ.62.1372}. 
\label{fig:dielectric}
}
\centering
\end{figure*}

The room-temperature, experimentally measured dielectric functions for our benchmark materials
are shown in Figs.~\ref{fig:dielectricIR} and~\ref{fig:dielectric}. 
We separately plot the dielectric functions for energies above and below $\omega=0.2 \, \eV$ to highlight the different processes relevant at high and low frequencies:
at small energies, 
the dielectric behavior of the materials is set by the lattice vibrations and electronic dynamics within each band,
while at higher energies electronic inter-band transitions become relevant~\cite{2005fspm.book.....Y}.
Our materials can be further classified in two groups:
polar and non-polar.
Sapphire, $\alpha$-quartz, and GaAs are polar materials (\textit{i.e.}, materials that have oppositely charged ions in the primitive cells).
We present their dielectric functions in the bottom panels of Figs.~\ref{fig:dielectricIR} and~\ref{fig:dielectric}.
Ge and Si are non-polar, and their dielectric functions are shown in the upper panels of the same figures.

Polar materials have phonons that react strongly to light due to dipole interactions. 
The effect of these interactions in the dielectric function is evident at energies 
$\omega \leq0.2\,\eV$ (\textit{c.f.}~Fig.~\ref{fig:dielectricIR} bottom-left panel),
where we see strong peaks on both the real and imaginary part of the dielectric functions,
which are due to resonant conversion of photons into single optical phonons. 
Non-polar materials do not have optically active phonon modes that can be resonantly produced from photon absorption.
Thus, 
such materials respond more weakly to light and correspondingly,
they only present small peaks in $\textrm{Im} \epsilon(\omega)$ (\textit{c.f.}~Fig.~\ref{fig:dielectricIR} upper-right panel),
which are due to multi-phonon processes.
It is also relevant to point out that in the infrared, multi-phonon processes allow dielectrics to be absorbing up to energies that are considerably larger than single-optical phonon energies. For instance, while optical phonons in Si have energies close to $60\,\textrm{meV}$, three-phonon processes involving one transverse-optical (TO) and two longitudinal-optical (LO) phonons in Si allow for photon absorption up to at least $\sim 160\,\textrm{meV}$~\cite{2005fspm.book.....Y}. 

In the range of energies above the lattice modes but much below the electronic bandgaps of all our dielectric materials,
$0.2 \, \eV \lesssim \omega \ll E_g$,
their dielectric functions are approximately real and constant. 
This constant is called the high-frequency dielectric function, $\epsilon_{\infty}$.
The high-frequency dielectric function depends mostly on the electron dynamics within bands,
and for our different materials is $\epsilon_{\infty}^{\textrm{Si}}=12$, $\epsilon_{\infty}^{\textrm{Ge}}=16$ ,$\epsilon_{\infty}^{\textrm{SiO}_2}=2.4$, $\epsilon_{\infty}^{\textrm{A}_2\textrm{O}_3}=3.1$, and $\epsilon_{\infty}^{\textrm{GaAs}}=16$
  \cite{PhysRevLett.20.550,1985hocs.book.....P}. 
Note that $\epsilon_{\infty}$ is larger in Si, Ge, and GaAs than in quartz and sapphire. 
The reason is that Si, Ge and GaAs are covalent materials, 
where electrons are shared between atoms.
Since these electrons are loosely bound to the lattice,
the material is then highly polarizable, leading to their large $\epsilon_{\infty}$~\cite{1977PhT....30R..61A}.
Quartz and sapphire, on the other hand, have ionic bonds, in which case electrons are strongly bound to each oppositely charged ion and the material is hard to polarize, leading to small $\epsilon_{\infty}$.
The larger $\epsilon_{\infty}$ in Si, Ge, and GaAs can also be explained by means of their small bandgaps.
In materials with small bandgaps, higher order perturbation theory effects on photon propagation in the material come with small-energy denominators, which lead to more polarizability~\cite{1977PhT....30R..61A}.
Note also that in the range $0.2 \, \eV \lesssim \omega \ll E_g$, 
$\textrm{Im}\,\epom \simeq 0$ so all our materials are transparent, 
as photons are too energetic to convert into either a single or multiple phonons, 
but too soft to excite electrons to the conduction band.

For higher photon frequencies, above the bandgap energies of each material $\omega > E_g$,
direct and indirect electronic inter-band transitions become energetically allowed.
Correspondingly, in Fig.~\ref{fig:dielectric}, we observe strong absorption peaks in $\textrm{Re}\,\epom$ and  $\textrm{Im}\,\epsilon(\omega)$ due to these transitions.

Throughout this section, we have presented optical data at room temperature, 
which we found to be most widely available for different materials at different photon energies.
Lowering the material's temperature has two effects on the dielectric functions. 
First, it leads to $\mathcal{O}(1)$ changes in the values of the dielectric constant itself.
For instance, in the infrared band, the dielectric function of Si decreases by a factor of $\sim$2 when lowering the temperature from room-temperature to 20~K~\cite{Johnson_1959}.
Second, it changes other material properties that affects its optical behavior, such as the bandgap. 
A change in the bandgap lead to shifts of the dielectric function's dependence on photon frequency.
A discussion on the dependence of the bandgap energy of semiconductors can be found in~\cite{VARSHNI1967149}, 
while references to measurements of dielectric properties of materials at different temperatures can be found in~\cite{1985hocs.book.....P}.


\textbf{Cherenkov Radiation in Detector Materials.}
Inspecting Figs.~\ref{fig:dielectricIR} and \ref{fig:dielectric}, 
we see that in all our materials there are several ranges of photon energies with $\textrm{Re}\,\epom > 1$, 
where Cherenkov radiation due to the passage of charged particles is possible. 
In Si and Ge, 
Cherenkov photons can be emitted with energies $\omega \lesssim 4 \, \eV$.  
In GaAs, photons are emitted with $\omega \lesssim 5 \, \eV$, 
with the exception of a very narrow region of parameter space right above its optical phonon resonant mode, around $\omega \sim 0.035 \, \eV$, where the material becomes very absorptive.
A material becoming absorptive results in a decrease of the real part of the dielectric function (related to the extinction coefficient $\kappa$ by $\textrm{Re}\,\epom=n^2-\kappa^2$, \textit{c.f.}~Eq.~\eqref{eq:epom}), 
leading to the breaking of the necessary condition for Cherenkov radiation $\textrm{Re}\,\epom > 1$.
In $\alpha$-quartz and sapphire, the allowed energy ranges are $ \omega \lesssim 14\,\eV$ and $\omega \lesssim 15\,\eV$, respectively, 
excepting narrow strips of energies above optical modes. 
The maximal frequency at which Cherenkov radiation can be emitted in a material is  strongly correlated with its bandgap, since for energies above the bandgaps, 
materials become strongly absorptive due to inter-band transitions, 
and the Cherenkov condition is thus broken. 
As a consequence, 
insulators, which have large bandgaps, 
generically allow for more energetic Cherenkov radiation than semiconductors.

\begin{figure*}[t!]
\centering
\includegraphics[width=\mywidth]{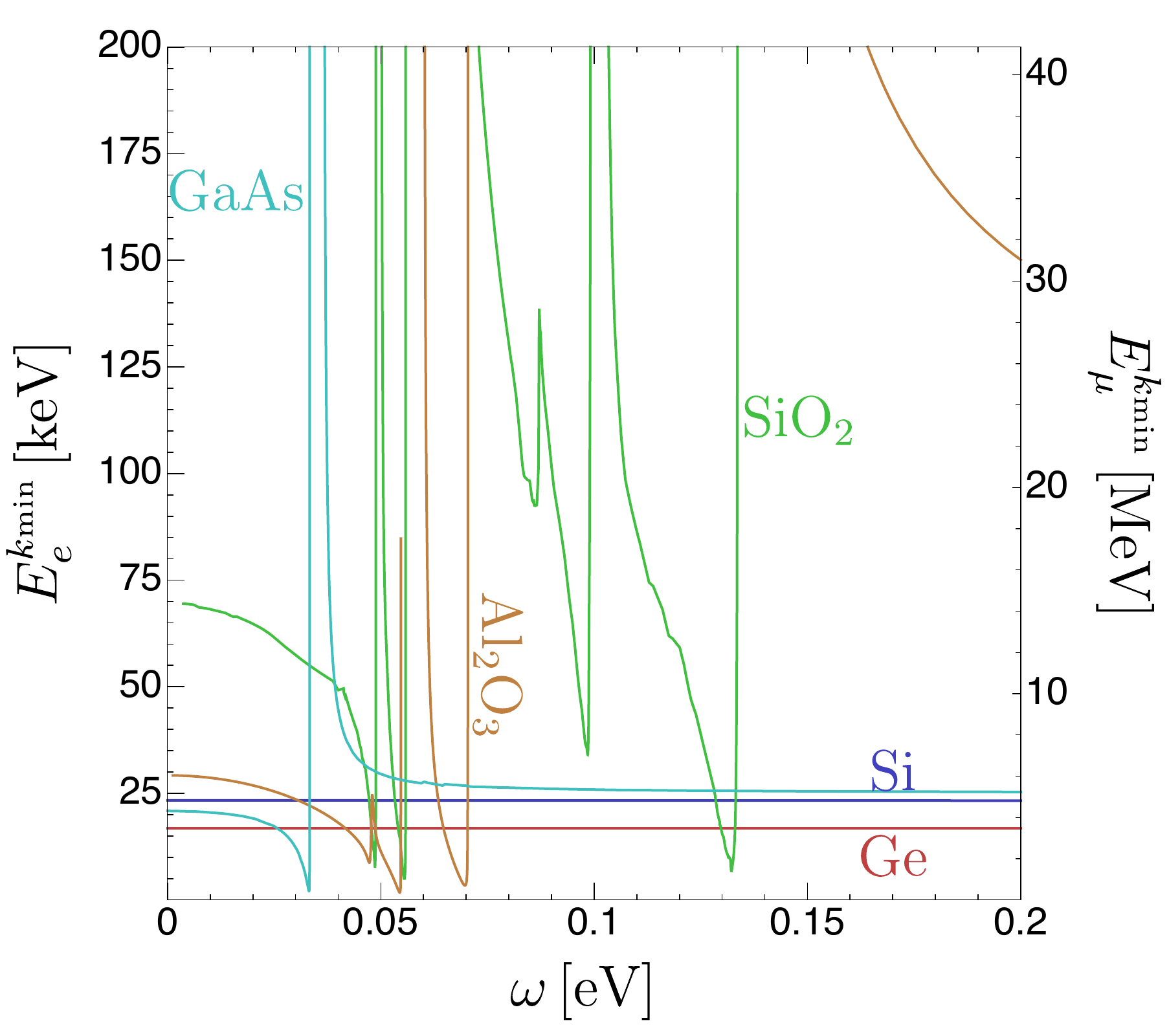}
\quad
\includegraphics[width=\mywidth]{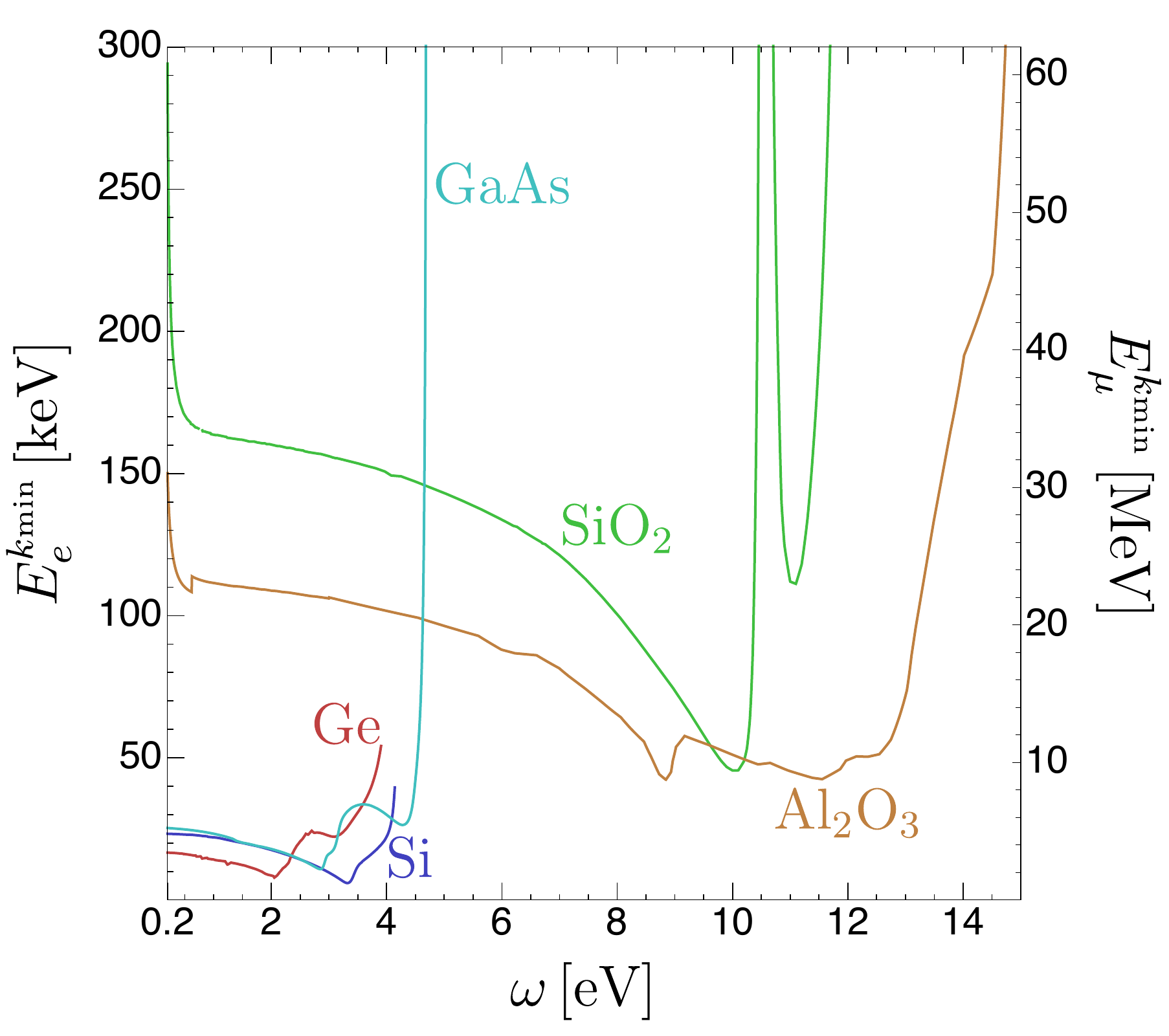}
\caption{
Minimal electron and muon kinetic energies required to emit Cherenkov photons of frequency $\omega$ obtained from Eq.~\eqref{eq:cherenkovcon}, in the frequency ranges $0 \leq \omega \leq 0.2\,\eV$ (\textbf{left}) and $0.2\, \leq \omega \leq 15\,\eV$ (\textbf{right}), 
in Si, Ge, $\alpha$-quartz, sapphire and GaAs.
The jumps in the minimal kinetic energies in the range $\omega\leq 0.2\,\eV$ for quartz, sapphire, and GaAs are due the strong  dielectric response of polar materials for photon energies close to the lattice modes. 
The jump in the minimal energy for quartz at $\sim 11 \, \eV$ is due to an accidental coincidence of the refraction index and the extinction coefficient at those energies, $n\sim\kappa$, 
that leads to a small value of $\textrm{Re}\,\epom$, \textit{c.f.}~Eq.~\eqref{eq:epom}. 
\label{fig:Ekel}
}
\centering
\end{figure*}

\begin{figure*}[th!]
\centering
\includegraphics[width=\mywidth]{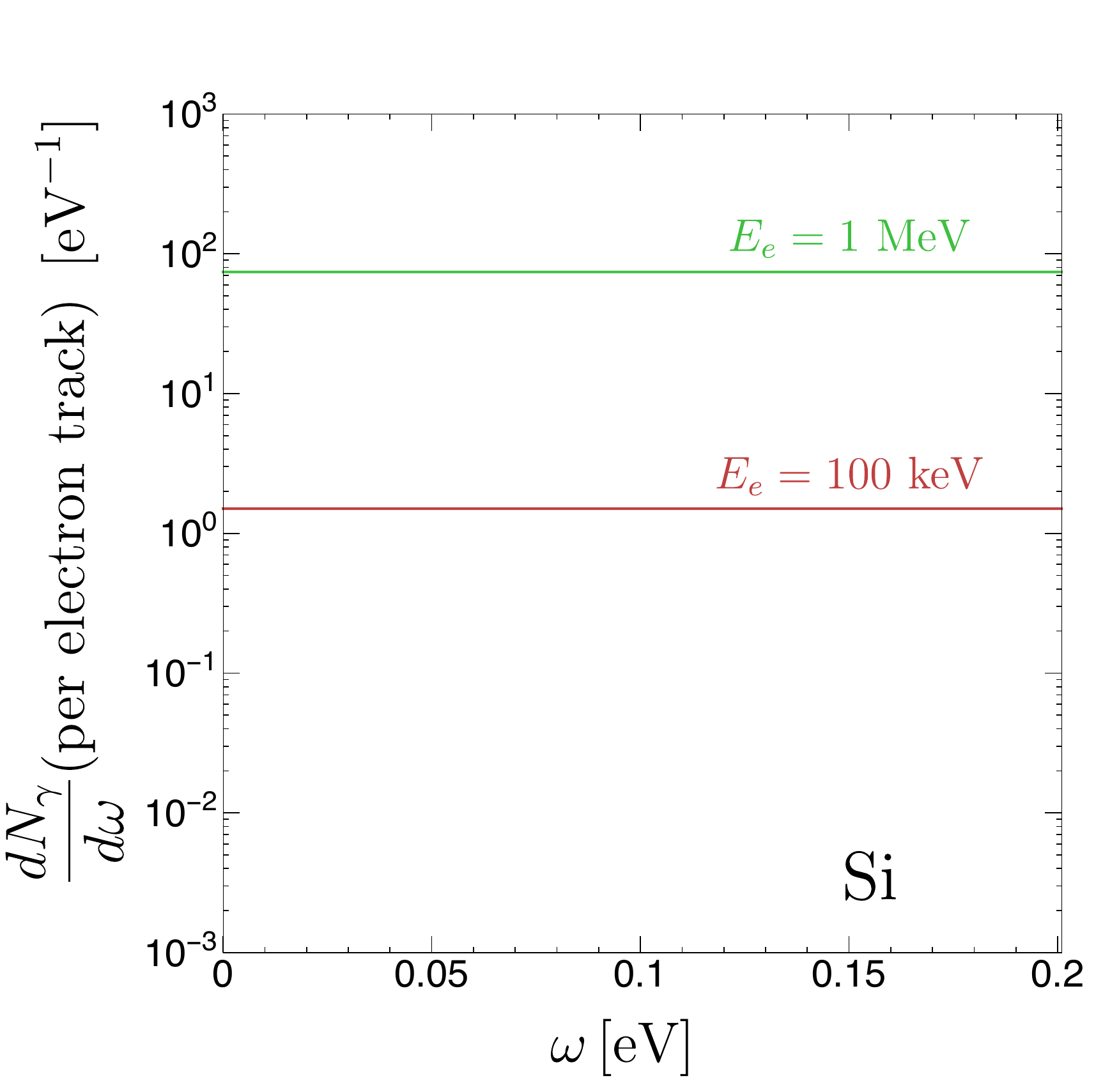}
\quad\quad
\includegraphics[width=\mywidth]{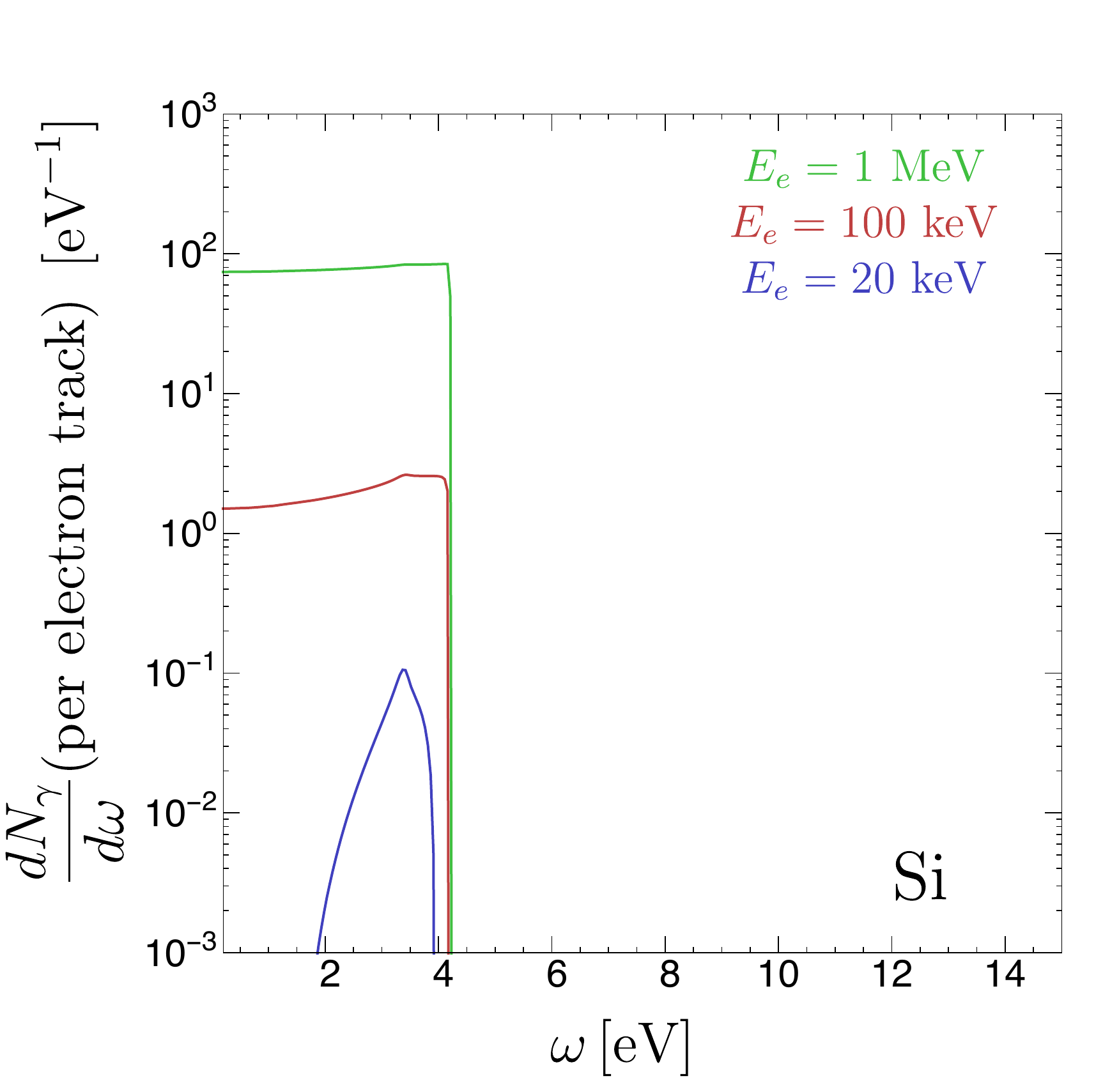}\\
\includegraphics[width=\mywidth]{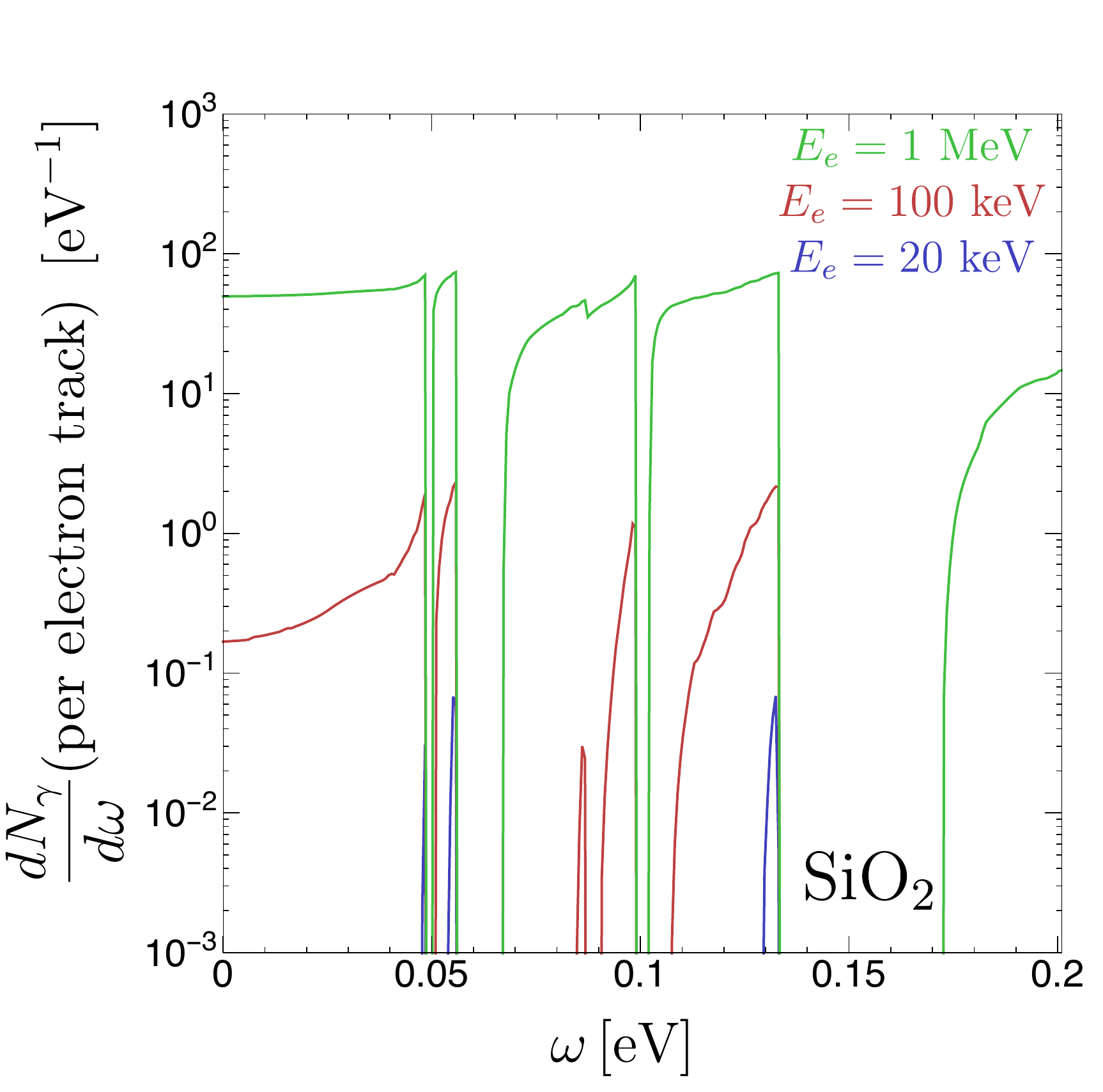}
\quad\quad
\includegraphics[width=\mywidth]{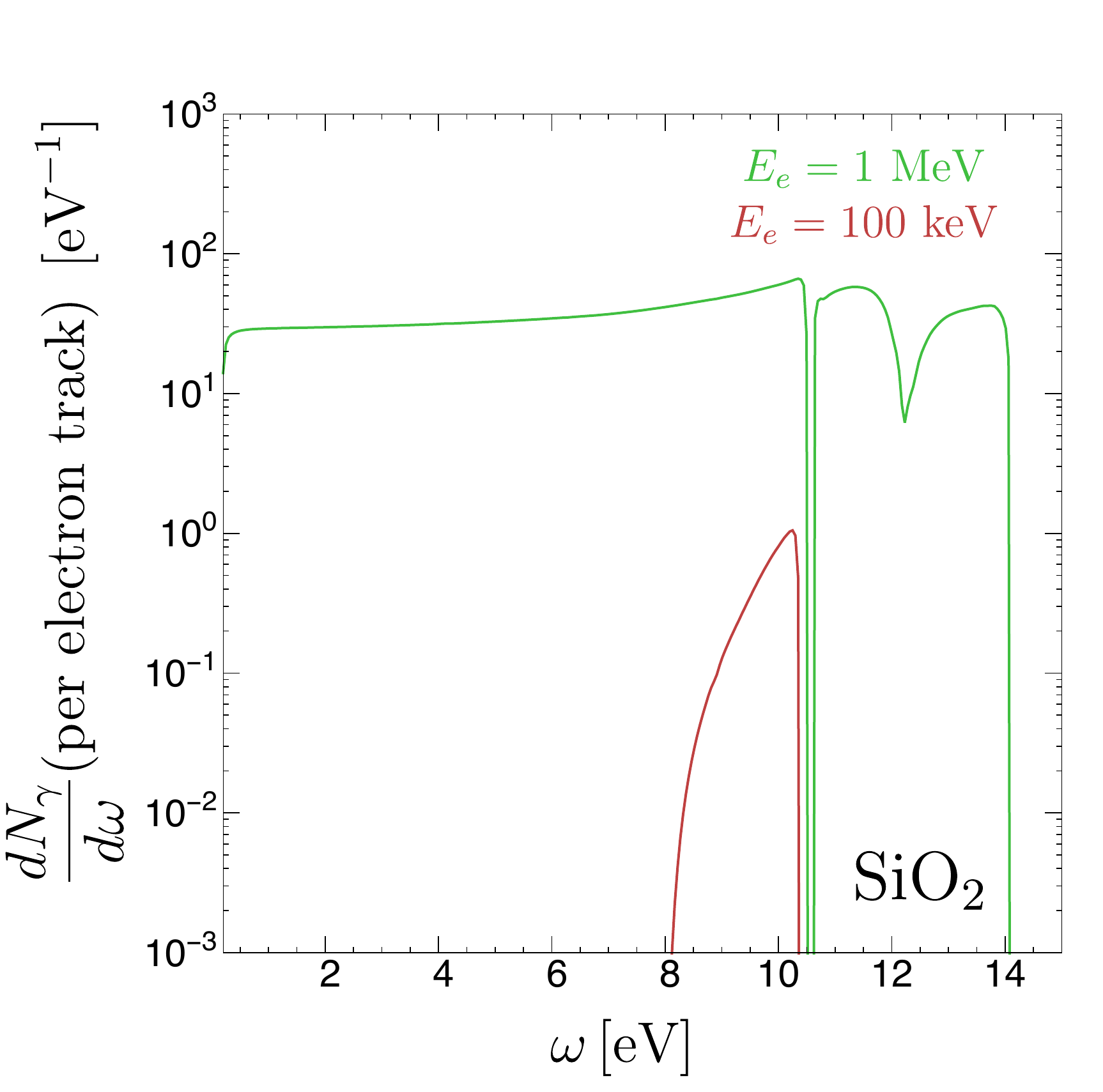}
\caption{
Cherenkov radiation spectrum Eq.~\ref{eq:singletrackspectrum} of a single electron track in Si (\textbf{top}) and $\textrm{SiO}_2$ (\textbf{bottom}). 
We show the emission spectrum for three different electron energies, $E_e=20 \kev$ (blue), $E_e=100 \kev$ (red), and $E_e=1 \textrm{MeV}$ (green).  For Si, there is no emission above $\sim$4~eV, since Si becomes highly absorptive. The abrupt changes in the spectra for $\textrm{SiO}_2$ at low energies correspond to peaks in its dielectric function due to a strong optical phonon response. 
\label{fig:spectrum}} 
\centering
\end{figure*}

Given $\textrm{Re}\,\epom > 1$, 
a minimal kinetic energy for electrons and muons is required in order for them to emit Cherenkov photons as they pass through the dielectrics, 
which can be obtained from Eq.~\eqref{eq:cherenkovcon}. 
We present the minimal energies in Fig.~\ref{fig:Ekel} for $\omega \leq 0.2\,\eV$ (left panel) and $\omega > 0.2\,\eV$  (right panel), 
for both electrons and muons.
In Si and Ge, electrons (muons) with energies as low as $\sim$10~keV ($\sim$2~MeV) 
emit Cherenkov radiation, 
while the threshold energies required in $\alpha$-quartz and sapphire are significantly larger, 
due the comparatively smaller dielectric function of these materials
(except in narrow strips of energies close to the lattice modes).

In order to illustrate the Cherenkov radiation rate, spectrum, and the dependence on the energy of the track in concrete scenarios, 
let us consider the case of an electron with initial energy $E_e$ passing through a piece of Si or $\alpha$-quartz.
An electron passing through a material loses energy at a rate set by the stopping power, $dE/dx$, which we obtain from \cite{NIST}. 
Thus, the spectrum of Cherenkov photons emitted by the electron before being stopped in the material is given by 
\begin{equation}
\frac{dN_\gamma}{d\omega}({\textrm{per electron track}}) = \int_{E_{\textrm{min}}}^{E_e}dE  \frac{d^2N_\gamma}{d\omega dx} \bigg[\frac{dE}{dx}\bigg]^{-1}  \quad ,
\label{eq:singletrackspectrum}
\end{equation}
where $\frac{d^2N_\gamma}{d\omega dx}$ is given by Eq.~\eqref{eq:cherenkov}.
Note that as the electron loses energy in the material, at some point its velocity falls below the threshold condition Eq.~\eqref{eq:cherenkovcon}
and no more Cherenkov radiation is emitted, so the integral is cutoff at the minimum energy $E_{\textrm{min}}$ that satisfies the aforementioned condition.

We present the spectrum for a single electron track for Si and $\alpha$-quartz in Fig.~\ref{fig:spectrum} for different initial energies of the electron.
Consider first the case of a relativistic electron with initial energy $E_e=1\, \textrm{MeV}$.
From the figures, we see that in this case Cherenkov radiation is emitted in a wide range of energies in both Si and $\alpha$-quartz.
As discussed above, the main difference between the two materials is that more energetic radiation is possible in $\alpha$-quartz than Si, given its larger bandgap.
In addition, at low energies $\omega \lesssim 0.2 \,\textrm{eV}$, we clearly see the effect of the optically active lattice modes of $\alpha$-quartz on the radiation spectrum. 
In particular, right above the material's lattice modes, where the material becomes absorptive, the emission of Cherenkov radiation is suppressed. 
The spectrum in Si, on the other hand, is rather flat for all energies below the bandgap.

\begin{figure}[t!]
\centering
\includegraphics[width=\mywidth]{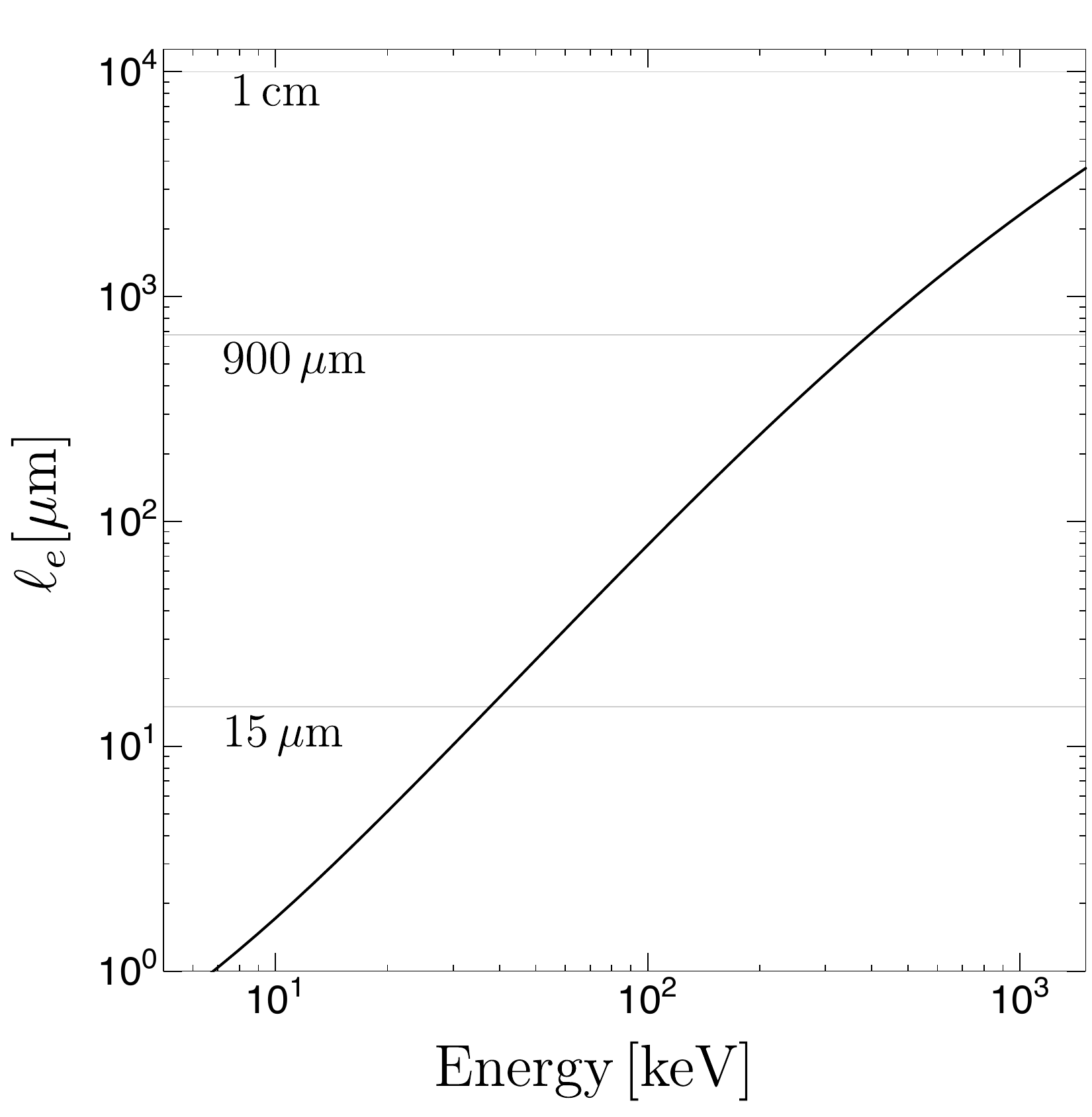}\\
 \caption{Electron mean range Eq.~\eqref{eq:meanrange} in Si, obtained from~\cite{NIST}.
Horizontal lines indicate an electron mean range of 15~$\mu$m and 900~$\mu$m, which corresponds to the size of $1$ pixel and 60~pixels, respectively, of the Skipper-CCDs used by SENSEI (see Sec.~\ref{sec:experiments-current}). 
 \label{fig:meanrange}}
 \centering
 \end{figure}

As we go towards lower electron energies, we observe two important effects. 
First and most importantly, we see a decrease of the Cherenkov radiation rate. 
The reason is that electrons with smaller initial energies have shorter track lengths,
and thus lead to less radiation. 
The length of an electron track can be characterized by its mean range, defined as the distance over which the electron travels before losing all its energy due to collisions,
\begin{equation}
\ell_e = \int_0^E \frac{dE}{dE/dx} \quad .
\label{eq:meanrange}
\end{equation}
We show the electron mean-range in Si in Fig.~\ref{fig:meanrange} (for other materials the track-length is simply inversely proportional to their density), where we clearly see that less-energetic electrons have shorter tracks.

The second effect of the electron track energy on the Cherenkov spectrum, 
is that when considering less-energetic electrons, 
materials either do not allow Cherenkov radiation, or they do it only in specific frequency ranges where $\textrm{Re}\,\epom$ is large enough so that the Cherenkov condition is satisfied despite the electron's small velocity. 
 For instance, for $E_e=20\,\textrm{keV}$, Cherenkov photons are emitted in $\alpha$-quartz only at energies right below its lattice modes $\omega \sim 0.05 \, \eV$ and $\omega \sim 0.13 \, \eV$, 
 where the dielectric function is both very large due to the dynamics of its polar lattice, 
and the material is not strongly absorptive (differently from the case of energies right above the lattice modes discussed before).

\begin{figure}[t!]
\centering
\includegraphics[width=\mywidth]{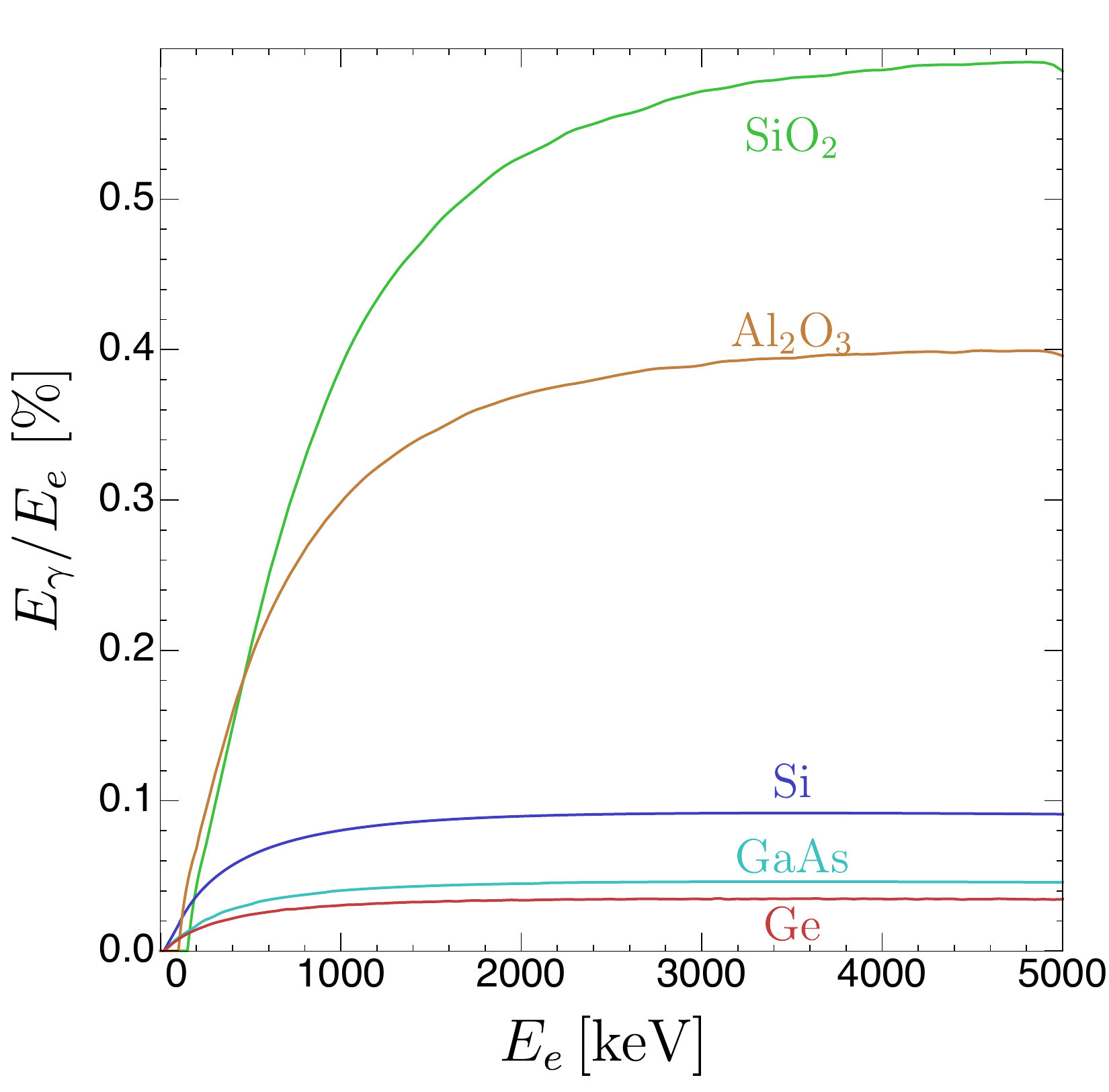}\\
 \caption{Percentage of energy lost as Cherenkov radiation by an electronic track with energy $E_e$ in different materials. The energy lost as Cherenkov radiation is obtained by calculating the total energy radiated into Cherenkov photons up to the point where the track is stopped due to interactions with the material.
  \label{fig:fractionenergy}}
 \centering
 \end{figure}

It is important to note that even if Cherenkov photons are emitted abundantly by energetic charged particles going through dielectrics, only a very small fraction of the track's energy is lost to Cherenkov radiation. To illustrate this point, in Fig. \ref{fig:fractionenergy} we show the percentage of the energy lost to Cherenkov radiation in different materials, for an electron with initial kinetic energy $E_e$ that travels in the material until losing all its energy (the situation for muons is qualitatively similar). We observe that in all materials, less than a percent of the electron's energy is lost to Cherenkov radiation. 
As we will discuss in Sec.~\ref{sec:recombination}, for the plotted range of energies, most of the electron's energy is instead lost to ionization, \textit{i.e.}, in the creation of electronic excitations in the material. Energy lost to radiation via other processes, such as bremsstrahlung, is also sub-leading for the plotted range of energies~\cite{PhysRevD.98.030001}. 

\begin{figure*}[t!]
\centering
\includegraphics[width=\mywidth]{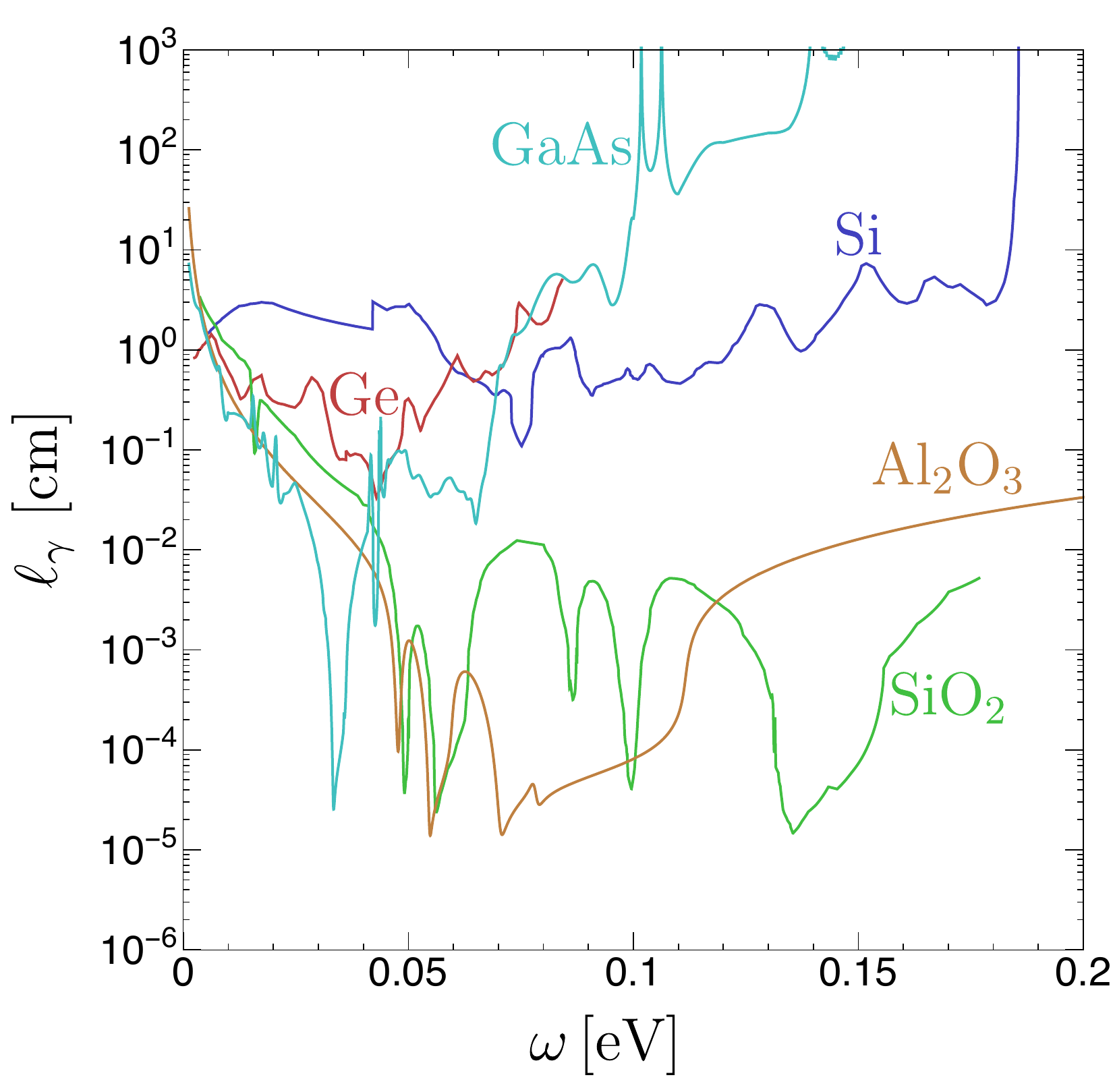}
\quad
\includegraphics[width=\mywidth]{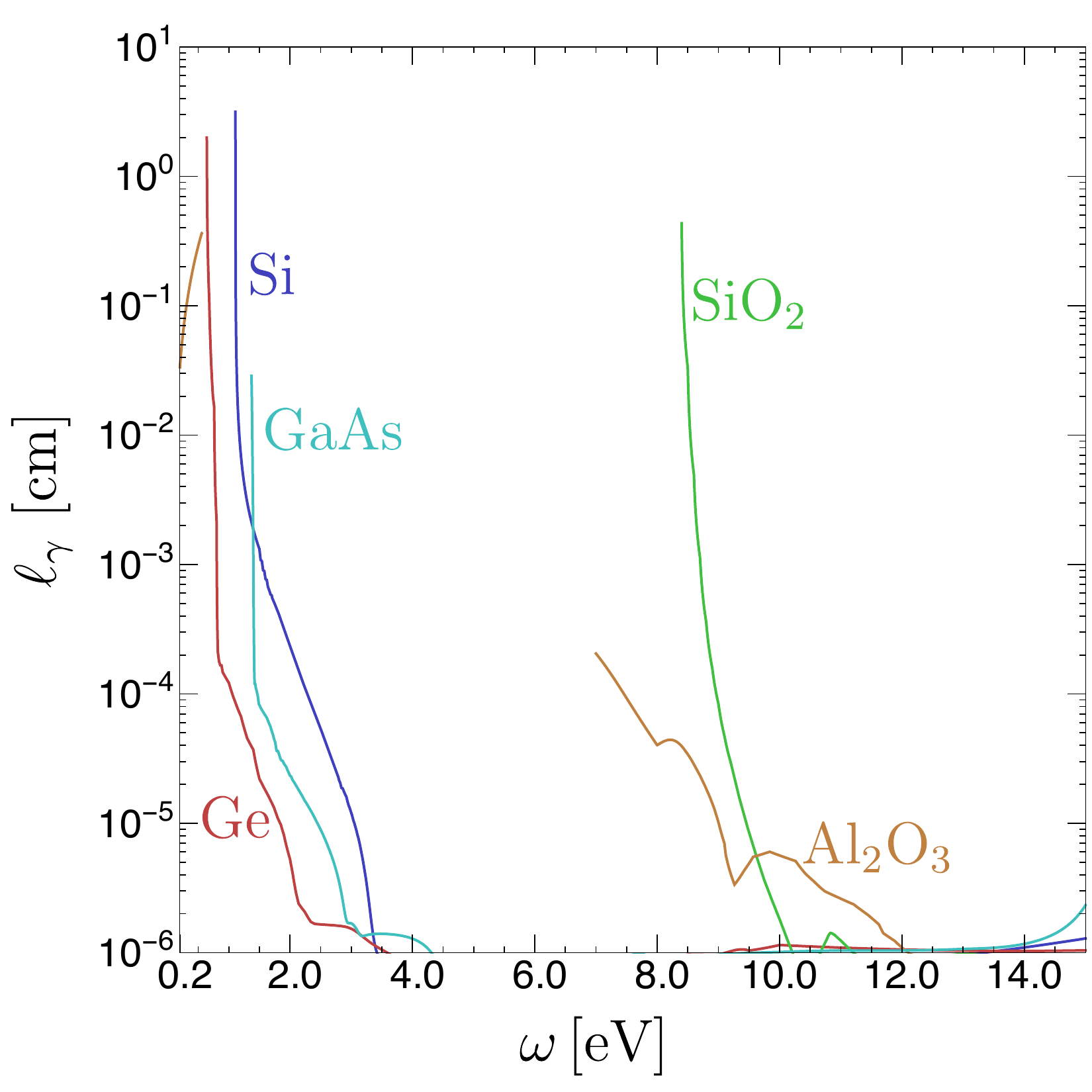}
\caption{Photon attenuation length, Eq.~\eqref{eq:photonmfp}, for photon energies $0\leq \omega \leq 0.2 \, \eV$ (\textbf{left}) and $0.2\leq \omega \leq 15 \, \eV$ (\textbf{right}), 
in different materials at room temperature.
Data has been taken from the same references given in Figs. \ref{fig:dielectricIR} and \ref{fig:dielectric}.
At energies very close to each material's bandgap the attenuation lengths presented in this figure must be only taken as approximate.
Precise band-edge absorption coefficients can be found in ~\cite{macfarlane1958fine,macfarlane1957fine,sturge1962optical,french1990electronic,godmanis1983exciton}. 
For some materials, lack of data does not allow us to indicate precise photon attenuation lengths at some energies, so gaps in the figures are noticeable. 
For instance, for Ge a gap in coverage in the range $0.2\, \textrm{eV} \lesssim \omega \lesssim 0.6 \, \textrm{eV}$ is left between the data measured in \cite{1954PhRv...93..674C} and \cite{dash1955intrinsic}.
\label{fig:mfpIR}}
\centering
\end{figure*}

After the Cherenkov photons are emitted, the distance that they travel from the charged track before being absorbed is set by the photon attenuation length $\ell_\gamma$,
\begin{equation}
\ell_\gamma \equiv \frac{1}{2\kappa\omega}= \frac{1}{\omega \sqrt{2|\epom|-2\textrm{Re}\, \epom}} \,,
\label{eq:photonmfp}
\end{equation}
where $\kappa$ is defined in Eq.~\eqref{eq:refraction} and  in the last equality we made use of Eq.~\eqref{eq:epom}. 
When $\textrm{Im}\,\epom=0$, the attenuation length is infinite, and the material is non-absorptive.
We present the attenuation lenghts for the different materials in Fig.~\ref{fig:mfpIR} for $\omega \leq 0.2 \, \eV$ (left panel) and $\omega \geq 0.2\,\eV$ (right panel).
For energies $\lesssim 0.2\,\eV$, 
in the range of the lattice modes,
photons may convert into phonons in distances as short as fractions of $\mu$m in $\alpha$-quartz and sapphire.
By contrast, in Ge and Si, photons generically travel much further, of order fractions of centimeters, before depositing their energy into phonons,
since these materials are non-polar and their lattice reacts more weakly to light.
For higher energies, 
the right panel of Fig.~\ref{fig:mfpIR} shows that photons with energy well above the bandgaps have short absorption lengths as they can excite electrons into the conduction band.
In contrast, photons with energies very close to each material's bandgap can travel over macroscopic distances before being absorbed, as inter-band transitions are then phase-space suppressed.
Photons with energies much below the bandgaps but above lattice modes are not absorbed.
A photon that is not converted into an electron-hole pair, a phonon, or is not absorbed via other mechanisms such as the breaking of excitons, 
reaches the material surfaces and either escapes, 
or bounces on the surfaces back into the material. 
Finally, we note that in semiconductors, these properties can be significantly affected by doping, 
see \textit{e.g.}~\cite{8847339}.


\subsection{Cherenkov Photons as a Background to Dark-Matter Searches}
\label{sec:cherenkovDM}

Equipped with our understanding of Cherenkov radiation in non-conducting materials, 
it is now easy to understand the relevance of this effect in typical low-energy threshold dark matter experiments.
A typical detector setup has a semiconducting or insulating target that reads energy depositions into \eh-pairs or phonons. 
In addition to the dielectric target, other dielectric and non-instrumented materials are usually found near the detector, such as
holders for the target, electronic connectors, epoxy glue, and insulating materials to cover cables. 
Surrounding all these elements, 
a metallic shielding (usually copper) is in place.
Some high-energy charged tracks are expected to pass through the dielectrics of the detector, 
either due to particles (usually gammas, betas, or muons) that penetrate the shielding, or due to radioactivity from impurities in the detector component material. 
These tracks lead to Cherenkov photons,
which may then convert their energy into phonons or into \eh-pairs in the detector target, 
mimicking the dark matter signal.

It is likely that many \eh-pairs or phonons created by Cherenkov photons will be removed by vetoing the charged track responsible for the photons.
However and as discussed above, 
photons may travel macroscopic distances away from the originating charged tracks, 
before depositing their energy into an electron-hole pair or phonons.
Such long-lived photons can lead to events avoiding vetoes in two ways.
First, in experiments such as SENSEI, vetoes are applied based on the distance at which the \eh-pair is created away from tracks.
If this distance is large, the \eh-pair event is not vetoed, so long-lived Cherenkov photons can constitute a background. 
Secondly, at most other experiments, long-lived Cherenkov photons may originate from auxiliary non-instrumented material, escape such materials, and make it into the detector. 
In this case, the charged track and the corresponding Cherenkov-induced \eh\ or phonon event cannot be easily vetoed.
We will see concrete examples of these possibilities in Sec.~\ref{sec:experiments-current}. 

It is also straightforward to check that Cherenkov photons are expected to be abundant within typical detector setups. 
For example, consider the Cherenkov radiation from a \textit{single} $200 \,\kev$ electron (corresponding to a velocity $v \simeq 0.7$).
For concreteness, we take Si as the medium, 
but the exercise can be easily repeated for other dielectric materials, and we will see further examples in later sections.
Since current low-energy threshold experiments are sensitive to energy depositions down to $\sim$1~eV,
we consider the emission of Cherenkov photons in the energy range $1\,\eV \leq \omega \leq 2\,\eV$.
At such frequencies, the dielectric function of Si is approximately real and given by $\epsilon \sim 14$, 
so
$v \epsilon^2 \sim 7$. 
To estimate the Cherenkov rate we may then neglect the second term in Eq.~\eqref{eq:cherenkov}, 
so the number of emitted photons is simply  
\begin{equation} 
N_{\gamma} \sim \alpha \times \Delta \omega \times \Delta x
\end{equation} 
where  $\Delta \omega=1\eV$ is the photon energy interval and $\Delta x$ is the electron track length.
To estimate the typical charged track length, 
we use the electron mean range $\ell_e$, 
presented in Fig. \ref{fig:meanrange}.
From the figure, we see that the mean range of a $200\, \kev$ electron in Si is $\sim 200\, \mu \textrm{m}$.
Thus, the number of Cherenkov photons emitted by a single charged electron track is
\begin{equation}
N_{\gamma} \sim 8 \,  \bigg[\frac{\Delta \omega}{1 \eV}\bigg] \bigg[\frac{\Delta x}{200\, \mu \textrm{m}}\bigg] \quad .
\label{eq:naiveestimate}
\end{equation}
Multiple charged tracks are expected in typical detectors due to imperfect shielding and radioactivity.
For example, in the SENSEI run in the MINOS cavern~\cite{Barak:2020fql}, 
hundreds of charged tracks go through the CCD per g-day of exposure.
Multiplying the result in Eq.~\eqref{eq:naiveestimate} with the number of tracks, we see that thousands of $\mathcal{O}(\eV)$ Cherenkov photons are expected per g-day.

 
\section{Transition Radiation}
\label{sec:transition}

In the previous section, we discussed the spontaneous emission of photons by charged particles going through homogeneous, non-conducting material. 
In the presence of inhomogeneities, 
additional radiation arises as tracks encounter interfaces via the Cherenkov-transition effect, 
also referred to as transition radiation.
~\\
\subsection{Transition Radiation: Theory}

\begin{figure}[t!]
\centering
\includegraphics[width=5cm]{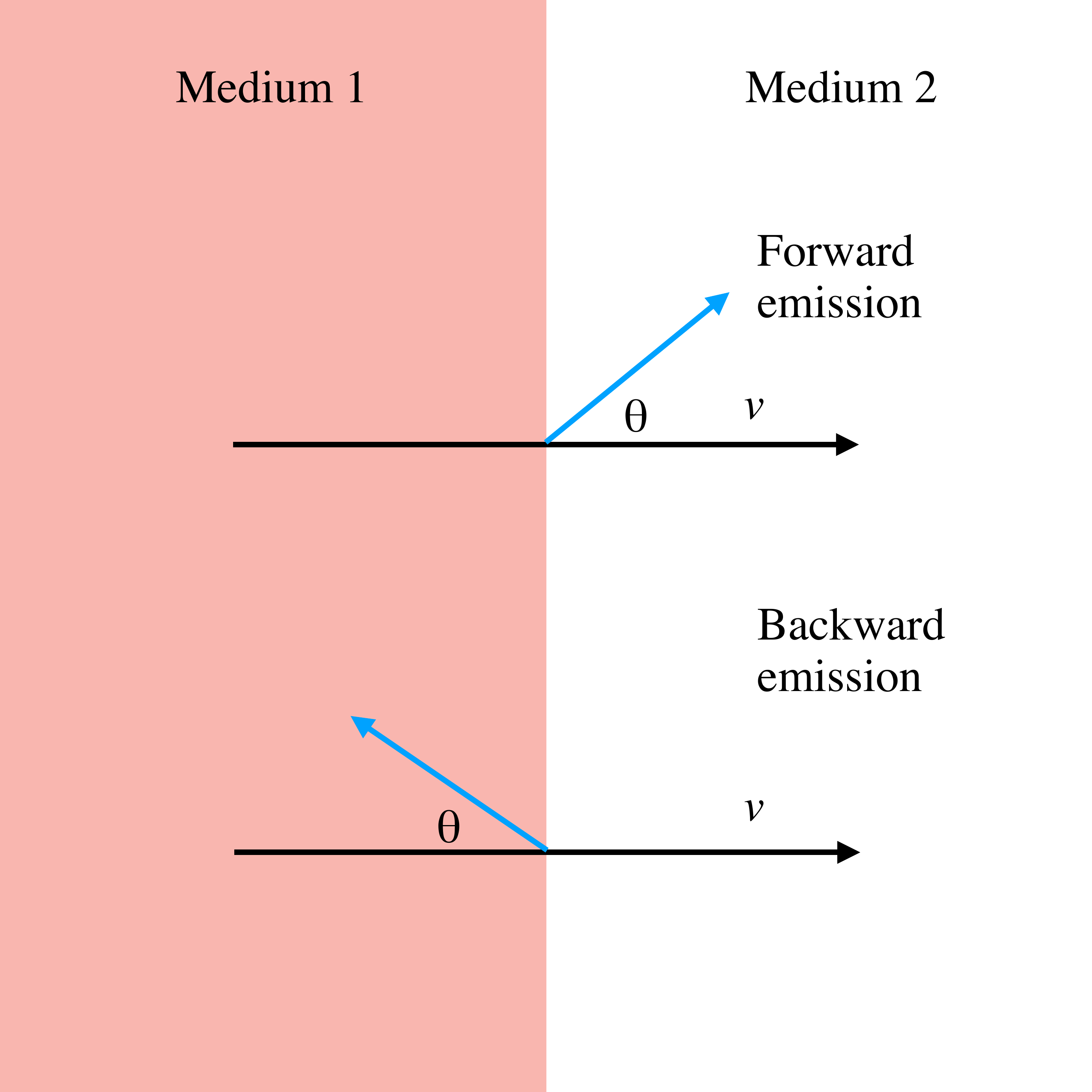}
 \caption{Polar angle conventions for forward and backward transition radiation emission.
By definition, $\theta \in [0,\pi/2]$ in both cases.} 
 \label{fig:TRschem}
 \centering
 \end{figure}

The main features of transition radiation can be discussed within its simplest realization,
which occurs when a charged particle transitions between two different, semi-infinite and homogeneous media, 
which we refer to as medium 1 and 2.
These media can be conductors or insulators.
Assuming that the particle goes from medium 1 to 2 with a constant velocity $v$ that is normal to the surface separating the media, 
the forward radiation spectrum, as observed from medium 2, 
is \cite{ter1972high,ginzburg1979several}
\begin{widetext}
\begin{equation}\label{eq:TR_general}
\frac{d^2N_\gamma}{d\omega d\Omega}= \frac{\alpha v^2}{\pi^2\omega}\sqrt{|{\epsilon_2}|}\sin^2\theta\cos^2\theta \left|\frac{(\epsilon_2-\epsilon_1)(1-v^2 \epsilon_2-v\sqrt{\epsilon_1-\epsilon_2 \sin^2\theta})}{(1-v^2\epsilon_2\cos^2\theta)(1-v\sqrt{\epsilon_1-\epsilon_2 \sin^2\theta})(\epsilon_1\cos \theta+\sqrt{\epsilon_1\epsilon_2-\epsilon_2^2 \sin^2\theta})}\right|^2,
\end{equation}
\end{widetext}
where $d\Omega=2\pi \sin\theta d\theta$ is the solid-angle differential, 
$\theta$ is the polar angle measured relative to the normal to the surface (see Fig.~\ref{fig:TRschem}), 
and $\epsilon_{1,2}$ are the complex and frequency-dependent dielectric functions of the two semi-infinite media.
Transition radiation is also emitted in the backward direction. 
The corresponding spectrum, as seen from medium 1 and with $\theta$ being the polar angle measured from the normal pointing towards medium 1 (see Fig.~\ref{fig:TRschem}), is obtained by exchanging $\epsilon_1$ and $\epsilon_2$, and setting $v$ to $-v$ in Eq.~\eqref{eq:TR_general}. 

For normal incidence on the interface, transition radiation is polarized in the plane specified by the velocity of the charged particle and the direction of the radiation.
While transition radiation is an interface effect, 
the emission is not localized and happens instead over a formation length from the surface.
This formation length is of order of the photon wave-length for non-relativistic particles, 
but is significantly enhanced in the relativistic case; 
see~\cite{ginzburg2013theoretical} for discussions on the formation zone. 
Finally, we point out that a generalized expression for the radiation spectrum for particles going through interfaces at arbitrary incidence angles can be found in~\cite{zrelov1978some}.

From Eq.~\eqref{eq:TR_general} we see that Cherenkov-transition radiation arises at leading order in $\alpha$, similarly to the conventional Cherenkov effect seen in homogeneous media.
In fact, the Cherenkov-transition effect can be regarded as a generalization of the conventional Cherenkov effect that accounts for the inhomogeneities in the media. 
The relation between the two processes is manifested in the poles of the emission spectrum Eq.~\eqref{eq:TR_general}.
For non-absorptive media with real dielectric functions, the singularities happen due to Cherenkov emission in medium 2 at the Cherenkov angle $\cos^2\theta_{\textrm{Ch}}=1/v^2\epsilon_2$,
and when the factor $1-v\sqrt{\epsilon_1-\epsilon_2 \sin^2\theta}$ vanishes,
which corresponds to the Cherenkov radiation emitted in medium 1 and refracted into medium 2~\cite{zrelov1978hybrid}.\footnote{In  \cite{zrelov1978hybrid}, Cherenkov-transition radiation is called hybrid radiation.}
These poles are kinematically accessible only when the Cherenkov condition Eq.~\eqref{eq:cherenkovcon} is fulfilled. 
When Cherenkov radiation is possible, integrating the spectrum Eq.~\eqref{eq:TR_general} over angles gives a divergent result, 
as it includes the infinite amount of radiation emitted by the tracks as it goes through each one of the two semi-infinite homogeneous media~\cite{zrelov1978hybrid,deraad1978interference,ginzburg2013theoretical}.
In order to extract the radiative effect due to the interfaces alone, 
one may define a finite ``pure'' transition radiation contribution by subtracting the divergent pieces of the radiation along the Cherenkov angles.
We refer the reader to~\cite{deraad1978interference} for a detailed discussion on how to extract systematically the ``pure'' transition contributions and also regulate the infinite conventional Cherenkov pieces,
so that a physical spectrum is obtained. 
For tracks with velocities falling below the Cherenkov emission threshold,
expression Eq.~\eqref{eq:TR_general} is finite and uniquely due to pure transition radiation.
Note that pure transition radiation can be emitted for any track velocity and in both conducting and non-conducting materials.\footnote{
This simplified scenario where no conventional Cherenkov radiation is possible is the most widely discussed in the literature, see \textit{e.g.} \cite{Jackson:1998nia,ginzburg1979several,landau2013electrodynamics}.}
Finally, for interfaces separating only absorptive media, where the dielectric functions are complex, or separating absorptive media and vacuum (such as a metal-vacuum interface),
expression \eqref{eq:TR_general} leads to a finite spectrum for all track velocities, as the poles are regulated by the absorption of the Cherenkov radiation in the material, and since Cherenkov radiation is not supported in vacuum.

 \begin{figure*}[t!]
\centering
\includegraphics[width=8cm]{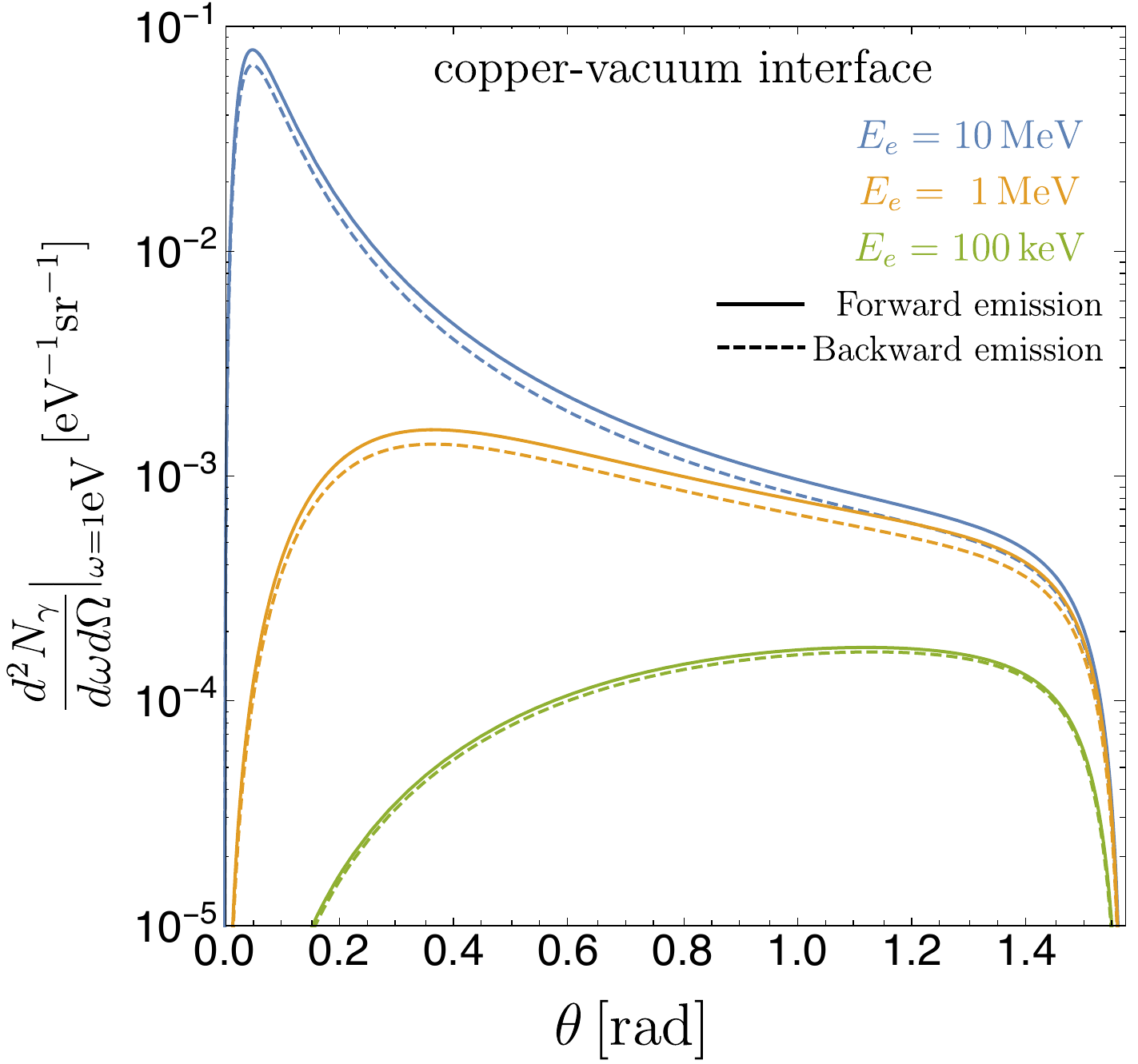}\quad\includegraphics[width=8.1cm]{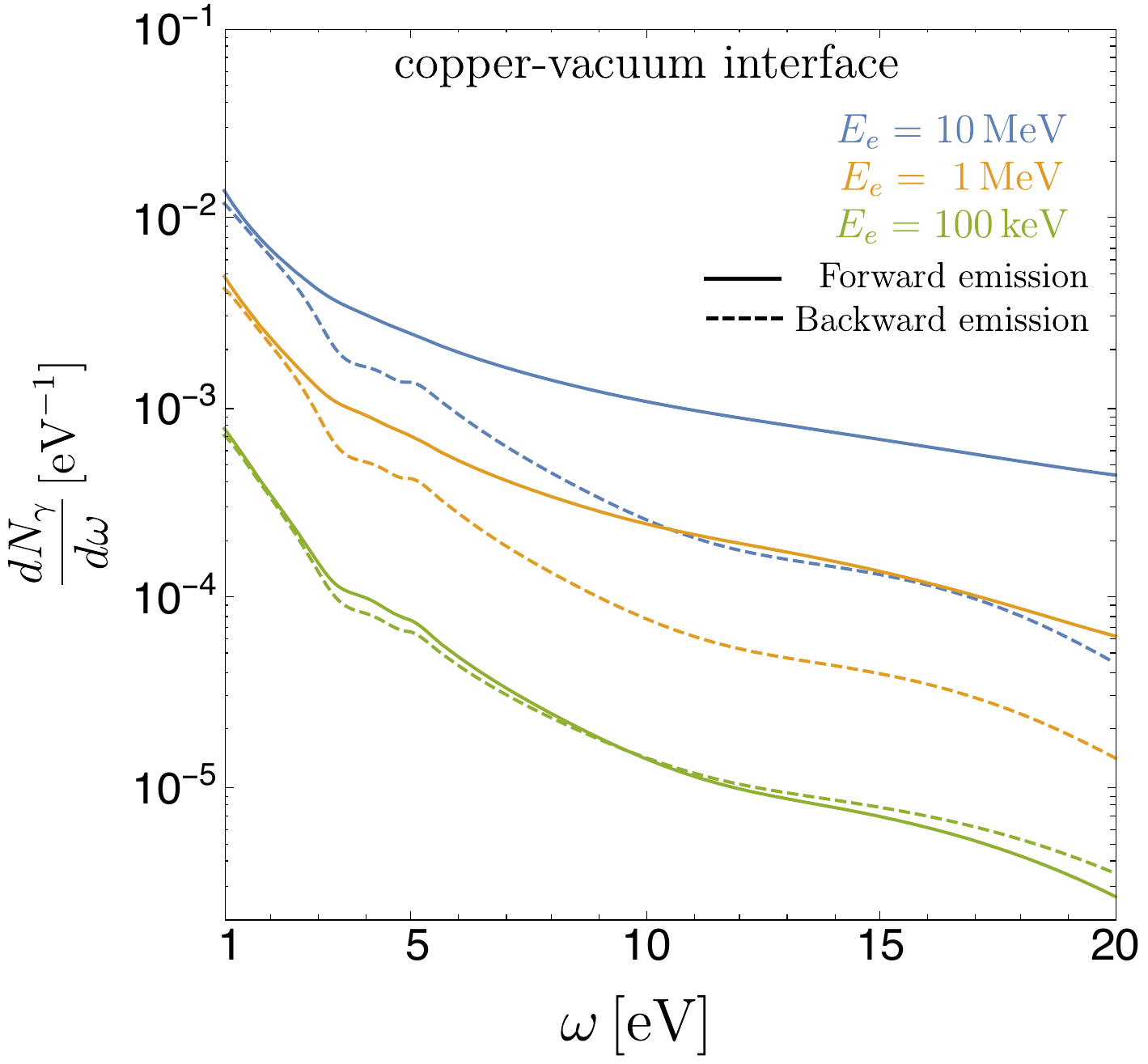}
 \caption{The angular spectrum  $d^2N_\gamma/d\omega d\Omega$ at $\omega=1$~eV  (\textbf{left}) and the energy spectrum $dN_\gamma/d\omega$ (\textbf{right}) of transition radiation, for an electron passing a copper-vacuum interface.
 We present the spectra for three choices of electron energies: $E_e=10$~MeV (blue), $E_e=1$~MeV (orange), and $E_e=100$~keV (green). Solid lines show the rate for forward emission (electron moving from copper to vacuum), while dashed lines show the rate for backward emission (electron moving from vacuum to copper). 
 \label{fig:TRplots}
} 
 \centering
 \end{figure*}
 
\subsection{Transition Radiation as a Background to Dark-Matter Searches}

In order to illustrate the relevance of transition radiation for dark matter detectors, 
let us now consider the simple case of a single relativistic electron track going through a copper-vacuum interface. 
This situation is typically encountered in experiments, 
since the detector target is usually enclosed in a copper vessel, with a vacuum gap left in-between the target and vessel (for concrete examples, see Sec.~\ref{sec:experiments-current}).
We consider two cases: first, the forward emission spectrum for a track going from copper (medium 1) to vacuum (medium 2), 
and second, the backward emission spectrum for a track going from vacuum (medium 1) to copper (medium 2).
Using Eq.~\eqref{eq:TR_general} and the dielectric function of copper from~\cite{Cudata}, we obtain the transition radiation spectrum, and we present its angular and frequency dependence in Fig.~\ref{fig:TRplots}.
From Fig.~\ref{fig:TRplots} (left panel), we see that the angular configuration of the spectrum at $\omega=1~\eV$ depends strongly on the track velocity, 
with a more forward (backward) spectrum observed at higher boost for tracks going from copper to vacuum (vacuum to copper). 
Such behavior can be understood by noting that at such frequencies, 
the dielectric function of copper is large, $|\epsilon_{\textrm{Cu}}(\omega=1\eV)|\sim 10^2\gg 1$, 
so Eq.~\eqref{eq:TR_general} can be approximated to
\begin{eqnarray}\label{eq:TR_special}
\frac{d^2N_\gamma}{d\omega d\Omega}
\sim
\frac{\alpha v^2}{\pi^2\omega}\frac{\sin^2\theta}{(1- v^2\cos^2\theta)^2} \quad.
\end{eqnarray}
From Eq.~\eqref{eq:TR_special}, we see that in this limit, the transition radiation rate is independent of the precise value of the copper dielectric function (as long as the condition $|\epsilon_{\textrm{Cu}}(\omega=1\eV)|\sim 10^2\gg 1$ is satisfied), and the  radiation in vacuum is peaked at a forward or backward angle, as measured from the track, that is inversely proportional to the tracks' boost, $1/\gamma$ ($\gamma=1/\sqrt{1-v^2}$).
The expression Eq.~\eqref{eq:TR_special} also indicates that
transition radiation from a vacuum-copper interface shows an infrared dominated spectrum, 
which can also be seen in Fig.~\ref{fig:TRplots} (right panel).
 

\section{Luminescence and Phonons from Recombination}\label{sec:recombination}

Cherenkov and transition radiation discussed in the two previous sections correspond to the direct emission 
of photons by charged particles passing through a material.  
Tracks can also excite electronic transitions in the material, which can produce low-energy quanta (photons and phonons) when electrons return to the ground state. 
Photons emitted by the excited material, instead of directly by the charged track, are called luminescence.

In this section, we focus on a special class of electronic excitations,
which is the creation of \eh-pairs in non-conducting media, 
a process which in this context is also referred as ionization. 
This process is the main mechanism by which electrons from radioactivity and cosmic-ray muons deposit energy in solid-state detector materials~\cite{PhysRevD.98.030001}. 
In this case, phonons or photons are obtained by the relaxation and recombination of the excited pairs via a variety of mechanisms 
that we will now explore.

Radiative and non-radiative recombination are material dependent processes, 
but some of the most important dynamics relevant for a wide variety of media can be illustrated by focusing on the case of recombination in semiconductors. 
In what follows, we will concentrate on three common semiconductors found in dark-matter detectors: Si, Ge, and GaAs.

\subsection{Recombination of \eh-pairs: Theory}

\begin{figure}[t!]
\centering
\includegraphics[width=6cm]{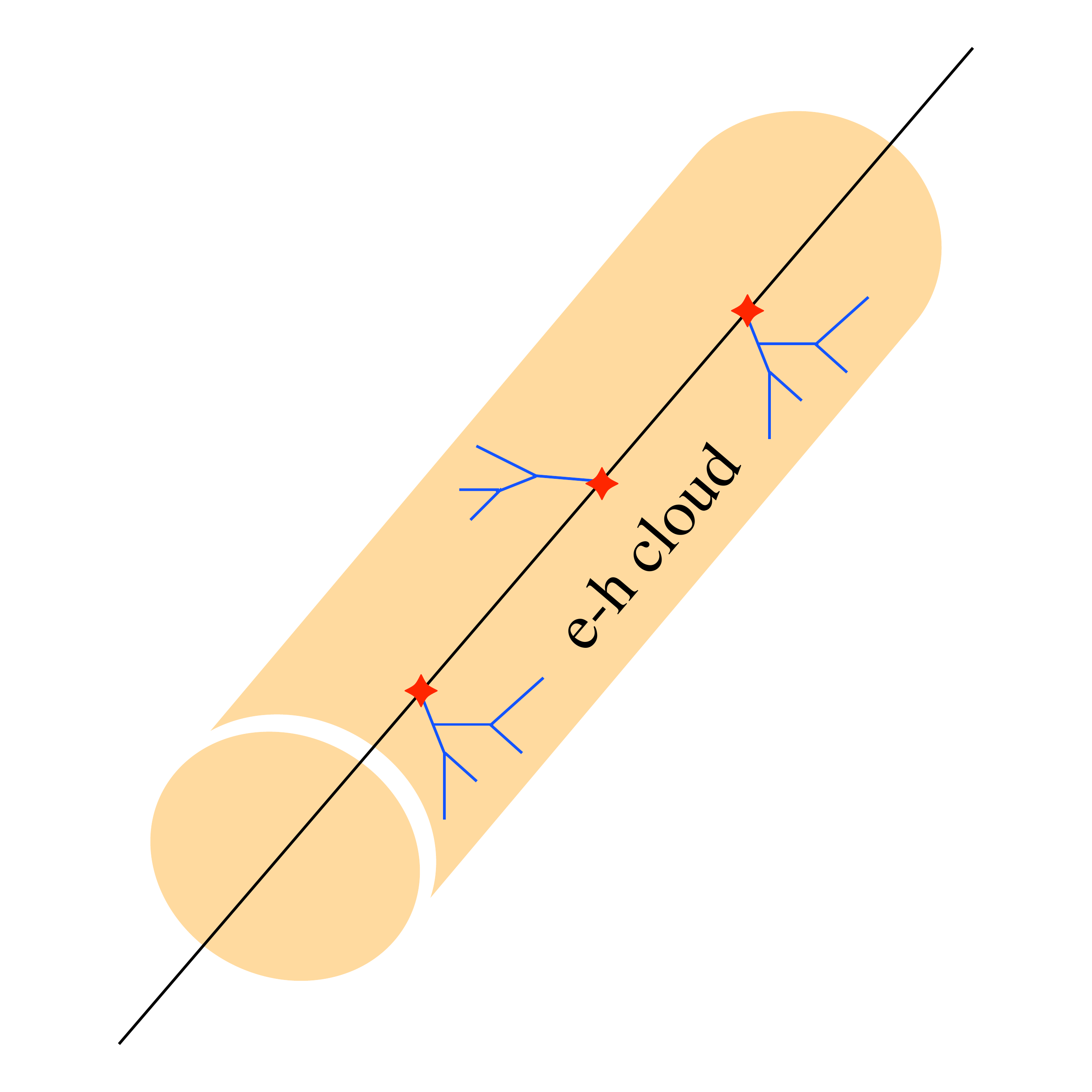}\\
 \caption{Schematic depiction of the creation of \eh-pairs by a track, shown as a black line, as it passes through a material. 
Hard scatters of the track, shown by the red stars, create secondary energetic electrons that subsequently create more \eh-pairs. 
The secondary electrons and \eh-pairs are represented by the blue lines, 
and the cloud that they form is represented in yellow.
 \label{fig:ehcloud}
 } 
 \centering
 \end{figure}

\textbf{Electron-hole Dynamics.}
A charged track passing through a material loses energy both by direct radiative processes and by the creation of excited \eh-pairs. 
The energy loss to \eh-pairs happens due to hard scatters that lead to energetic secondary electrons in the material, 
which subsequently create multiple \eh-pairs via phonon-mediated scattering~\cite{shockley1961problems}.
The multiplication of the secondary electrons happens over their mean free path, determined by the aforementioned phonon-mediated scattering in the material (which, in Si, is of order $10^{-2}\,\mum$~\cite{shockley1961problems}), and stops when the energy of the hardest electron participating in the scatterings falls below the energy required to create a new pair. 
As a result of this process, the track leaves behind a dense cloud of several \eh-pairs, shown schematically in Fig. \ref{fig:ehcloud}.

\begin{figure}[t!]
\centering
\includegraphics[width=9cm]{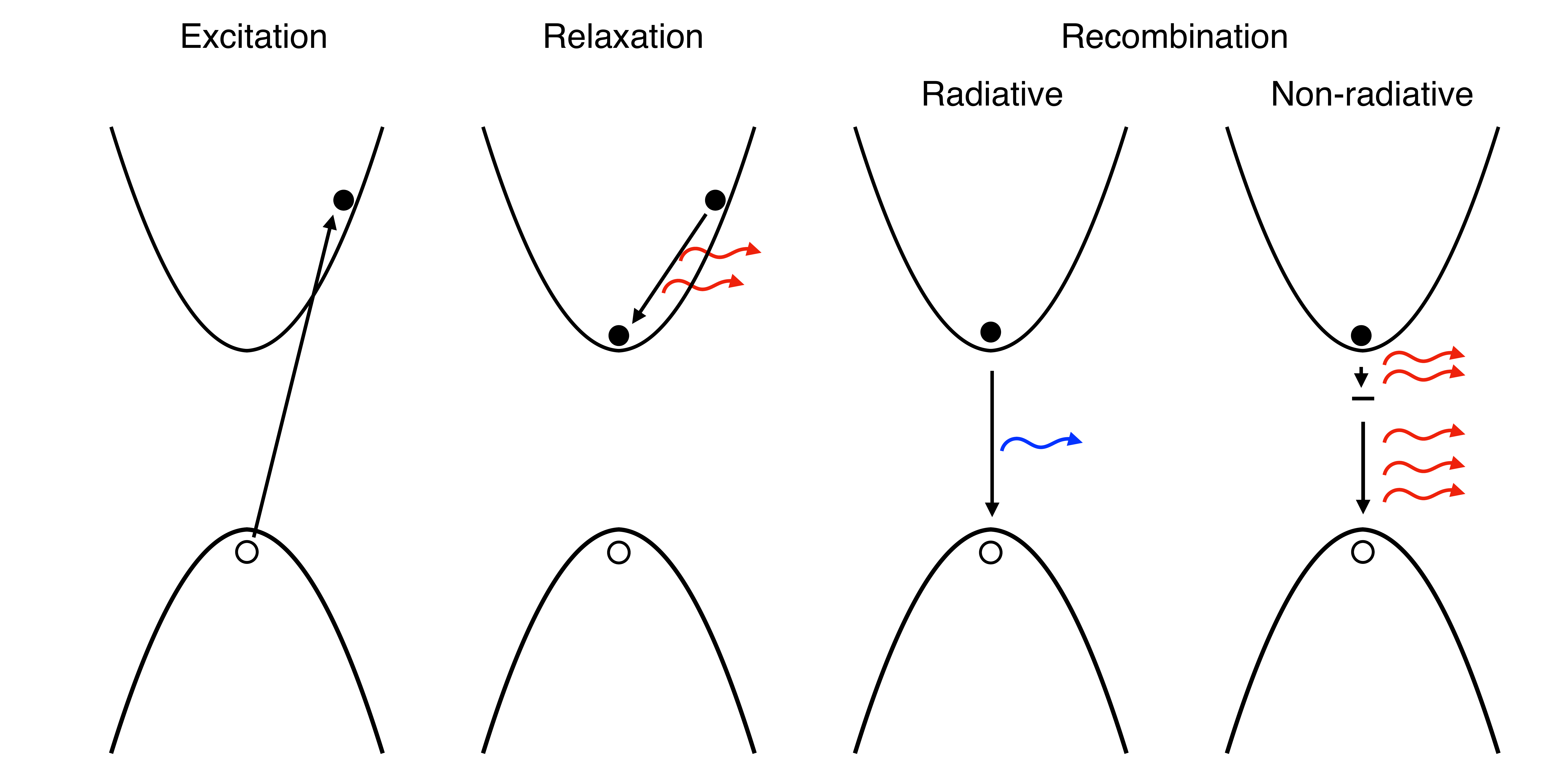} 
 \caption{Simplified band-structure representation of typical \eh\ excitation, relaxation, and recombination processes.
 From left to right, in the first step a scattering event excites an electron (black circle) from the valence band into the conduction band, leaving a hole (empty circle) behind. 
In the second step, the electron releases its excess energy in the form of phonons (red arrows), and moves to the bottom of the conduction band. 
Recombination of the electron and hole may then happen directly across the bandgap via the emission of a near-bandgap photon (blue arrow). 
Another possibility shown in the rightmost panel of the figure, is that there is an available energy level between the valence and conduction bands, referred to as a ``trap'', due to impurities in the material. 
In this case, the electron is first captured by the trap, and may subsequently recombine with the hole, with the released energy usually going into multiple phonons. While we show a direct bandgap, our schematic description is valid for both direct and indirect gap materials. 
 \label{fig:recombination}
 } 
 \centering
 \end{figure}

The total amount of \eh-pairs created by the passage of the track is characterized by the mean ionization energy, $\varepsilon$, 
which is defined as the average amount of energy that the hard track needs to leave in the material to create each pair.
We present values of the mean ionization energy $\varepsilon$ for our benchmark semiconductors in Table~\ref{tab:semicondprop}.
The mean ionization energy is larger than the material's bandgap, since part of the deposited energy goes into excess residual energy of the generated pairs. 
The excess energy can be released as the \eh-pairs go to the band edges by emitting phonons~\cite{alig1975electron}.
The \eh\ excitation process and the relaxation of an electron to the conduction band edge are schematically shown in Fig.~\ref{fig:recombination}. 

In terms of the mean ionization energy, 
the total number of pairs created by a track traveling a distance $L$ in the material is given by
\begin{equation}
N_e=N_h=\int_{0}^L dx  \, \varepsilon^{-1}\, \frac{dE}{dx}\bigg|^{\textrm{ionization}} \quad ,
\label{eq:NeNh}
\end{equation}
where $\frac{dE}{dx}\big|^{\textrm{ionization}}$ is the stopping power due  to ionization.

The dominant process setting the stopping power in a material depends on the track energy.
In typical dark-matter experiments, 
especially those well-shielded and underground,
the majority of the tracks correspond to electrons arising from radioactivity, 
which have energies  below a few MeV.
For such range of energies, 
tracks deposit their energy in non-conducting media overwhelmingly by creating \eh-pairs~\cite{PhysRevD.98.030001}, 
and $\frac{dE}{dx}\big|^{\textrm{ionization}}$ in Eq. \eqref{eq:NeNh} can be replaced by the full stopping power $dE/dx$, first introduced 
in Sec.~\ref{subsec:cherenkov-theory}. 
When considering highly energetic tracks, such as cosmic muons with very large boosts, most of the energy loss is due to radiation instead, 
in which case the stopping power only due to ionization must be used in Eq.~\eqref{eq:NeNh}. 

After the track passes by and the creation of pairs stops, 
the subsequent dynamics of the \eh-cloud left behind can be captured by the continuity equation,
\begin{equation}
\frac{ \partial n_{e,h}}{\partial t} = -\nabla \cdot \vec{j}_{e,h} - \Gamma_{e,h}\,,
\label{eq:recomasterequation}
\end{equation}
where $n_{e,h}$ is the electron or hole number density. 
The first term on the right-hand side is the electron and hole current, $\vec{j}_{e,h}$, 
which accounts for their motion.
The second term $\Gamma_{e,h}$ is the electron and hole disappearance rate density, 
which in the bulk material is due primarily to recombination due to defects (``traps''), direct recombination across the bandgap
and Auger recombination. 
We will show below that the rates for these three processes are proportional to one, two, and three powers of the \eh-densities,
so they are dominant at low, intermediate, and high \eh-concentrations, respectively (see, \textit{e.g.}, \cite{ruff1993spectral} for a discussion of the dominant modes in Si).
We show a schematic depiction of some of the possible recombination processes in Fig. \ref{fig:recombination}.
In the following paragraphs, we discuss each one of the processes involved in the dynamics of the \eh pairs in detail.
Note that for brevity we do not take into account the formation and diffusion of excitons (bound states of \eh-pairs), which needs to be considered in a careful treatment.  This is especially important below room temperature~\cite{greensilicon,corkish1996excitons}. 

\begin{table}[t]
 \begin{tabular}{|c|c|c|c|}
 \hline
 & Si &  Ge   & GaAs  \\
 \hline
 $E_g$ [eV] \cite{VARSHNI1967149} & 1.11 & 0.66 & 1.43 \\
 $\varepsilon$  [eV]  \cite{ryan1973precision,bertuccio2002electron,alig1975electron,klein1967simple}& 3.63 & 2.8 & 4.57 \\
 $\mu_e\,\,[\!\cm^2/\textrm{V}/\s]$ \cite{Hu2009ModernSD} & 1400 & 3900 & 8500 \\
$\mu_h\,\,[\!\cm^2/\textrm{V}/\s]$ \cite{Hu2009ModernSD} &  470 & 1900 & 400 \\
$B\,\,[\!\cm^3/\s]$ \cite{doi:10.1002/pssa.2210210140,5244084,nuese1972future}& $ 10^{-14}$ & $3.4 \times 10^{-14}$ & $7.2 \times 10^{-10}$ \\
$a_e\,\,[\!\cm^6/\s]$ \cite{huldt1971band,doi:10.1063/1.89694,steiauf2014auger} & $2.8 \times 10^{-31}$ & $2\times 10^{-32}$ & $1.7 \times 10^{-31}$ \\
$a_h\,\,[\!\cm^6/\s]$ \cite{huldt1971band,doi:10.1063/1.89694,steiauf2014auger} & $9.9 \times 10^{-32}$ & $1.1 \times 10^{-31}$ & $2.4 \times 10^{-30}$ \\
 \hline
\end{tabular}
\caption{
Electronic properties of Si, Ge and GaAs at room-temperature and low doping levels. 
$E_g$ is the bandgap, $\varepsilon$ the mean ionization energy, 
$\mu_{e,h}$ the electron and hole mobilities, 
$B$ the radiative recombination coefficient, 
and $a_{e,h}$ the electron and hole Auger coefficients.
For our estimates in SENSEI's Si Skipper-CCDs (see Sec.~\ref{subsec:SENSEI}), which operate at 135~K, we will use $\varepsilon=3.75\ \eV$~\cite{Rodrigues:2020xpt}.
}
\label{tab:semicondprop}
\end{table}

\textbf{Diffusion and Drift Currents.}
The electron and hole currents due to electric-field induced drifts and diffusion are given by~\cite{Hu2009ModernSD}
\begin{eqnarray}
\vec{j}_{e} &=& -n_e \mu_e \vec{E} - D_e \nabla n_e \quad ,\\
\vec{j}_{h} &=& n_h \mu_h \vec{E} - D_h \nabla n_h \quad ,
\end{eqnarray}
where $\vec{E}$ is the electric field, and $\mu_{e,h}$ and $D_{e,h}$ are the mobility and diffusion constants, respectively, for the electron or hole. The mobility and diffusion constants are related to each other via the Einstein equation \cite{Hu2009ModernSD}
\begin{equation}
D_{e,h}= \frac{ \mu_{e,h}  T}{e} \quad ,
\label{eq:einstein}
\end{equation}
where $e=\sqrt{4\pi \alpha_{\textrm{em}}}\simeq0.3$ is the electric charge and $T$ is the material's local temperature.\footnote{
Unless specified otherwise, we work in natural units. 
In SI units, a factor of the Boltzmann constant $k_B$ needs to be introduced on the right hand side of \ref{eq:einstein}.} In deriving Eq.~\eqref{eq:einstein}, 
it is assumed that the semiconductor is non-degenerate so that Boltzmann statistics for the energy distributions of the electrons and holes can be used. 
A generalized version of the Einstein relation accounting for Fermi statistics can be found in~\cite{melehy1965diffusion}.

The mobility and diffusion constants depend on both temperature and doping levels \cite{2005fspm.book.....Y,dorkel1981carrier}. 
We summarize their room-temperature, low doping-level values for our benchmark materials in Table~\ref{tab:semicondprop}. 
The mobilities are also related to the material's conductivity by \cite{Hu2009ModernSD}
\begin{equation}
\sigma = e(\bar{n}_e \mu_e + \bar{n}_h \mu_h) \quad ,
\label{eq:conductivity}
\end{equation}
where $\bar{n}_{e,h}$ are the background electron and hole densities discussed in Appendix~\ref{app:semiconductors}.

Finally, we define the characteristic timescales for diffusion and drift processes as the periods over which the \eh-densities change by an e-fold due to transport, $|(\partial\log{n_{e.h}}/\partial t)^{-1}|$. 
The timescales are given by
\begin{eqnarray}
\nonumber 
\tau_{e,h}^{\textrm{drift}}
&\equiv&
\frac{1}{\mu_{e,h}}\frac{n_{e,h}}{|\nabla \cdot ({n_{e,h} \vec{E})|} } \quad ,  \\
\tau_{e,h}^{\textrm{diff}}
&\equiv&
\frac{1}{D_{e,h}}\frac{n_{e,h}}{|\nabla^2 n_{e,h} |}   \quad .
\label{eq:timediffusion}
\end{eqnarray}
Note that the drift time can alternatively be estimated from Ohm's law, $J\equiv e n_{e,h} {v}_{e,h}=\sigma E$, where $v_{e,h}$ is the drift velocity of the carriers. 
Using the material's conductivity Eq.~\eqref{eq:conductivity} in Ohm's law, 
the drift carrier velocity is found to be $v_{e,h}=\mu_{e,h} E$.
Thus, the timescale for the carrier to transverse a distance $d$ is $\tau_{e,h}^{\textrm{drift}}\sim d/ (\mu_{e,h} E)$.
This timescale coincides with the one in Eq.~\eqref{eq:timediffusion} upon replacement of the gradient by the inverse of the typical length-scale $\nabla \cdot ({n_{e,h} \vec{E})}  \sim n_{e,h} E/d$.

\textbf{Band-to-band Recombination.} 
The first mechanism by which electrons and holes can disappear is by band-to-band (``direct'') recombination across the bandgap. 
This process happens radiatively, via the emission of a near-bandgap photon, 
as shown in Fig.~\ref{fig:recombination}.
The corresponding rate density is given by~\cite{5244084}
\begin{equation}
\Gamma_{e}^{\textrm{direct}}=\Gamma_{h}^{\textrm{direct}}=B (n_e n_h-\bar{n}_e \bar{n}_h) \quad ,
\label{eq:recomb}
\end{equation}
where $\bar{n}_{e,h}$ are the background carrier concentrations discussed in Appendix~\ref{app:semiconductors}, and $B$ is the radiative recombination coefficient, discussed below. 
The rate in Eq.~\eqref{eq:recomb} is proportional to the product of $n_e$ and $n_h$ as both an electron and a hole are required for band-to-band recombination, 
and vanishes when the densities reach the background values, 
at which point the material stops emitting photons.\footnote{In the literature, usually the product $\bar{n}_e \bar{n}_h$ is replaced by the intrinsic carrier density $n_i^2$ (see \textit{e.g.}, \cite{5244084}), 
as both quantities are always equal, see Appendix~\ref{app:semiconductors}.}

The radiative recombination coefficient can be related to the light absorption coefficient, defined as the inverse of the photon attenuation length $\alpha\equiv \ell_{\gamma}^{-1}$ (\textit{c.f.}~\ref{eq:photonmfp}), using detailed balance.
The result is~\cite{5244084} 
\begin{eqnarray}
\nonumber B &\equiv& \int_{0}^{\infty} d\omega \frac{dB}{d\omega} \\
\nonumber
&=& \frac{1}{n_i^2} \int_{0}^{\infty} d\omega \frac{\alpha(\omega,T) n^2(\omega,T) \omega^2}{\pi^2}\frac{1}{\exp(\omega/T)-1} \quad, \\
\label{eq:B}
\end{eqnarray}
where $n(\omega,T)$ is the refraction index, and $n_i$ is the intrinsic carrier concentration given in Eq.~\eqref{eq:intrinsic} of 
Appendix~\ref{app:semiconductors}.
In Eq.~\eqref{eq:B},
$dB/d\omega$ is the differential recombination coefficient, which determines the recombination radiation spectrum. 
 The expression for the recombination coefficient above must be taken only as approximate, 
as it neglects important corrections from the long-range Coulomb forces between electrons and holes. 
We refer the reader to~\cite{doi:10.1002/pssa.2210210140} for a detailed treatment of these effects.
Note that the recombination coefficient depends on temperature both explicitly  
and via its dependence on the dielectric properties of the material and the intrinsic carrier density.
A discussion of the temperature dependence of this coefficient in Si can also be found in~\cite{doi:10.1002/pssa.2210210140,bruggemann2017radiative}. 
In this material, the radiative recombination coefficient increases by roughly an order of magnitude as the temperature is decreased from room temperature down to 100 Kelvin, and by four orders of magnitude if the temperature is decreased to 20 Kelvin. 
This sharp increase of the radiative rate at low temperatures is due to exciton-mediated recombination \cite{doi:10.1002/pssa.2210210140}.

\begin{figure}[t!]
\centering
\includegraphics[width=\mywidth]{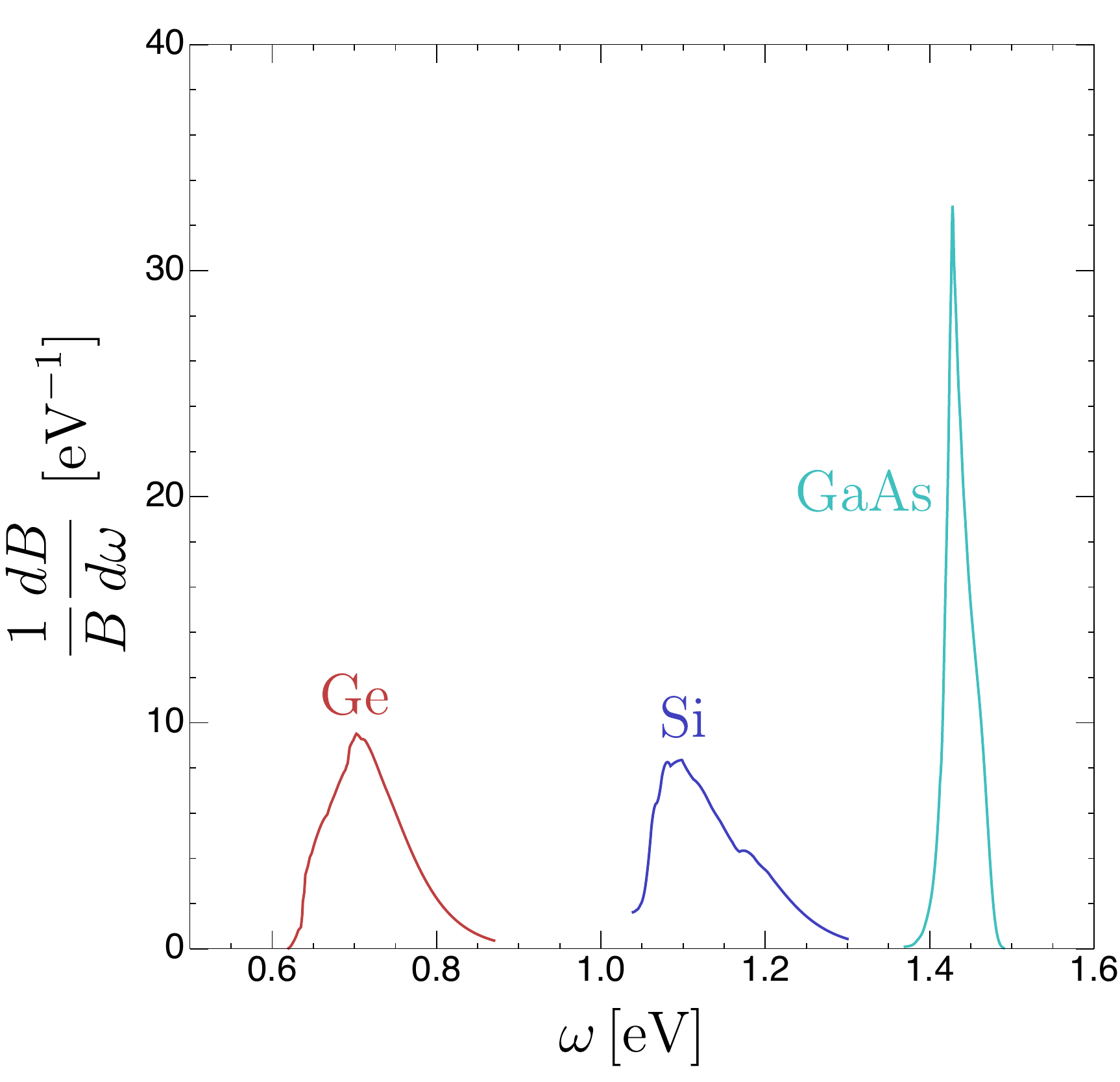}
\caption{
Normalized spectrum of photons $(dB/d\omega)/B$ emitted by the recombination of \eh-pairs in Ge, Si, and GaAs, 
obtained using Eq.~\eqref{eq:B} and the radiative absorption coefficients presented in~\cite{macfarlane1958fine,macfarlane1957fine,sturge1962optical}.  
We assume zero doping and room temperature.  
\label{fig:plotB}
}  
\centering
\end{figure}

We present the room-temperature radiative recombination coefficients for our benchmark semiconductors in Table~\ref{tab:semicondprop}. 
From the table we see that the recombination coefficient in GaAs is significantly larger than in Si and Ge. 
This is due to the fact that GaAs is a direct-bandgap semiconductor, while Si and Ge are indirect-bandgap materials. 
In indirect-bandgap semiconductors, radiative recombination is suppressed as it must be assisted by the emission of a phonon.
We also show in Fig.~\ref{fig:plotB} the normalized recombination spectrum, $(dB/d\omega)/B$, for Ge, Si, and GaAs at zero doping and at room temperature. 
From the figures and as expected, we see that the radiative recombination spectrum is peaked around each material's bandgap. 
Inspecting expression \ref{eq:B}, 
we see that the peak arises since the absorption coefficient is small at energies much below the bandgap, 
while at energies much above it the recombination spectrum is exponentially suppressed. 
The bandgap and therefore the position of the radiative recombination spectrum peak depend both on temperature and doping levels. 
The position of the peak is thus expected to shift towards lower frequencies when increasing doping levels, and towards higher frequencies when decreasing temperatures~\cite{van1987heavy,VARSHNI1967149}.

As for the diffusion and drift processes discussed in the previous section, we may define a timescale for the band-to-band recombination rate. 
The corresponding timescale is given by
\begin{equation}
\tau^{\textrm{direct}}_{e,h} \equiv \frac{ n_{e,h} }{\Gamma^{\textrm{direct}}} \quad .
\label{eq:timedirect}
\end{equation}

\textbf{Trap-assisted Recombination.} 
Impurities in the material lead to energy levels located in-between the valence and conduction bands. 
Such levels are referred to as ``traps'', 
as they can capture electrons from the conduction band or holes from the valence band.
We show a schematic representation of the trapping of an electron in Fig. \ref{fig:recombination}.
Differently from band-to-band recombination where an \eh-pair is eliminated, 
here only an electron \textit{or} a hole disappears by going into the trap.

Trap-assisted recombination is described by the Shockley-Read-Hall (SRH) theory~\cite{PhysRev.87.835}. 
In the SRH theory, 
the trap-assisted recombination-rate density is 
\begin{eqnarray}
\nonumber \Gamma_{e}^{\textrm{trap}}&=&c_{e} n_t \big[ (1-f_t)n_{e}- f_t n_{et}\big] \quad, \\
\Gamma_{h}^{\textrm{trap}}&=&c_{h} n_t \big[ f_t n_{h}- (1-f_t) n_{ht}\big] \quad,
\label{eq:traprecomb}
\end{eqnarray}
where $n_t$ is the density of impurities leading to traps, $c_{e,h}$ are trapping coefficients (discussed below),  
$f_t$ is the probability that a trap is occupied by an electron.
In Eq. \eqref{eq:traprecomb} we have assumed that the material is non-degenerate (see \cite{PhysRev.87.835} for a discussion of degenerate systems), so that $f_t$ is described by Boltzmann statistics, 
\begin{equation}
f_t=\exp\big[-(E_t-E_F^t)/T\big] \quad ,
\end{equation}
where $E_t$ is the trap energy level, and $E_F^t$ is the Fermi level or chemical potential of the trap, 
which sets its occupation probability.
In Eq.~\eqref{eq:traprecomb}, the quantities $n_{et}$ and $n_{ht}$ are defined as 
\begin{eqnarray}
\nonumber
n_{et}&\equiv& 
N_C \exp\big[(E_t-E_C)/T \big] \quad ,
\\
n_{ht}&\equiv& 
N_V  \exp\big[(E_V-E_t)/T \big] \quad ,
\end{eqnarray}
where $N_{C,V}$ are the conduction and valence band density of states given in Eq.~\eqref{eq:NCNV} in Appendix~\ref{app:semiconductors} and Table~\ref{tab:semicondprop2}, 
and $E_{C,V}$ are the conduction and valence band energies.
Note that the rates in Eqns.~\eqref{eq:traprecomb} have terms proportional to $n_{e} n_t$ or $n_{h} n_t$, indicating that trap-assisted recombination requires only a trap and an electron \textit{or} a trap and a hole. 
The second terms on the right-hand sides of Eqns.~\eqref{eq:traprecomb} ensure that the recombination rates vanish in 
thermal equilibrium,
where the chemical potential of the trap is equal to the equilibrium Fermi level of the semiconductor, 
whose determination is discussed in Appendix~\ref{app:semiconductors}.

Traps are divided in two types: deep and shallow. 
Deep traps are those for which the energy levels are well-separated from both the conduction and valence bands. 
They are usually associated with metallic impurities, such as gold, zinc, and copper. 
Shallow traps, on the other hand, 
are close to either the valence or conduction band, 
and can be due to dopants such as phosphorus (n-type) and boron (p-type). 
A summary of the energy levels of different traps for Si, Ge, and GaAs can be found in~\cite{semicond}.

Trap-assisted recombination can be either radiative (accompanied by photon emission) or non-radiative (by phonon emission).
For deep traps, recombination in semiconductors happens usually non-radiatively via multi-phonon processes~\cite{TYAGI1983577,5244084,pankove1975optical,landsberg2003recombination}. 
Non-radiative deep-trap coefficients are typically of order $c^{\rm deep}_{e,h}=10^{-6}\-10^{-9} \, \cm^3/\s$~\cite{5244084}.
Tables of radiative recombination coefficients by shallow traps can be found in~\cite{doi:10.1063/1.91234} and~\cite{5244084}. 
They are typically of order $c^{\rm shallow}_{e,h}=10^{-13} \, \cm^3/\s$,
much smaller than the ones for deep traps. 
Thus for intrinsic semiconductors, 
trap-assisted recombination happens mostly via deep-traps, 
and non-radiatively.
Moreover, radiative trap-assisted recombination dominates over its non-radiative counterpart in large-bandgap materials such as phosphors, 
where it is hard to dispose the large amount of bandgap energy as phonons \cite{5244084}.

The timescale for trap-assisted recombination is 
\begin{equation}
\tau^{\textrm{trap}}_{e,h}= \frac{ n_{e,h}}{\Gamma^{\textrm{trap}}} \quad .
\label{eq:timetrap}
\end{equation}
Since $\Gamma^{\textrm{trap}}$ is proportional to $n_h$ or $n_e$, 
the timescale is independent of the electron or hole number density, 
and is set instead by the trap density. 
Thus, as electrons and holes diffuse or drift, 
the trap-assisted recombination timescale stays constant.

\textbf{Auger Recombination.}
Finally, electrons and holes can recombine due to the Auger process, 
in which an electron in the conduction band gives its energy to another carrier by scattering and recombines with a hole, 
or a hole scatters with another hole and recombines with an electron.
In all cases, the process is three-body, and leads to the disappearence of both an electron and a hole.
The extra carrier in the process dissipates the energy liberated in the recombination in the form of phonons \cite{pankove1975optical}.
The corresponding rate density is given by
\begin{equation}
\Gamma^{\textrm{Aug}}=(a_n n_e + a_h n_h)(n_en_h -\bar{n}_e \bar{n}_h) \quad ,
\end{equation}
where $a_{e,h}$ are the Auger recombination coefficients, 
which are given in Table~\ref{tab:semicondprop} for our benchmark semiconductors, 
and $\bar{n}_{e,h}$ are the background carrier concentrations discussed in Appendix~\ref{app:semiconductors}.
The timescale for the Auger process is given by
\begin{equation}
\tau^{\textrm{Auger}}_{e,h} \equiv \frac{n_{e,h} }{\Gamma^{\textrm{Auger}}} \quad .
\label{eq:timeauger}
\end{equation}

\textbf{Other Recombination Channels.}
The processes discussed above are the basic mechanisms by which electrons and holes can recombine. Different combinations of the above processes, or slight variations of them, can lead to more complex recombination channels. Consider for example scintillation in GaAs. 
As discussed above, in intrinsic GaAs band-to-band radiative recombination happens efficiently, as 
GaAs is a direct bandgap semiconductor.
If, for instance, n-doping is added, a new radiative recombination channel opens up, as now a hole from the valence band can recombine with an electron provided by the donor by emitting a photon.
Doping can significantly increase the GaAs scintillation rate, as the rate is proportional to the number of electrons and holes available to recombine, \textit{c.f.} \eqref{eq:recomb}.
If the complexity is further increased by also adding p-doping to GaAs, which is the case studied in~\cite{Derenzo:2018plr}, then further recombination channels become available. For instance, an electron from the donor energy level can then recombine radiatively with a hole in the acceptor energy level.  Also, an electron in the conduction band can first transition to the donor energy level, and subsequently recombine with a hole in the acceptor level or valence band by emission of a photon~\cite{Derenzo:2018plr}.  Given the similarity of these processes with the ones already discussed in the previous paragraphs, their analysis can be performed by including additional recombination channels in Eq.~\eqref{eq:recomasterequation} similar to the ones already presented. 

\subsection{Recombination of \eh-pairs as a Background to Dark-Matter Searches}
\label{sec:recombestimates}

Recombination of \eh-pairs created by tracks passing through detector materials is an important source of backgrounds at sub-GeV dark-matter experiments.
One reason for the relevance of this process, is that the energy liberated in the recombination of a single \eh-pair is of the order of the material's bandgap, \textit{i.e.}, a few eV or less. 
Such energies coincide with the ones obtained from sub-GeV dark matter scattering in detectors,
so recombination photons or phonons naturally fall within the signal regions of dark-matter searches.
In addition, luminescence or phonon events from recombination can avoid track-related vetoes and effectively constitute backgrounds in a variety of ways.
As an example, and as discussed in Sec.~\ref{sec:cherenkovDM}, 
photons may avoid vetoes if they are long-lived in the material, 
a possibility that is very natural in band-to-band recombination, since the emitted photons are near bandgap.
Another possibility, 
which for brevity will not be discussed here in much detail,
is that vetoes based on the time-coincidence of the luminescence with the originating tracks can be avoided
if the return of the excited electrons to the ground state happens slowly, 
a phenomenon called long-lasting phosphorescence or ``afterglow''~\cite{wu2017long}. 
Phonons from non-radiative recombination, on the other hand, may avoid vetoes if they are generated in un-instrumented detector materials that are in contact with the detector target, such as holders and clamps, 
and subsequently make it into the target.

To illustrate how recombination can lead to large backgrounds rates at dark-matter experiments, 
we discuss the recombination of \eh-pairs created by the passage of a single electron track through a typical detector material.
For brevity, we focus here on possible backgrounds from radiative recombination only, 
and we briefly comment on the relevance of non-radiative recombination in Sec.~\ref{subsec:future-collective}.

We consider the case of a track passing through Si and take the track energy $E= 200$ keV. 
According to Fig.~\ref{fig:meanrange}, 
such a track travels $\ell_e\sim 200 \mum$ in Si before stopping.
Given the mean ionization energy in room-temperature Si, $\epsilon=3.63\,\eV$~\cite{ryan1973precision}, 
the track leads to 
\begin{equation}
N_{eh}=6\times 10^4
\label{eq:Neh}
\end{equation}
\eh-pairs distributed as a cloud around the track trajectory (a similar number of \eh-pairs is obtained at other temperatures). 
Assuming that the track is approximately straight, 
the shape of the cloud is cylindrical.
As discussed in the previous sections, the radius of the cylinder $r$ is expected to be a few times the electron and hole mean free paths, 
which in Si are of the order 
\begin{equation}
r \sim 10^{-2}\,\mum \quad .
\label{eq:cylradius}
\end{equation}
For our estimates below in what follows we take the cylinder radius to be equal to Eq. \eqref{eq:cylradius}.
The corresponding density of the \eh-cloud left by the track is then
\begin{equation}
n_e= n_h = \frac{N_{eh}}{\pi r^2 \, \ell_e} \sim 10^{18} \, \cm^{-3} \quad .
\label{eq:neh}
\end{equation}
After the passage of the track, the electrons and holes in the cloud move due to diffusion or are drifted by an electric field.
They also recombine,
leading to low-energy photons that contribute to backgrounds.
We now calculate the number of recombination photons emitted in the process, 
first for a track passing through high-purity Si and later for doped Si.

\textbf{Passage of a Track Through Undoped Si.}
For the case of undoped Si, 
most of the carriers found in the material are simply those left by the track.
In order to understand the fate of these electrons and holes,
we must compare the timescales for diffusion, drift, and different recombination mechanisms. 
Using Eqns.~\ref{eq:timediffusion}, \ref{eq:timedirect}, \ref{eq:timetrap}, and~\ref{eq:timeauger}, 
typical initial diffusion and recombination times are
\begin{eqnarray}
\nonumber
\tau^{\textrm{drift}}_0 &\sim& \frac{r}{\mu E} = 10^{-12} \,\s \, \bigg[\frac{10^3 \, \cm^2/\textrm{V}/\s}{\mu}\bigg]  \bigg[\frac{10^3\, \textrm{V}/\cm}{E}\bigg] \quad , \\
\nonumber
\tau^{\textrm{diff}}_0 &\sim& \frac{r^2}{D} = 10^{-13} \,\s \, \bigg[\frac{10 \, \cm^2/\s}{D}\bigg] \quad , \\
\nonumber
\tau^{\textrm{direct}}_0 &=& \frac{1}{B \, n_{e,h}} = 10^{-4}\,\s \, \bigg[\frac{10^{-14} \, \cm^3/\s}{B}\bigg]
 \quad ,
\\
\nonumber
\tau^{\textrm{trap}}_0 &=& \frac{1}{c^{\rm deep} n^{\rm deep}_t} = 10^{-6}\,\s 
\, 
\bigg[\frac{10^{-9} \cm^3/\s}{c^{\rm deep}}\bigg] \bigg[\frac{10^{15} \cm^{-3}}{n^{\rm deep}_t}\bigg] ,
\\
\tau^{\textrm{Auger}}_0 &=& \frac{1}{a n_e^2} = 
10^{-5}\,\s 
\, \bigg[\frac{10^{-31} \cm^{6}/\s}{a}\bigg]
 .\label{eq:timescales}
\end{eqnarray}
where the subscript $``0"$ emphasizes that these are the initial timescales right after the passage of the charged track;
we discuss the evolution of the timescales below. 
In Eq.~\eqref{eq:timescales}, we have taken typical room-temperature values for the diffusion constant and recombination coefficients from Eq.~\eqref{eq:einstein} and Table~\ref{tab:semicondprop}.
In order to obtain the electric drift time, 
we assumed that an electric field value $E$ similar to the one applied across the SENSEI CCD to drift electrons and holes towards the readout stage, 
which is also within a few orders of magnitude of the electric field applied at the SuperCDMS HVeV detector target to induce the Neganov-Trofimov-Luke  effect \cite{Barak:2020fql,Amaral:2020ryn}.
We also took a deep-trap density of $n^{\rm deep}_t=10^{15}\,\cm^{-3}$ and a trap recombination coefficient of $c^{\rm deep}=10^{-9} \, \cm^3/\s$, which are typical values in Si~\cite{ruff1993spectral}.
However, in ultra-pure Si, the trap-assisted recombination can be as large as $\sim 10^{-2}$ s.\footnote{We thank Steve Holland for referring to us this value.}   
Inspecting the timescales in Eq.~\eqref{eq:timescales}, 
it is clear that immediately after the passage of the track, 
the fastest process in action is \eh-diffusion. 
If the electric field is present, the second-fastest process is movement of the pairs by electric drift.
After carrier movements, the fastest process is trap-assisted recombination, 
which, as discussed previously, happens non-radiatively. 

Now, as the electrons diffuse radially away from the track trajectory,
both the density and the density-gradients of the electrons and holes decrease. 
Correspondingly, the diffusion time increases as the inverse density, $\tau^{\textrm{diff}}\sim 1/n_{e,h}$, starting from its initial value in 
Eq.~\eqref{eq:timescales}.
The trap-assisted recombination time, however, is independent of the \eh-density and stays constant. 

From the above discussion, a clear picture of the evolution of the \eh-cloud arises.
Consider first the case of the absence of an electric field.
After the passage of the track, electron-holes quickly diffuse away from the track.
The electron and hole densities and density gradients drop, 
so diffusion slows down until the diffusion timescale becomes comparable to the trap-assisted recombination time, 
$\tau^{\textrm{diff}} \sim \tau^{\textrm{trap}} \sim 10^{-6}\,\s $.
Finally, after a few trap-assisted recombination times, 
most of the electron-holes disappear by recombining non-radiatively.
On the other hand, if an homogeneous electric field across the material is present,  
carriers first diffuse, and then are drifted to the surfaces of the material. 
If, for instance, an electronics readout stage or electronics connections to ground are located at the surfaces, 
the carriers leave the material. 
Otherwise
they eventually recombine close to the surfaces of the material into phonons. 

Even if most of the electrons and holes recombine into phonons, 
a small fraction of them recombine radiatively. 
The radiative recombination rate, which is proportional to the density of electrons times the one of holes (\textit{c.f.}~Eq.~\eqref{eq:recomb}), is large right after the passage of the track, when these densities are the highest.
As electrons and holes diffuse or drift, the radiative recombination rate decreases.
Thus, we can obtain an estimate of the number of pairs that recombine radiatively by considering only the process over the first diffusion time.
The total number of radiated photons is then
\begin{eqnarray}
\nonumber N_\gamma &\sim& N_{eh} \, \frac{\tau^{\textrm{diff}}}{\tau^{\textrm{direct}} } \\
 \nonumber &\sim& 6\times 10^{-5} \, \bigg[\frac{10 \, \cm^2/\s}{D} \bigg] \bigg[ \frac{B}{10^{-14} \cm^3/\s}\bigg]  . \\
 \label{eq:recoestimate}
\end{eqnarray}
Comparing the estimate in Eq.~\eqref{eq:recoestimate} with the amount of photons created via the Cherenkov effect by a single track,  Eq.~\eqref{eq:naiveestimate}, 
we find that radiative recombination is a sub-leading source of photons in undoped Si.
However,
three important observations elucidate the relevance of radiative recombination. 
First, while the overall number of recombination photons from a single track is small,
\textit{all} these photons are near-bandgap (\textit{c.f.}~Fig.~\ref{fig:plotB}) 
and therefore long-lived, so they can deposit their energy in the detector far from their originating track.
As already discussed in Sec.~\ref{sec:cherenkovDM},
such photons may efficiently constitute backgrounds in, \textit{e.g.}, CCD-based detectors, since they can avoid track-related vetoes. 
Second, while the rate for radiative recombination is small in high-purity Si, 
the situation is drastically different in the doped scenario, 
which we discuss below.
Third, the radiative recombination coefficient is significantly enhanced at cryogenic temperatures \cite{bruggemann2017radiative}, and in high-purity semiconductors trap-assisted phonon-mediated recombination is suppressed. In these setups, radiative recombination can be a significant source of backgrounds. The enhancement of the radiative recombination coefficient at low temperatures is particularly relevant, as many dark matter detectors are in fact operated cryogenically.  

\textbf{Passage of a Track Through Doped Si.}
When doping is included, 
in addition to the \eh-pairs created from the passage of the track,
a large background density of electrons or holes from the dopant already exists in the material.
This enhances radiative recombination of the pairs left by the track,
as these electrons or holes can recombine with the large number of background carriers donated by the dopant.

For concreteness, let us consider the case of n-doped Si with an n-doping density $n_d=10^{18} \cm^{-3}$.
In this case, holes left by the track recombine with the background electrons from the dopant,
and the timescales for the corresponding hole recombination channels are
\begin{eqnarray}
\nonumber \tau^{\textrm{direct}} &=& \frac{1}{B n_e} = 10^{-4} \, \s  \,\bigg[\frac{10^{-14} \, \cm^3/\s}{B}\bigg] \quad , \\
\nonumber \tau^{\textrm{trap}} &=& \frac{1}{c^{\rm deep} n^{\rm deep}_t} = 10^{-6}\,\s \, 
\bigg[\frac{10^{-9} \cm^3/\s}{c^{\rm deep} }\bigg] \bigg[\frac{10^{15} \cm^{-3}}{n^{\rm deep}_t}\bigg] 
 , \\
\nonumber \tau^{\textrm{Auger}} &=& \frac{1}{a n_e^2} = 
10^{-5}\,\s 
\, \bigg[\frac{10^{-31} \cm^{6}/\s}{a}\bigg] \quad . \\
\label{eq:timescales2}
\end{eqnarray}
Differently from the undoped scenario, these timescales do not evolve in time, as they only depend on the electron density, which is set by the dopant and not affected by the diffusion of the holes left by the track.
As a consequence,
we can obtain the total number of recombination photons by simply multiplying the radiative recombination rate with the lifetime of the holes, which is set by the fastest recombination process in Eq.~\eqref{eq:timescales2}, \textit{i.e.}, by trap-assisted recombination.
The total number of photons obtained from recombination for our single electron track is then given by
\begin{eqnarray}
\nonumber
N_\gamma &\sim& N_{eh} \frac{\tau^{\textrm{trap}}}{\tau^{\textrm{direct}}} \\
\nonumber
&\sim& 
6 \times 10^{2} \,
\bigg[\frac{10^{-9} \cm^3/\s}{c^{\rm deep}}\bigg] \bigg[\frac{10^{15} \cm^{-3}}{n^{\rm deep}_t}\bigg]
\\ 
& & \quad \times \bigg[ \frac{B}{10^{-14} \cm^3/\s}\bigg] \quad .
\label{eq:recoestimate2}
\end{eqnarray}
The large number of photons obtained in our estimate clearly shows the relevance of radiative recombination as a source of backgrounds in dark-matter detectors.
Note that this number is also many orders of magnitude larger than the one in Eq.~\eqref{eq:recoestimate} obtained for undoped Si, 
and also larger than the number of photons with $\sim$1~eV energies obtained from the Cherenkov process by the same track, 
Eq.~\eqref{eq:naiveestimate}.
In the next section, we will demonstrate that radiative recombination in SENSEI is an important background, since the CCD backside is un-thinned and has a few-$\mu$m-thick layer with high phosphorus doping.


 \section{Radiative Backgrounds at Current Dark-Matter Experiments}
 \label{sec:experiments-current}
 
In this section, we give an overview of how backgrounds from Cherenkov radiation, recombination, and transition radiation are realized in recent results from SENSEI at MINOS~\cite{Barak:2020fql}, SuperCDMS HVeV~\cite{Amaral:2020ryn}, and EDELWEISS~\cite{Arnaud:2020svb}.  
We will also briefly discuss the results from CRESST-III~\cite{Abdelhameed:2019hmk,Abdelhameed:2019mac}, the EDELWEISS-Surf Ge bolometer~\cite{Armengaud:2019kfj}, and the SuperCDMS Cryogenic PhotoDetector (CPD)~\cite{Alkhatib:2020slm}.  
In Sec.~\ref{sec:experiments-future}, we will discuss upcoming and proposed experiments.  

Our selection of experiments is representative of a variety of detection strategies and environmental characteristics. 
For example, the data from SENSEI at MINOS and SuperCDMS-HVeV were taken at a shallow underground site or on the surface, respectively, with little shielding~\cite{Barak:2020fql,Amaral:2020ryn},
while the data from EDELWEISS are taken deep underground in a well-shielded environment~\cite{Arnaud:2020svb}.  
All experiments have excellent timing resolution with the exception of SENSEI, 
which instead has excellent position resolution and very little timing information. 
Detectors with good timing resolution will not be affected by radiative backgrounds that are generated in the target itself, since such photons will be detected at the same time as  the high-energy charged track. 
For these detectors, radiative backgrounds are observed as low-energy events if they originate from un-instrumented materials nearby the target. 
At SENSEI and other CCD-based detectors, on the other hand, 
backgrounds photons can be generated both in the target and surrounding material. 

We will see that a precise calculation of the Cherenkov radiation, recombination, and transition radiation backgrounds 
is in general rather complicated.  
However, our discussion will make it clear that these processes undoubtably contribute to the low-energy backgrounds 
at current and future experiments. 


\subsection{SENSEI at MINOS}
\label{subsec:SENSEI}

The SENSEI experiment uses Si Skipper-Charged-Coupled devices to probe dark matter that interacts with electrons~\cite{Tiffenberg:2017aac,Crisler:2018gci,Abramoff:2019dfb,Barak:2020fql}.
Dark matter scattering in a $\sim$1.9~g Skipper-CCD excites electrons from the Si valence band into the conduction band, 
creating \eh-pairs.
The excited pairs are quickly drifted to the Skipper-CCD's surface via an applied electric field, where they are stored until being read out.  

We will focus on the results reported by SENSEI in~\cite{Barak:2020fql}.  
These results are based on data collected during 24~days in early 2020 at the MINOS cavern at Fermilab. 
During the 24~days, the Skipper-CCD was exposed 22 times, for a period of 20-hours each time, 
which is followed by a 5.15 hour reading period.
Each such 20-hour exposure plus readout is called a ``CCD image.'' 
The Skipper-CCD has a thickness of $675 \mum$ and a total area of size $\sim 9.4~{\rm cm} \times 1.6~{\rm cm}$. 
The latter area is divided into a two-dimensional array of pixels of size $15 \mum \times 15 \mum$, with each pixel having the same thickness as the CCD. 
Signal events are categorized by the number of \eh-pairs that they create within sets of pixels defined by the search analysis ($1e^-$-events, $2e^{-}$-events, etc). 

To mitigate backgrounds arising from high-energy events such as charged tracks,
SENSEI imposes several selection criteria (``cuts'') on the data.
One of the most important cuts is the ``halo mask.'' 
High-energy events leave tracks of energy depositions in the CCD, 
in the form of multiple \eh-pairs along the track.
To discriminate such background events from the hypothetical dark matter signal,
all the pixels that have a distance to a track smaller than some ``halo mask radius" (measured in number of pixels)
are removed from the analysis.  
For quoting a final single-electron event rate, 
a benchmark halo mask radius of $60$ pixels is chosen, 
outside of which the rate of single-electron events is reported to be \Rs $\sim (450\pm 45)/\textrm{g-days}$.
Several hypotheses to explain these events have been discussed in~\cite{Barak:2020fql}, 
but no currently known background is able to account for them.

 \begin{figure}[t!]
\centering
\includegraphics[width=6cm]{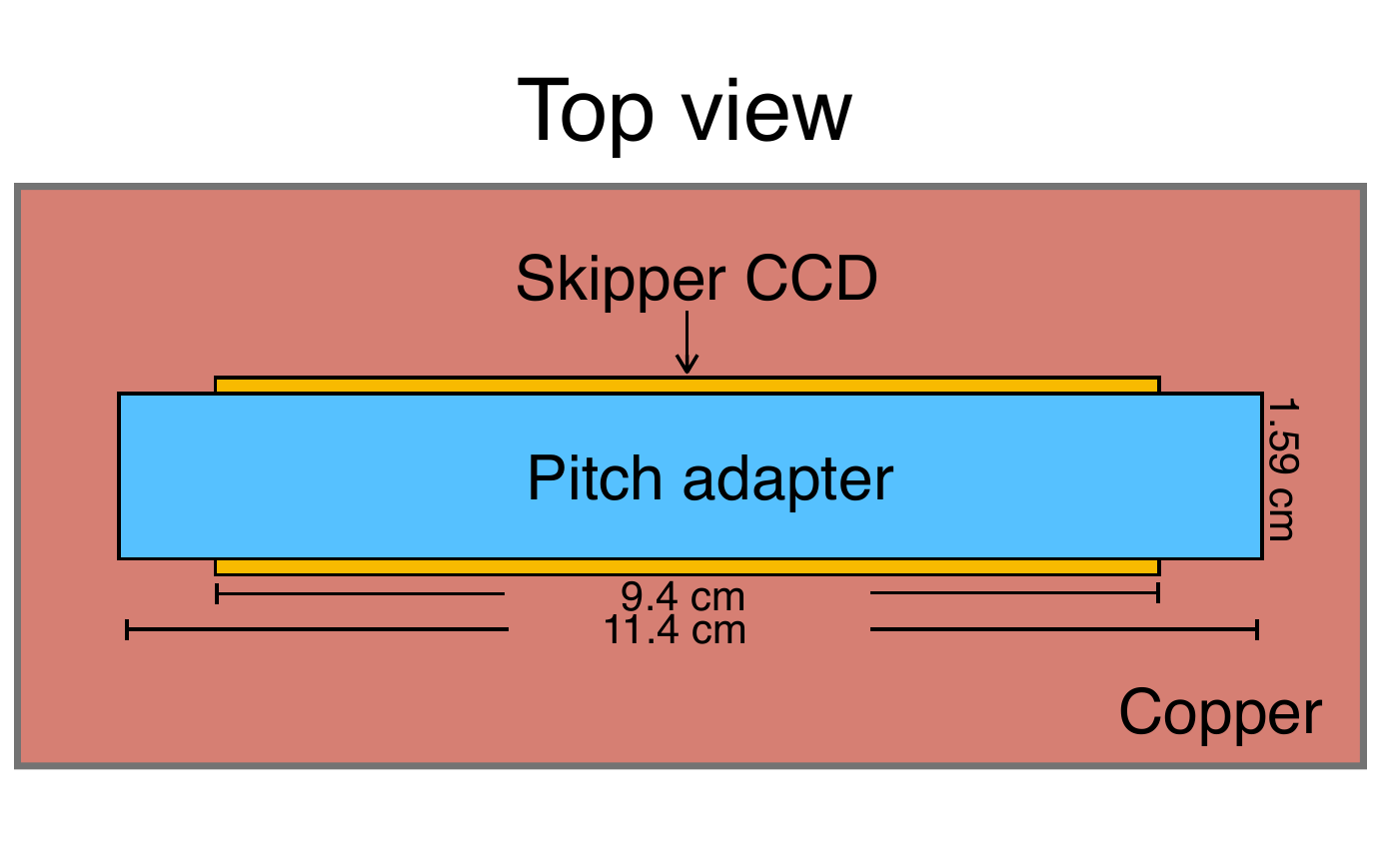}\\
\includegraphics[width=6cm]{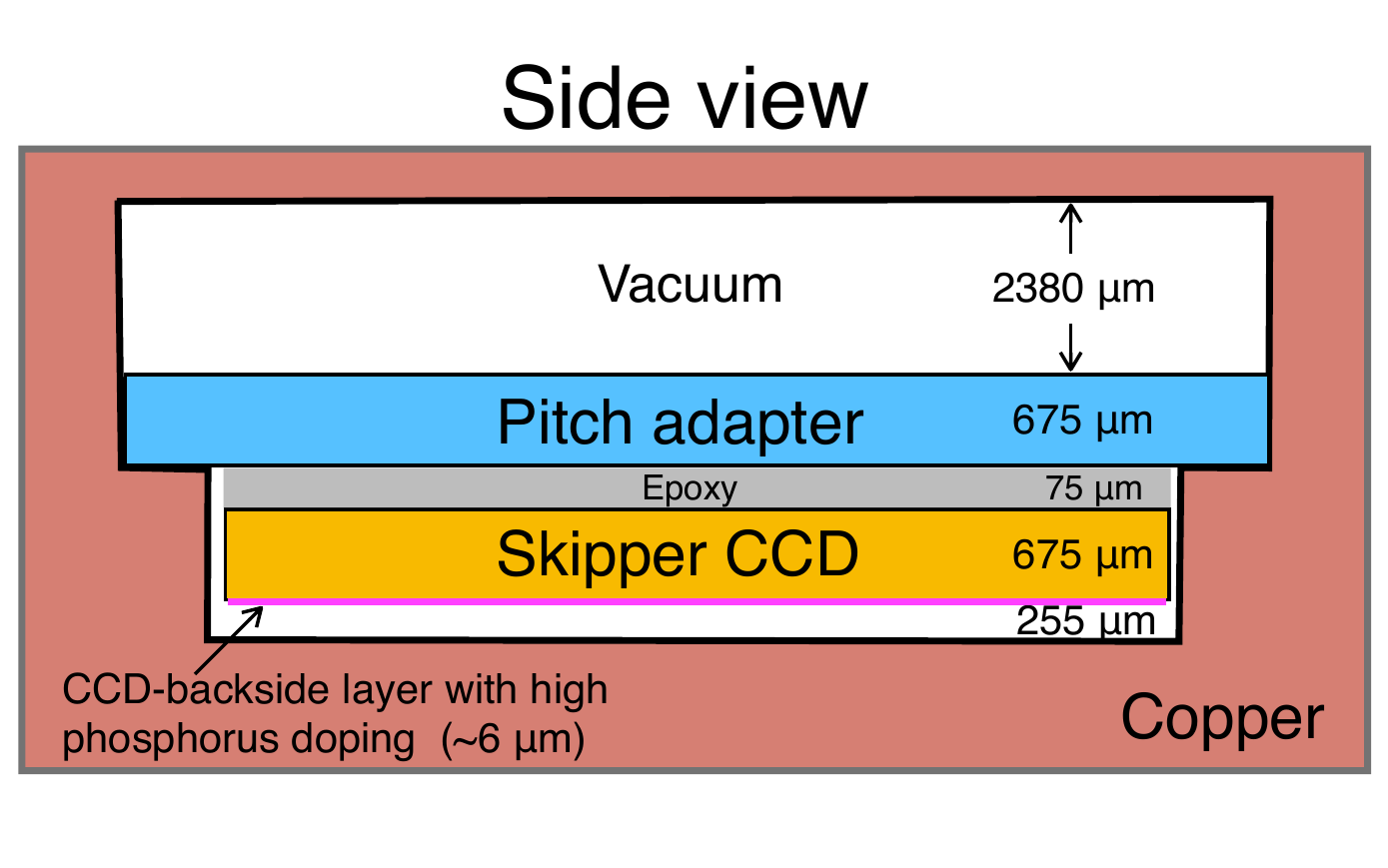}
 \caption{Schematic depiction of the SENSEI Skipper-CCD detector module in the MINOS cavern, a shallow underground site, for taking 
 the data described in~\cite{Barak:2020fql}. 
The active part of the detector is the Skipper-CCD (orange). A Si pitch adapter (blue) is glued with epoxy (gray) to the Skipper-CCD. 
This is placed inside a copper module (red). 
The CCD backside contains a few $\mu$m-thick layers that is heavily doped with phosphorus (magenta).  
We omit several details of the detector, 
including a series of layers with a thicknesses of order $\sim$1~$\mu$m or below that exist in-between the Skipper-CCD and the epoxy. 
The entire copper module shown is surrounded by lead bricks and placed inside a copper vessel, 
and then surrounded with additional lead shielding.
 \label{fig:schematics}
 } 
 \centering
 \end{figure}

Here we propose two novel explanations for the origin of the single-electron events: 
we argue that a significant fraction of them is due to the emission of (i) Cherenkov and (ii) recombination photons that are generated as charged particles pass through the detector.\footnote{As we will see, the SENSEI data provides important clues as to the origin of the single-electron events.  We are grateful to several members of the SENSEI Collaboration for numerous useful discussions of the characteristics of the data and the details of the detector setup, including Mariano Cababie, Juan Estrada, Yonatan Ben Gal, Daniel Gift, Sravan Munagavalasa, Aman Singal, Javier Tiffenberg, Sho Uemura, and Tomer Volansky.} 
Transition radiation will also contribute single-electron events, but 
the corresponding rate at SENSEI is negligible compared to the rates obtained from Cherenkov and recombination photons.\footnote{We note that bremsstrahlung contributes negligibly to the rate.} 

We present a simplified schematic of the SENSEI detector in Fig.~\ref{fig:schematics}. 
The main parts of the detector are the CCD and a ``pitch adapter'', 
both made of Si,
and the epoxy-glue between them. 
The backside of the CCD contains a $\mathcal{O}(\textrm{few}~\mu\textrm{m})$-thick layer of Si that is highly doped with phosphorus.  
These parts are placed in a copper housing.

 \begin{center} \textbf{Cherenkov Photons} \end{center}

Cherenkov photons are emitted by tracks as they pass through the CCD, pitch adapter, or epoxy glue.
For brevity, we consider here only Cherenkov radiation arising from the CCD, 
which, given its dimensions compared with the pitch and epoxy-glue, should already provide an $\mathcal{O}(1)$ fraction of all Cherenkov photons.

\begin{figure}[t]
\centering
\includegraphics[width=6cm]{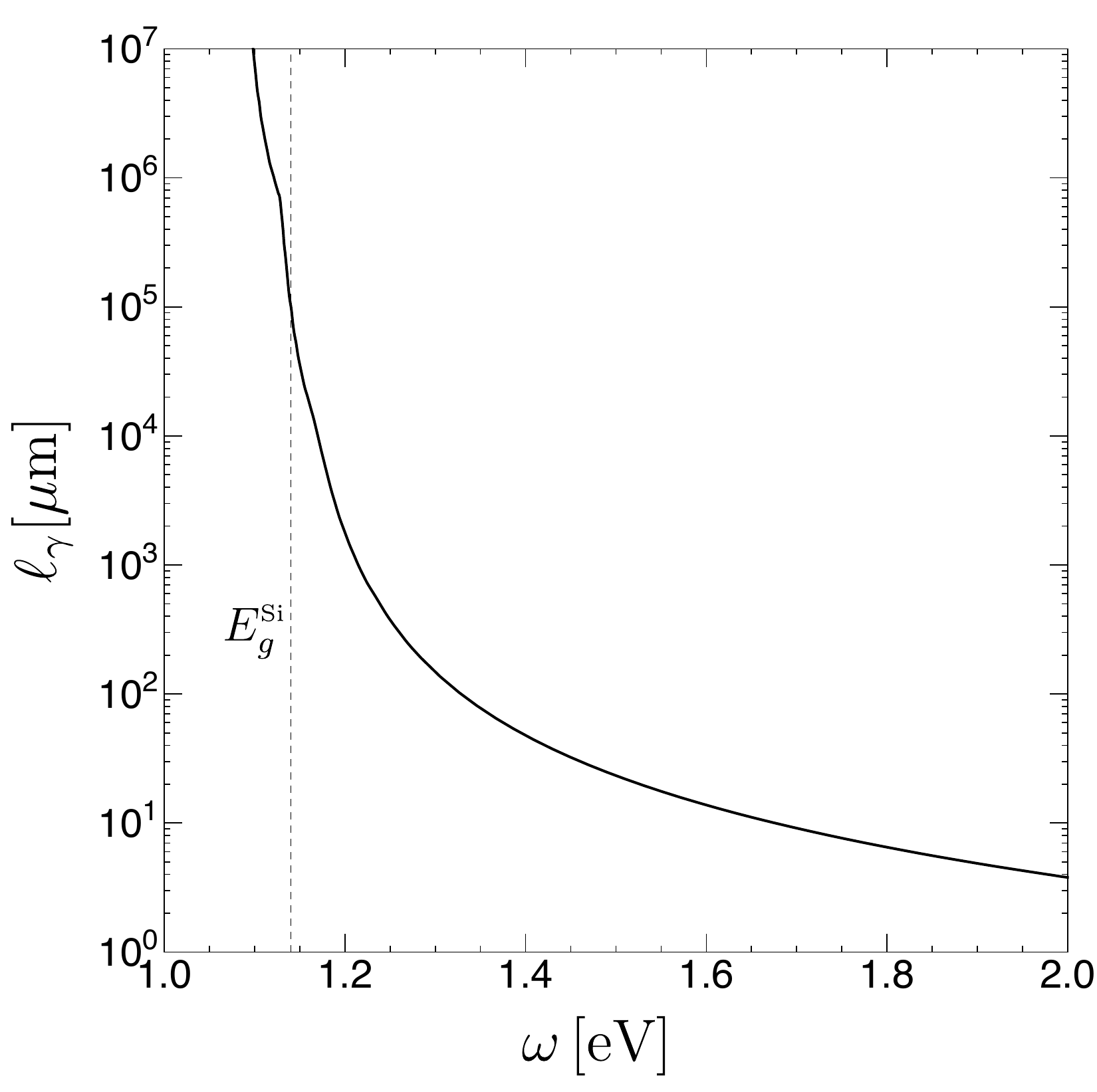}\\
\caption{Photon attenuation length in Si for energies close to the indirect Si bandgap, $E_g^{\textrm{Si}}=1.14 \, \eV$, 
at the SENSEI operating temperature $T=135\, \textrm{K}$.
The data is taken from a fit in~\cite{RAJKANAN1979793}, 
which allows for single phonon-assisted transitions.
 \label{fig:photonabs}
 } 
 \centering
 \end{figure}

In order for the Cherenkov photons to be registered as events after applying the halo mask cut, 
they must travel at least 60~pixels (900~$\mu$m) away from their originating track before converting into \eh-pairs.
As discussed in Sec.~\ref{subsec:cherenkov-theory}, 
photons have a large mean-free path in the dielectric material if they have energies near or below its bandgap. 
In Fig.~\ref{fig:photonabs}, we show the photon mean-free path in Si at the SENSEI operating temperature, $T=135\,\textrm{K}$. 
From the figure, we see that only photons with energy $\omega  \lesssim 1.2 \, \eV$  can travel more than 60 pixels\,=\,900 $\mum$ radially away from the track.  
On the other hand, only photons with frequency above $\omega \gtrsim 1.1 \, \eV$ are energetic enough to create \eh-pairs in Si~\cite{RAJKANAN1979793}.
Thus, 
we can obtain an order of magnitude estimate of the number of Cherenkov events by calculating the total number of  Cherenkov photons in the frequency range $1.1 \leq \omega  \leq 1.2 \, \eV$, which are emitted by all the charged tracks passing through the Skipper-CCD.
The charged tracks and their energies have been measured by SENSEI, and are reported in~\cite{Barak:2020fql}.
With this procedure, we find that the number of single-electron events expected at SENSEI from Cherenkov backgrounds is (see Appendix~\ref{app:SENSEI} for more details)
\begin{equation}
R_{1e^-}^{\textrm{Cherenkov}} \sim 500/\textrm{g-day} \quad .
\label{eq:senseicherenkov}
\end{equation}
This compares favorably with the observed rate, $R_{1e^-}= 450/\textrm{g-day}$, although as we discuss below, this result should only be seen as a rough estimate. 

\begin{figure}[t!]
\centering
\includegraphics[width=6cm]{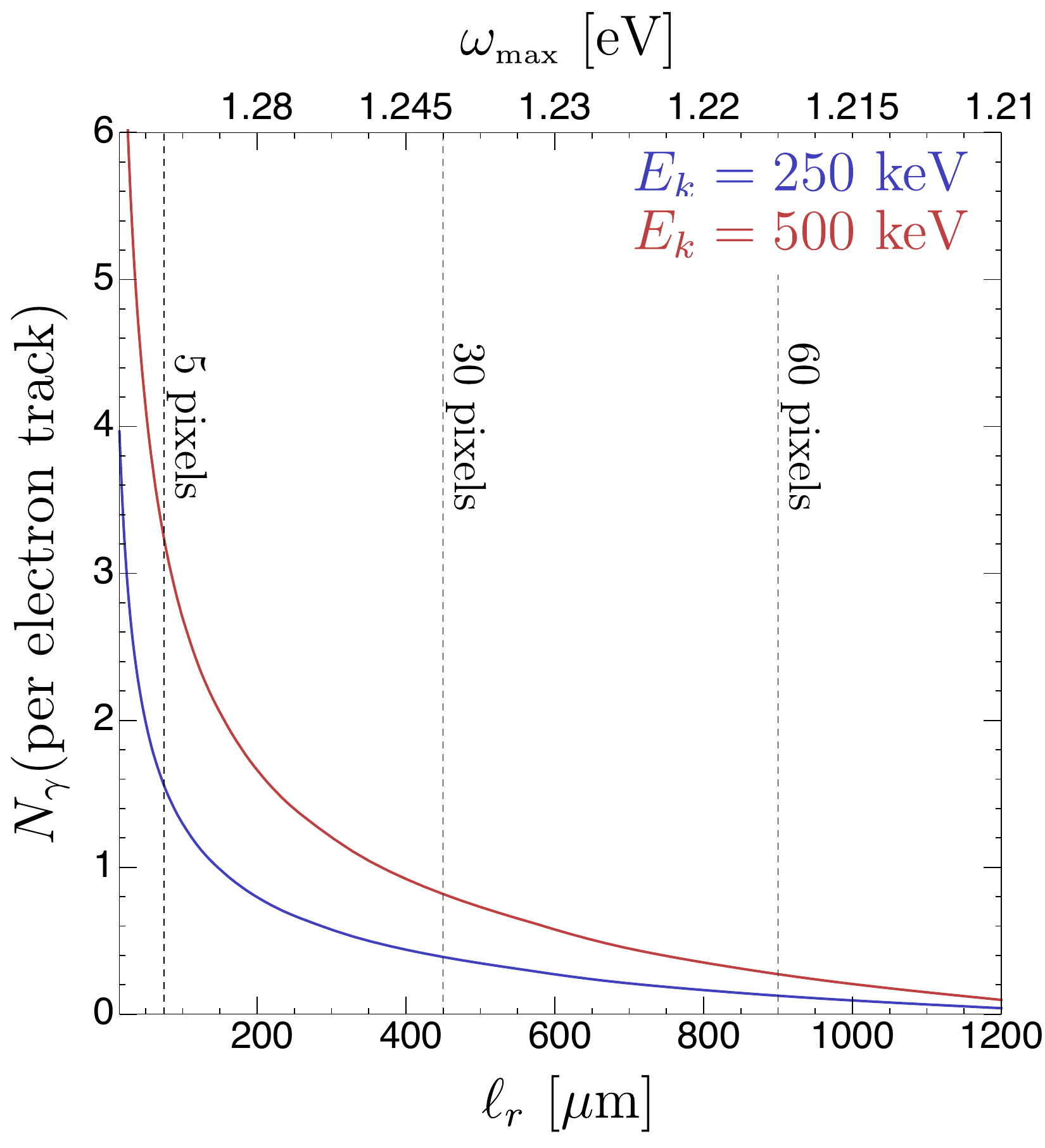}
 \caption{
Total number of photons produced by a single high-energy electron track passing through Si,
 which travel radially away from the track for a distance of at least $\ell_r$.
 Such photons have a mean-free path satisfying $\ell_{\gamma}\geq \ell_r/\sin \theta_{\textrm{Ch}}$.  
 The electron's kinetic energy is $E_k=250$~keV (blue) or $E_k=500$~keV (red). 
 The top $x$-axis shows the maximum frequency that allows  the photon to travel at least $\ell_r$. Vertical lines show the dimensions corresponding to 5, 30, and 60~pixels in the Skipper-CCD at SENSEI. 
 \label{fig:photondistance}}
 \centering
 \end{figure}

Further evidence in favor of our hypothesis comes from analyzing the Cherenkov event rate as a function of the halo mask radius.
Besides the benchmark halo radius of 60 pixels, 
SENSEI also reports rates for other halo mask radii.
They find that the single-electron event rate increases for smaller halo mask radii, \textit{i.e.}, when including events closer to the tracks.
As an example, when reducing the halo radius from its benchmark value to $\sim$5~pixels, 
the single-electron event rate was measured to increase to $\sim$900/g-days.
Our hypothesis also accounts for the increase of the rate as the halo mask is decreased,
since most of the emitted Cherenkov photons are converted into \eh-pairs close to their originating track. 
This is illustrated in Fig.~\ref{fig:photondistance},  
where we show the number of Cherenkov photons that arise from one electron track
and that travel \textit{at least} a distance $\ell_{r}$ radially away from the track before converting into an \eh-pair. 
Such photons would thus pass the cut corresponding to a halo mask radius of order $\ell_r$.
From the figure, we clearly observe that the number of expected Cherenkov-induced \eh-events increases as smaller halo mask radii are considered.

 \begin{center} \textbf{Luminescence from Recombination of \eh-pairs} \end{center}
  
A second possible contribution to the backgrounds observed by SENSEI are photons that arise from the recombination of \eh-pairs, which are created as charged tracks pass through the CCD, epoxy, or pitch adapter. 
As discussed in Sec.~\ref{sec:recombestimates}, each track generates tens of thousands of \eh-pairs when passing through these materials, 
which can recombine either radiatively or non-radiatively. 
The rates for the different recombination mechanisms are material-dependent. 
The CCD and pitch adapter found in SENSEI are made mostly out of high-purity, undoped Si. 
In addition, while there is no electric field in the pitch adapter, there is a non-zero electric field across the bulk of the CCD.  
In both cases, we know from Sec.~\ref{sec:recombestimates} that only a tiny fraction of the \eh-pairs recombine radiatively (\textit{c.f.}~Eq.~\eqref{eq:recoestimate}, 
naively suggesting that recombination is not a relevant source of backgrounds for SENSEI.  

Very importantly, however, the backside of SENSEI's CCD (as seen in Fig.~\ref{fig:schematics}) has a few-$\mu$m 
thin layer of phosphorus-doped Si, where the doping concentration is as high as $10^{20}\,\cm^{-3}$ and where there is no electric field. 
Even if this region is small, in doped materials radiative recombination is extremely efficient, 
as discussed in Sec.~\ref{sec:recombestimates}. 
As a consequence, as tracks pass through the doped layers, they lead to a large number of recombination photons, 
which can travel to and be absorbed in the bulk of the CCD, and contribute to the observed single-electron events. 
Moreover, most of the recombination photons are near-bandgap and thus have a large absorption length, 
so they can avoid the halo-mask cut. 

As with the Cherenkov events, to estimate the number of single-electron events now due to recombination, we calculate the total number of photons emitted in the range of frequencies $1.1\, \ev \leq \omega \leq 1.2 \, \ev$.
Using the radiative recombination fraction in Eq.~\eqref{eq:recoestimate2}, the spectrum Eq.~\eqref{eq:B}, and the number of tracks observed at SENSEI, we obtain (for details see Appendix \ref{app:SENSEI})
\begin{equation}
R_{1e^-}^{\textrm{recombination}} \sim 500/\textrm{g-day} \quad .
\label{eq:senseirecomb}
\end{equation}
Our estimate clearly indicates that radiative recombination in the doped layers of the CCD is an important source of backgrounds at SENSEI, and may explain a large fraction of the observed single-electron event rate. 

Finally, we comment briefly on the possible luminescence of epoxy. 
After the passage of a charged particles through the epoxy, it is possible that some of the deposited energy is re-emitted as radiation, which can produce events in the CCD.  
The frontside of the CCD (which faces the epoxy) contains a poly-Si layer with a thickness of about 0.6~$\mu$m.  Photons need to travel through this layer and be absorbed in the Si bulk in order to be registered as an event.  Only photons with an energy below $\sim$2.8~eV can traverse the frontside layer.  It is possible that the epoxy will luminesce below this energy~\cite{ALLEN198297}.  It may be possible, even if unlikely~\cite{Ramanathan:2020fwm}, that such events occasionally create a (single-pixel) 2$e^-$-event. 
A detailed study of the luminosity and scintillation efficiency of epoxy is needed to characterize these possible backgrounds more precisely.  

 \begin{center} \textbf{Transition Radiation} \end{center}
 
Transition radiation from the copper-vacuum, vacuum-CCD, vaccum-pitch-adapter, and other interfaces can produce photons across a wide range of energies.\footnote{Transition radiation was recently suggested to cause some of the excess events in SENSEI~\cite{Robinson:2020zec}. We believe, however, that it is subdominant to Cherenkov radiation and recombination photons.}  However, in order for photons to be registered as an event, they first need to penetrate into the bulk of the CCD, which requires their energy to be close to the bandgap.  Most of the CCD frontside is covered by the pitch adapter and epoxy, requiring photons that are generated at the copper-vacuum interface to be within 0.1~eV of the bandgap.  Even the small regions that are uncovered have a 0.6~$\mu$m-thick poly-Si layer (see discussion above), requiring photons to have an energy below 
 $\sim$2.8~eV to traverse into the bulk.  On the backside, there is a $\sim$6~$\mu$m layer of (n-doped) Si and poly-Si, requiring the photon energies to be below $\sim$1.8~eV in order to be registered in the bulk.  
 We find that transition radiation contributes much less than Cherenkov radiation or recombination to the single-electron event rates.  
 More details are provided in Appendix~\ref{app:SENSEI}. 
 
 \begin{center} \textbf{Summary} \end{center}
  
Summing both our Cherenkov and recombination event-rate estimations, we see that both of these processes can easily generate over 1000 events per g-day, 
a number that is a factor of $\sim$2 larger than the measured rate of 450/g-day.
Our calculations, however, must only be taken as order-of-magnitude estimates, as we have neglected a myriad of details that must be accounted for to calculate a precise background event rate.
In order to refine our estimates, 
a detailed simulation of these two sources of backgrounds in the SENSEI detector is required,
which will be done in a companion paper~\cite{SENSEI-radiative-to-appear}.  
In that reference, 
we will discuss several aspects that are crucial for obtaining a realistic background rate estimate, 
which include: (i) the precise distribution of the electron and muon tracks in the detector, 
(ii) all data analysis cuts, 
(iii) the exact geometry and properties of the detector materials, 
(iv) reflection and transmission of Cherenkov and recombination photons on detector surfaces, 
(v) the detailed properties of the layers of doped material in the CCD, and 
(vi) absorption of Cherenkov and recombination photons in the phosphorus-doped CCD layers and the epoxy.
There we confirm that our very rough estimates Eq.~\eqref{eq:senseicherenkov} and Eq.~\eqref{eq:senseirecomb} are within  a factor of  $\mathcal{O}(1-10)$ of a much more detailed background simulation.  
In~\cite{SENSEI-radiative-to-appear}, we will also discuss the 2\eh-pair event rate.


\subsection{SuperCDMS HVeV Above Ground}\label{subsec:SuperCDMSHVeV}

The SuperCDMS high-voltage eV-resolution (HVeV) detectors \cite{Agnese:2018col,Amaral:2020ryn} aim to detect dark matter via its scattering with electrons in a Si target.
The \eh-pairs created in the scattering process are drifted towards the surfaces of the Si target by an applied bias voltage, 
creating phonons along their way via the Neganov-Trofimov-Luke effect~\cite{Neganov:1985khw,Luke:1988}.
In this way, the energy deposited by the scattering event is amplified, so that a small energy resolution can be achieved.
The phonon energy is measured by Quasiparticle-trap-assisted Electrothermal-feedback Transition edge sensors (QETs)  attached to the surface of the Si detector. 

In our discussion, we will focus on the detector setup and results described in~\cite{Amaral:2020ryn},\footnote{We expect an analogous discussion of radiative backgrounds to be relevant also for the detector and data described in~\cite{Agnese:2018col} (see also~\cite{Romani:2017iwi,Hong:2019zlm}. The precise detector setup differs however from the one in~\cite{Amaral:2020ryn}, and hence would require a dedicated discussion and analysis that is beyond the scope of this paper.}  
which correspond to data collected during 7 days in 2019 in a surface laboratory at Northwestern University using a $0.93 \textrm{ g}$ Si crystal target. 
Given that the detector is near-surface, 
it is exposed to a large number of backgrounds from cosmic high-energy events.
One important tool available at SuperCDMS to discriminate signal events from backgrounds is the detector's excellent timing resolution (of order $10~\mu$s), which allows events that coincide with the passage of energetic tracks to be vetoed.
More specifically, physics events are identified using a pulse shape, which is of the order of $100~\mu\textrm{s}$.
Note that this is in sharp contrast with the strategy followed by SENSEI discussed previously, where there is little timing information and backgrounds are instead discriminated using information regarding the location of energy depositions in the CCD.

The final dark matter search dataset observed by SuperCDMS is a spectrum of $n$-\eh-pair events registered within their pulse-shape time window, 
where $n$ ranges from 1 to 6~\cite{Amaral:2020ryn}.  
We show their results in Table~\ref{tab:SuperCDMS}. 
The observed single-\eh-pair events are likely dominated by charge leakage~\cite{NoahKurinsky-discussion}; 
however, the origin of the events with $n\geq2$ \eh-pairs is unknown~\cite{Amaral:2020ryn}.  
SuperCDMS tested the results with different bias voltages (100~V, 60~V), finding that the rate of events 
is similar for both voltages. 

\begin{table}[t]
  \begin{tabular}{|c| c  c|}
  \hline
 \multirow{2}{*}{~} & \multicolumn{2}{c|}{HVeV Rates (g-day)$^{-1}$} \\
 \cline{2-3}
& 100 V & 60 V \\
\hline
\hline
$R_1$&$(149\pm1)10^3$&$(165\pm2)10^3$  \\
 $R_2$ &$(1.1\pm0.1)10^3$&$(1.2\pm0.2)10^3$\\
$R_3$&$207\pm40$&$245\pm86$ \\
$R_4$ &$53\pm20$&$77\pm48$\\
$R_5$ &$16\pm11$&$20\pm25$\\
$R_6$ &$5\pm6$&$10\pm17$\\
    \hline
  \end{tabular}
  \caption{The rate of $n$ \eh-pairs, $R_n$, observed with the SuperCDMS HVeV detector for two different bias voltages~\cite{Amaral:2020ryn}.  The uncertainty on the quoted rates corresponds to $3\sigma$, assuming Poisson statistics.  
  \label{tab:SuperCDMS}}
\end{table}

We now argue that Cherenkov photons generated in the non-conductive material surrounding the HVeV Si detector can plausibly explain the number of events containing $n\geq2$ \eh-pairs, both in magnitude and in spectral shape.  We will also mention the contribution from transition radiation, which is, however, likely sub-dominant compared to that from Cherenkov photons.  In addition, luminescence from the materials surrounding the target could contribute to the backgrounds, but nothing precise can be said about this possibility without detailed measurements of the luminescence properties of these materials.\footnote{We thank Noah Kurinsky and Matt Pyle for very useful discussions of the HVeV detector and data.  When discussing with them it became clear that the SuperCDMS HVeV Collaboration had already been studying the hypothesis of photon production within the PCB holders as the source of the events containing $n\geq2$ \eh-pairs.  However, to our knowledge, no exact microphysical mechanism was identified by them.  We believe that we have now identified several plausible mechanisms.} 

\begin{figure}[t]
\centering
\includegraphics[width=6cm]{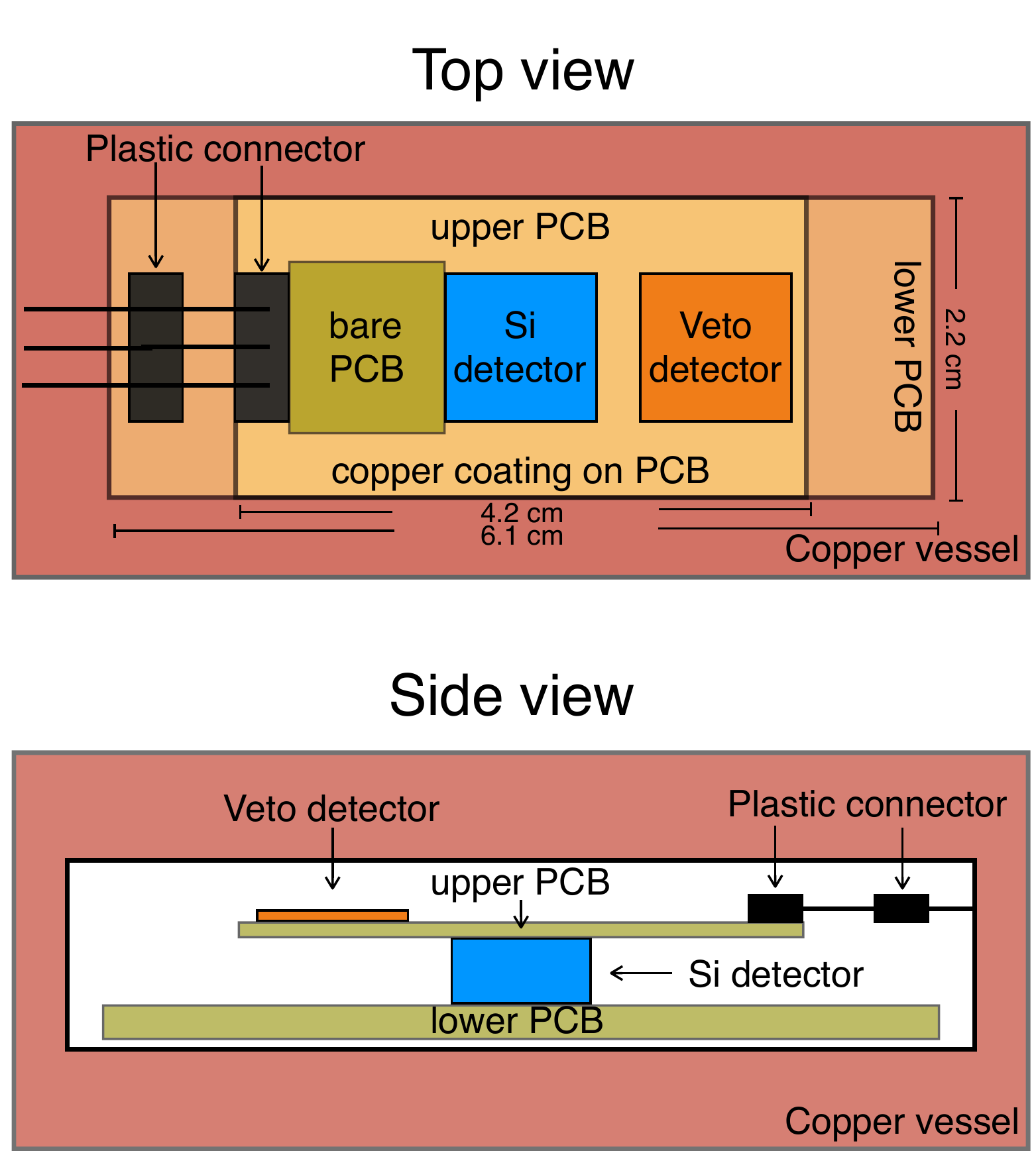}
 \caption{
 Schematic depiction of the SuperCDMS HVeV detector module that took surface data described in~\cite{Amaral:2020ryn}.  
 The active part of the detector is the ``Si detector'' (blue), which is clamped between two printed circuit boards (PCBs, gray-green) 
 that are partially covered with a layer of copper (two shades of tan color).  All surfaces of the two PCBs are coated with copper, except the area denoted as ``bare PCB'' on the front surface of the upper PCB from the top view, as well as the side surfaces. A Si ``veto detector'' is placed on top of the top PCB.  Two plastic connectors (black) and a flex cable (not shown) are also located in the copper housing (red). 
 \label{fig:SuperCDMS_setup}
 }
 \centering
 \end{figure}

A schematic layout of the main components of the HVeV detector module inside its copper housing is presented in Fig.~\ref{fig:SuperCDMS_setup}. 
The Si HVeV detector is clamped with metallic screws between two printed-circuit boards (PCBs) made of a material composed of woven fiberglass cloth with an epoxy resin binder. 

 \begin{center} \textbf{Cherenkov Photons} \end{center}
 
Cherenkov photons are generated by tracks passing through the PCBs, 
the Si detector, and other dielectrics found in smaller quantities within the copper vessel, such as plastic connectors.
Cherenkov photons produced by tracks passing through the target are easily removed by vetoing time-windows around the high energy event, as described above. 
That leaves us only with the Cherenkov photons due to tracks that pass exclusively through uninstrumented materials, such as the PCBs, which are not vetoed. 
Such photons may escape the un-instrumented materials and make their way into the detector (either directly or after reflecting one or more times off the copper or other surfaces inside the copper vessel), producing a low-energy background. 
We now provide an overview on how to estimate the number of photons created in this way, concentrating in photons originating in the PCB.
We refer the reader to Appendix~\ref{app:SuperCDMS-HVeV} for a more detailed explanation of our calculations.

The number and spectrum of Cherenkov photons created in the PCB can be obtained given its dielectric properties and the number of tracks going through it, the details of which are not available. 
We will assume that the PCB's dielectric properties are similar to the ``NEMA FR4 plate'' found at~\cite{PCB-FR4}, which is made of $61\%$ SiO$_2$, $15\%$ epoxy, and $24\%$ of bromine and oxygen.  
We then approximate the real part of the dielectric function, $\textrm{Re}\,\epom$, by the PCB's dominant SiO$_2$ component (see Sec.~\ref{subsec:cherenkov-theory}).
To determine if photons can escape the PCB, we estimate the PCB's photon absorption coefficient from the absorption coefficient of its  SiO$_2$ and epoxy components, ignoring possible contributions from bromine and oxygen.
We take the photon absorption length in SiO$_2$ from~\cite{1985hocs.book.....P} and epoxy from~\cite{doi:10.1002/app.33287}. 

To estimate the number of tracks passing through the PCB, we proceed as follows. 
For the incident high-energy muon track rate and energy spectrum we use the data in~\cite{Bogdanova:2006ex}. 
There is no public information available for the rate and spectrum of the high-energy electron background, 
so we simply assume that the spectrum is the same as that observed at SENSEI~\cite{Barak:2020fql},
 but normalized higher by a factor of 60 given that SuperCDMs is closer to the surface. 
This brings the background rate in line with the one observed in the SuperCDMS Cryogenic PhotoDetector (CPD)~\cite{Alkhatib:2020slm}, 
and should constitute a reasonable estimate of the background rate in the HVeV detector~\cite{NoahKurinsky-discussion}. 
In addition, we only consider tracks that pass sufficiently close to the edges of the PCB board, which lead to photons that have a large chance of escaping given the PCB's absorption coefficient.
It is very plausible that a large fraction of such tracks pass through the PCB without hitting instrumented parts of the detector and are thus not vetoed. 

 \begin{figure}[t]
\centering
\includegraphics[width=8cm]{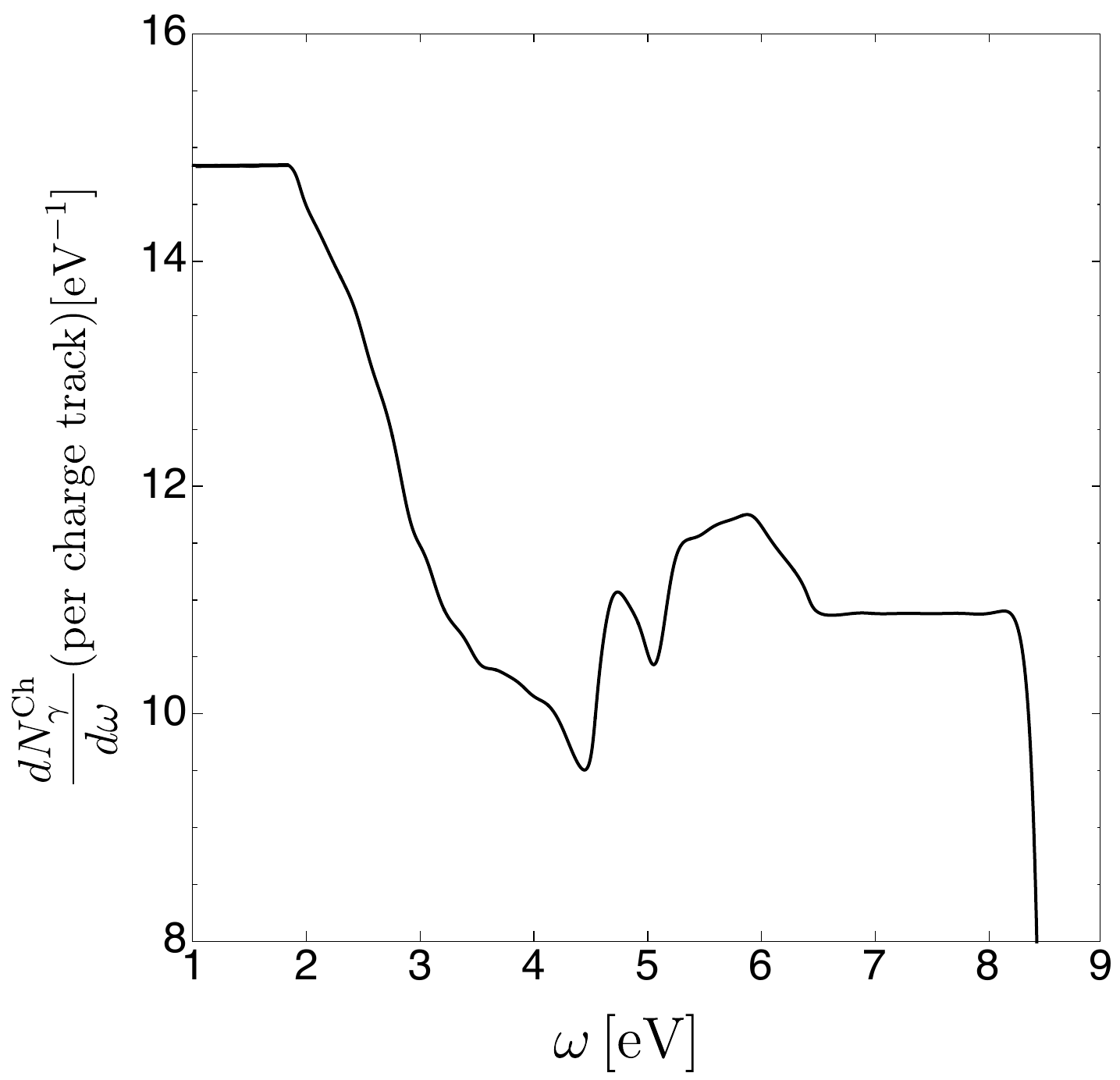}\\
\caption{The spectrum of Cherenkov photons that are able to escape the PCB, produced by an average single charged particle interacting in the PCB. 
The sharp drop in the spectrum at $\sim$8~eV is due to the $\textrm{SiO}_2$ component of the PCB, 
which becomes highly absorptive at photon frequencies close to and above the material's bandgap.
See Appendix~\ref{app:HVeV_Ch} for details. 
 \label{fig:HVeV_Ch_number}
 }
 \centering
 \end{figure}
 
With the above assumptions, we estimate that the number of Cherenkov radiation events per g-day of detector exposure from interactions of charged particles in the PCB (see Appendix~\ref{app:SuperCDMS-HVeV} for details) is  
\begin{equation}
N_\textrm{events}^{\textrm{Cherenkov}} \sim 2.8\times 10^4\textrm{/g-day}.
\label{eq:totalestimateSuperCDMs}
\end{equation}
Each of these events will contain some number of Cherenkov photons with a range of energies,
some of which may escape the PCB.
Given that the PCB has a frequency-dependent absorption coefficient, 
the photons that can escape the PCB inherit a characteristic spectrum, 
shown in  Fig.~\ref{fig:HVeV_Ch_number} for photons originating from a typical (average) track.
Some of the Cherenkov photons that exit the PCB are expected to reach the detector target, either directly, or after bouncing off various materials or the copper walls of the vessel. 
As shown in Fig.~\ref{fig:HVeV_Ch_number}, many of these photons have energies above the Si bandgap, so these will produce \eh-pairs if they are absorbed in the Si detector. 
The observed $n$-electron events are then obtained from the absorption of one or multiple Cherenkov photons, depending on the energy of each photon and the charge yield in Si.
Given the large number of photons obtained in our estimate as compared with the rates measured at SuperCDMS, 
it is very plausible that Cherenkov photons are an important source of backgrounds at the HVeV detector.

\begin{figure}[t]
\centering
\includegraphics[width=8cm]{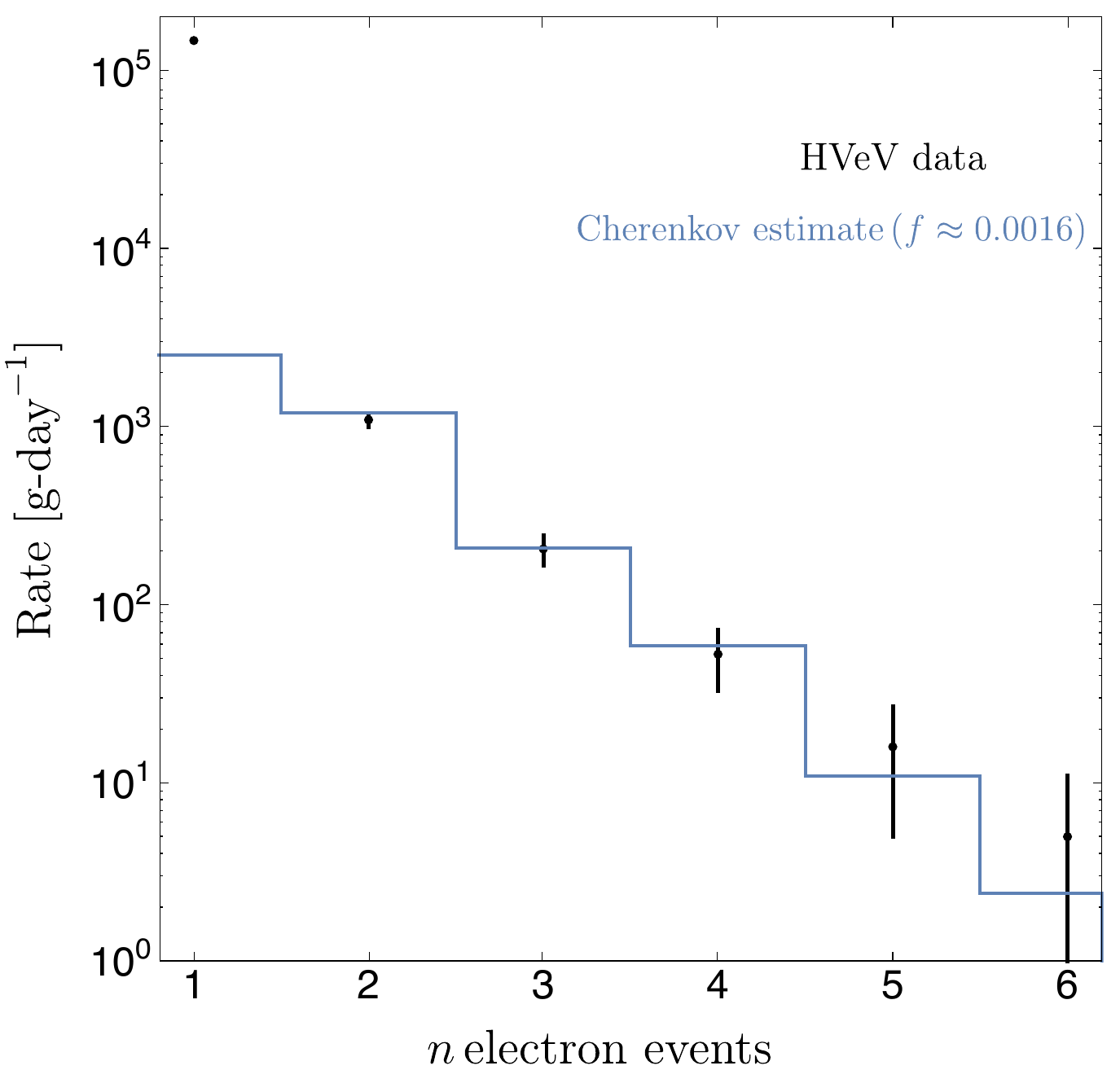}\\
\caption{The estimated $n$-electron event rate from Cherenkov radiation at SuperCDMS HVeV (blue) based on a simple model with several assumptions and a photon detection probability $f \approx 1.6\times 10^{-3}$ (see text for details). The experimental data with $100$~V bias voltage is shown in black with the error bars corresponding to the $3\sigma$ statistical uncertainty~\cite{Amaral:2020ryn}.   \label{fig:HVeV_Ch_rate}
 } 
 \centering
 \end{figure}

In order to calculate the actual rates of $n$-electron events from Cherenkov radiation, 
it is necessary to know the probability $f$ for a photon that has escaped the PCB to
reach the detector target without being absorbed by other detector parts. 
Calculating this detection probability requires a detailed detector simulation, which is beyond the scope of this work.
However, 
an alternative way to obtain this probability is to simply fit the predicted $n$-electron spectrum for a given $f$ to the observed SuperCDMS data, assuming that all the measured $n\geq 2$ events are due to Cherenkov photons (recall that the single-electron events are expected to come from charge-leakage).
With this assumption, and ignoring that $f$ is likely energy-dependent, 
we obtain $f\approx 1.6 \times 10^{-3} $. 
The smallness of the detection probability required to fit the data is indicative of the extremely large number of Cherenkov photons that are being created by tracks at the PCB.
With this one-parameter background model we can predict six observables, namely the Cherenkov-induced $n$-electron spectrum shape, $1\leq n \leq 6$. 
We show the corresponding spectrum shape in Fig.~\ref{fig:HVeV_Ch_rate}.
The figure clearly indicates that Cherenkov photons coming from the detector's PCB could plausibly precisely reproduce the observed event spectrum of the $n\geq 2$ electron events. 

While our hypothesis that Cherenkov radiation explains the observed low-energy events with $n\geq2$ \eh-pairs is plausible, we cannot definitively prove that all, or even most, of the observed events are from Cherenkov radiation.  
For this, we need to know i) the precise spectrum of the high-energy background, ii) the optical properties of the PCBs and all dielectric materials inside the copper vessel, and iii) a better understanding of the detection probability $f$ of each Cherenkov photon etc.  
This requires a careful measurement of the optical properties of the PCB and the other non-conductive materials inside the copper housing, as well as a detailed simulation.  

\ \begin{center} \textbf{Transition Radiation} \end{center}
 
We now discuss the contribution from transition radiation. When charged particles pass through the surfaces of the copper vessel, PCB, or other materials, without intersecting the instrumented target, they will generate photons from transition radiation.  These photons can be registered as an event once they reach the detector. Considering only the surfaces with the largest area, \textit{i.e.}, the inner wall of the copper vessel, we estimate that the total number of transition radiation events is 
\begin{equation} 
N_\textrm{events}^{\textrm{TR}} \sim 1000 \textrm{/g-day}. 
\label{eq:numbertransitionSuperCDMs}
\end{equation}
More details of our estimate can be found in Appendix~\ref{app:SuperCDMS-HVeV}. Comparing this number with $N_\textrm{events}^{\textrm{Cherenkov}}$ in Eq.~(\ref{eq:totalestimateSuperCDMs}) suggests that transition radiation provides a much smaller contribution to the observed event rate.   For example, assuming the same detection probability $f$ mentioned above, transition radiation contributes negligibly to $n$-electron events at SuperCDMS. However, since the spectrum of emitted photon is quite different than that of Cherenkov photons, the detection probability $f$ might not be the same, so we cannot draw a definite conclusion without a dedicated simulation. 
Moreover, the PCB consists of different materials that are woven together; it is possible that a track can make multiple transitions between different materials inside the PCB, thereby generating a large number of transition photons.  

The value of $N_{\rm event}^{\textrm{TR}}$ in Eq.~(\ref{eq:numbertransitionSuperCDMs}) is still very large, so that transition radiation should be carefully studied as a source of low energy backgrounds in this and future versions of SuperCDMS-HVeV.  
In particular, transition radiation could be an important background source even in the absence of non-conductive materials in the detector housing (which would remove Cherenkov radiation). 

\ \begin{center} \textbf{Luminescence} \end{center}

We are unable to say anything quantitative about possible luminescence backgrounds at SuperCDMS-HVeV.  It is possible that some low-energy events are generated if the PCB has a non-zero luminescence, but this requires detailed measurements of its properties.  
That being said, 
it is very important to point out that radiative recombination is expected to be significantly enhanced at cryogenic temperatures (see \textit{e.g.} \cite{bruggemann2017radiative} for the case of recombination in Si at low temperatures), so it could be a significant source of low energy photons.


\subsection{EDELWEISS at Modane}\label{subsec:edelweiss-MeV}

The EDELWEISS Collaboration published a first search for sub-GeV dark matter using a Ge cryogenic target in~\cite{Arnaud:2020svb}.  In this search, electrons produced in the Ge target are drifted with a bias voltage ($\sim$78~V) and amplified, creating phonons via the Luke-Neganov-Trofimov effect~\cite{Neganov:1985khw,Luke:1988}.  While the detection principle is similar to the SuperCDMS HVeV search discussed in the previous section,  an important difference is that the EDELWEISS Ge detector was operated in a well-shielded, underground, low-background environment at the Modane Underground Laboratory.  Nevertheless, a large event rate is observed, peaking at $\sim$$4\times 10^8$~differential-rate units\footnote{One differential-rate unit is defined as 1~event/kg/day/keV.}  (DRU) at an ionization energy equal to $\sim$4~eV, and flattening out at a rate of $\sim$10$^5$~DRU for ionization energies in the range $\sim$15~eV.  
The total exposure was $\sim$2.4~days. 
We will see below that given the quiet environment in which the detector is placed, 
such events cannot originate from high-energy tracks going through the materials nearby the target sensor. 
This is regardless of the mechanism by which the charged tracks produce radiation, 
may it be Cherenkov radiation, transition radiation, or luminescence. 
However, we will also show that Cherenkov backgrounds in particular, 
and perhaps even luminiscence due to \eh-pairs created by tracks, could still be large enough to limit the future sensitivity of 
this experiment to sub-GeV dark matter.  
Transition radiation constitutes likely only a small background, since many of the tracks that pass, \textit{e.g.}, the copper-vacuum interface will also hit the detector and be vetoed. 

 \begin{figure}[t!]
\centering
\includegraphics[width=0.5\textwidth]{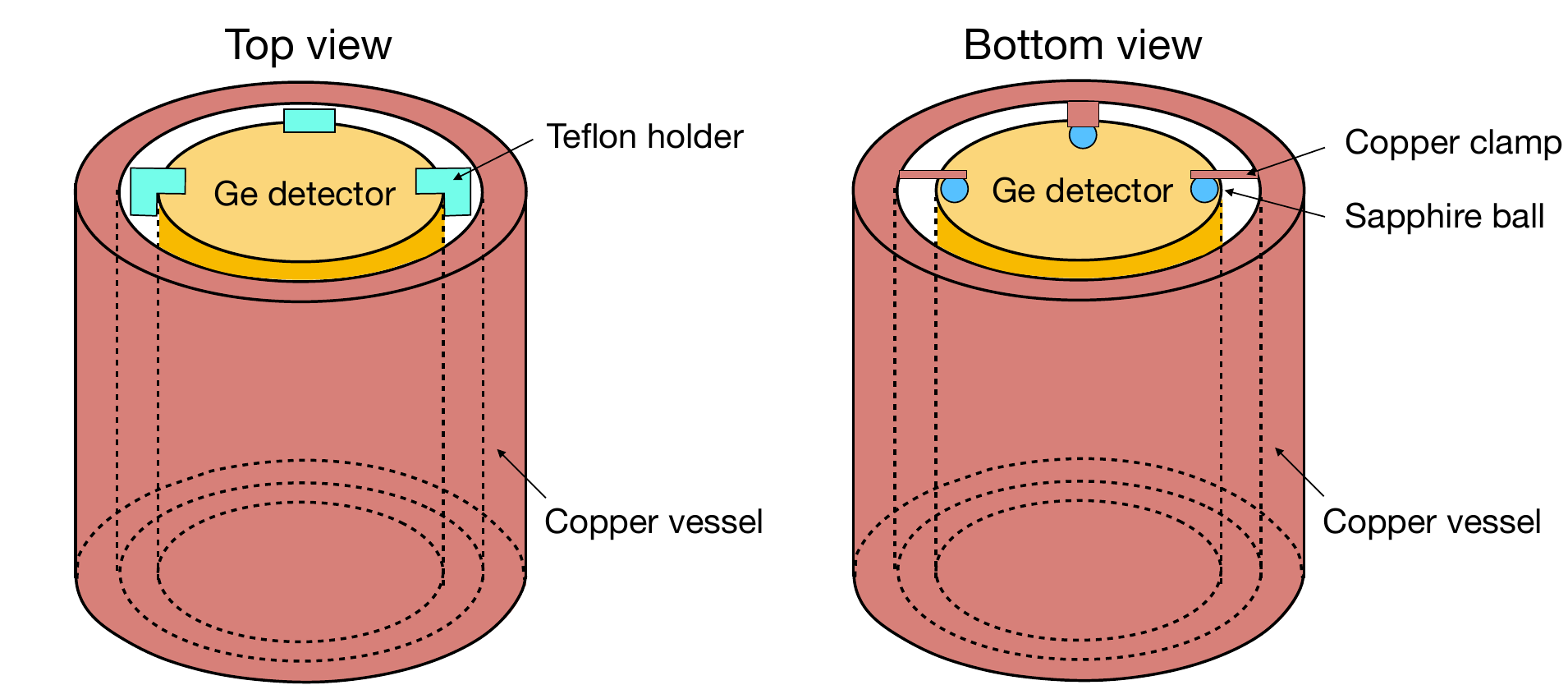}\\
 \caption{Schematic depiction of the EDELWEISS Ge detector module used to take the data described in~\cite{Arnaud:2020svb}.
The teflon holders each have a mass of 0.05~g ($\rho_{\rm PTFE}=2.2$~g/cm$^3$).  The sapphire balls have a diameter of 3.18~mm, and each have a mass of 0.067~g ($\rho_{{\rm Al}_2{\rm O}_3}=3.98$~g/cm$^3$) 
\label{fig:edelweiss-detector}
 }
 \centering
 \end{figure}

A schematic of the EDELWEISS detector is given in Fig.~\ref{fig:edelweiss-detector}.  
It consists of a 33.4~g Ge target that is contained inside a copper housing.  
The detector is clamped between three teflon (PTFE) holders on the one side and three sapphire spheres on the other side.  
High-energy tracks that interact with either the uninstrumented teflon holders or sapphire spheres can produce radiation. 
It is possible that the heat produced by the high-energy events that interact in the teflon is detected also in the EDELWEISS detector~\cite{JulesGascon-discussion}, so that radiation produced in the teflon could be vetoed; however, this does not seem to be possible with the events produced in the sapphire. 

Given the deep underground location of the detector, cosmic high-energy events are rare, 
so the tracks passing through the sapphire or teflon are mostly due to radioactivity intrinsic to these materials or from environmental radioactivity.
The most important source of radioactive backgrounds is the former,
which is due to the contaminants $^{238}$U and $^{40}$K~\cite{Scorza:2015vla}.  
The teflon holders have a measured $^{238}$U contamination of $10\pm5$~mBq/kg, while their $^{40}$K contamination is not reported. 
The contamination of the sapphire spheres is unknown.  
Assuming
that the $^{238}$U and $^{40}$K contamination of the teflon and sapphire is all equal at a rate of 10~mBq/kg, 
we estimate that the 33.4~g Ge detector in 2.4~days would see about 1.5~tracks from internal radioactivity in the sapphire and teflon holders.

Regarding the environmental gamma-ray backgrounds inside the EDELWEISS cryostat, we may use numbers from the 860~g Ge EDELWEISS-III detector that operated in a similar environment~\cite{Scorza:2015vla}. 
For this detector, the total integrated rate from 100~keV to 4~MeV is about 370~counts/kg/day~\cite{Scorza:2015vla,JulesGascon-discussion}.  
If this same rate is incident on the sapphire and teflon, then the number of tracks due to environmental radioactivity in the 2.4~days of exposure is about 0.3.  
Therefore, summing the contributions from high-energy particles from the ambient radioactive background and from impurities, we expect at most a couple of tracks passing through the teflon and sapphire during the detector's exposure, 
which would lead to at most two events. 
The EDELWEISS data shows, however, an event rate that is several orders of magnitude higher.  
We conclude that the observed backgrounds cannot be due to any type of radiation coming from tracks in the main un-instrumented materials surrounding the detector. 
 
Looking towards the future, 
it is worth pointing out that the radioactive high-energy tracks discussed above, even if rare, do still contribute a sizable low-energy background rate. 
Since both sapphire and teflon are dielectric materials,
tracks passing through them will lead to Cherenkov radiation. 
Cherenkov radiation in sapphire was discussed in Sec.~\ref{subsec:cherenkov-theory}, 
where we found that tracks with energies above $\sim$100 keV, 
which is in the typical energy range of radioactive tracks, 
lead to Cherenkov photons.
In addition, electrons and holes produced by tracks in sapphire have been shown to produce luminescence due to defects, leading to 
$\sim$3~eV photons~\cite{ghamnia2003luminescent}.
Regarding teflon, 
the dielectric constant of teflon is measured in~\cite{Yang:2008} to be about $\epsilon_{\rm PTFE}\sim 1.6$ in the energy range 0.8~eV to 3~eV, rising to $\epsilon_{\rm PTFE}\sim 2.1$ at an energy of 8.8~eV; 
this implies that Cherenkov photons can be generated from electrons with energies above $\sim$200~keV to 300~keV,
which again is an energy range that can be expected from radioactive events.  
Consequently, given the high-energy track rate discussed above, 
we would expect $\mathcal{O}$(1)~events/day associated with Cherenkov radiation from such tracks. 
A detailed modeling of the Cherenkov and luminescence backgrounds is therefore warranted to understand more precisely the implications for the future sensitivity of this detector setup to sub-GeV dark matter. 


\subsection{Radiative Backgrounds at Other Existing Experiments}

We briefly comment on radiative backgrounds in a few other experiments that observe low-energy events above what is expected from Compton and other radiogenic backgrounds.  
\begin{itemize}[leftmargin=*]\addtolength{\itemsep}{-0.5\baselineskip}

\item \textbf{CRESST-III}~\cite{Abdelhameed:2019hmk,Abdelhameed:2019mac}: 
Using a 23.6~g CaWO$_4$ cryogenic crystal target, the CRESST-III Collaboration achieved a nuclear recoil threshold of $30.1$~eV and an exposure (after cuts) of 3.64~kg-days.  The target was well-shielded and placed underground at the Laboratori Nazionali del Gran Sasso (LNGS).  An excess of events above that expected from radioactive backgrounds is observed at energies well below $\sim$200~eV.  

We find that it is unlikely for Cherenkov radiation, transition radiation, or luminescence to be an important background in 
the final dataset after cuts have been applied, since all holders of the CaWO$_4$ crystal target and the vast majority of non-conductive material inside the copper housing has been designed to scintillate if it interacts with a high-energy particle.  
In this way, while many Cherenkov and possibly recombination photons are generated inside the copper housing, it is easy to veto them by observing the accompanying scintillation from the high-energy event that generates the radiation.  

Research into the origin of the low-energy events is ongoing, and plausible hypotheses that are being investigated include cracking or micro-fracturing of the crystal or support holders that occur when the detectors are being tightly clamped~\cite{Astrom:2005zk}. 
Another possibility that deserves investigation is luminescence from the various detector materials that only occurs well after a track has passed through it (phosphorescence).

\item \textbf{SuperCDMS Cryogenic PhotoDetector (CPD)}~\cite{Alkhatib:2020slm}: 
Using a 10.6~g Si athermal phonon detector with an energy resolution of $\sim$5~eV and a raw exposure of 9.9~g-days, the SuperCDMS Collaboration obtained the strongest constraints on elastic dark matter-nucleon scattering for dark matter particle masses from 87 to 140~MeV.  The detector was operated on the surface, where it was subject to a large environmental background rate of $\sim$few~$\times 10^5$~DRU at $\mathcal{O}$(keV) energies.  A large number of events---much larger than that expected from background photons that Compton scatter in the target Si---are observed for  recoil energies below $\sim$150~eV.  

Radiative backgrounds may arise, however, from uninstrumented auxiliary materials.
The detector is clamped between six Cirlex holders (three on the top, three on the bottom), which can generate radiation when high-energy electrons and muons interact with them.  
The CPD detector operates without a bias voltage,
has very little position resolution (so radiated photons that interact at the surface of the detector likely cannot be vetoed), 
and has only a single detector inside the copper housing (so photons cannot be vetoed by using a coincident signal between two or more detectors).
We expect Cherenkov or recombination radiation to be dominantly generated by environmental background radioactive events incident on the Cirlex clamps (rather than from impurities in the Cirlex), and transition radiation to be generated from environmental backgrounds and cosmic-rays passing through the various inner-detector layers.  
Including muons from~\cite{Bogdanova:2006ex} and naively scaling up the number of tracks observed in the SENSEI MINOS data~\cite{Barak:2020fql} to the background rate observed in the CPD detector at $\mathcal{O}$(keV) energies, 
reveals that radiation from tracks passing through the Cirlex cannot make up for more than 10\% of the observed signal.  
Given the large observed event rate in the CPD detector, this implies that radiation from tracks, 
which could for instance arise via the Cherenkov effect, 
still constitutes a significant background in this experiment, 
even if in this case it is subdominant to whatever else is causing the low-energy events.  

\item \textbf{EDELWEISS-Surf Ge bolometer~\cite{Armengaud:2019kfj}}: 
Using a 33.4~g Ge cryogenic target operated with 
a neutron-transmutation-doped Ge thermal sensor, the EDELWEISS collaboration achieved a heat-energy resolution of 17.7~eV and constrained elastic dark matter-nucleon scattering for dark matter masses above $\sim$600~MeV.  
An excess above an approximately flat Compton background is observed for recoil energies below $\sim$150~eV.  
As discussed above in Sec.~\ref{subsec:edelweiss-MeV}, Cherenkov radiation can be generated from high-energy electrons and muons interacting with the teflon (PTFE) holders.  
Transition radiation can also be generated from the passage of high-energy events through various detector layers. 
A detailed investigation of what fraction of the observed excess events is from such radiative backgrounds is beyond the scope of this paper.  

\item \textbf{Optical Haloscopes~\cite{Baryakhtar:2018doz}}: 
Dark matter may be composed of light bosonic fields, which behave as a background dark matter oscillation. 
Using layers of dielectric materials, \cite{Baryakhtar:2018doz} proposes to enhance the conversion of the background dark matter, 
dark photons in particular, into visible photons. 
The visible photons can be focused into a small super-conducting nanowire detector using a lens.\footnote{We thank Masha Baryakhtar, Junwu Huang, Robert Lasenby, and Sae-Woo Nam for detailed information on the detector setup.}
In this setup, backgrounds can arise from tracks that pass through the lens, which emit a large number of Cherenkov photons that mimic the dark matter signal.

\end{itemize}


 \section{Radiative Backgrounds at Future and Proposed Experiments}\label{sec:experiments-future}

In this section, we discuss radiative background at the upcoming SuperCDMS SNOLAB 
experiment~\cite{Agnese:2016cpb}, comment briefly on upcoming and proposed large-exposure Skipper-CCD detectors (SENSEI at SNOLAB, DAMIC-M, and Oscura)~\cite{Castello-Mor:2020jhd,BRN-announcement2}, and also comment on proposed experiments based on athermal phonon detection and low-energy scintillation~\cite{Derenzo:2016fse,Knapen:2017ekk,Arvanitaki:2017nhi,Essig:2019kfe,Blanco:2019lrf,SPICE}. 

\subsection{SuperCDMS at SNOLAB}\label{subsec:SuperCDMS-SNOLAB}

The upcoming SuperCDMS at SNOLAB experiment will use two types of cryogenic detectors, denoted HV and iZIP, and two types of 
target materials, namely Si and Ge~\cite{Agnese:2016cpb}.  The goal is to operate the HV detectors with a bias voltage ($\sim$100~V) that will amplify 
ionization signals produced in the target material by drifting electrons and creating phonons via the Luke-Neganov-Trofimov effect~\cite{Neganov:1985khw,Luke:1988}.  
We estimate below the expected background from Cherenkov photons produced from non-conductive materials inside the detector housing for the HV detectors.  We will argue that the Cherenkov background may dominate over other previously considered backgrounds at low energies, but a careful simulation and more details about the optical properties of Cirlex are needed to estimate precisely this background.  Additional details about our estimate can be found in Appendix~\ref{app:SuperCDMS-SNOLAB}. 
We do not attempt here an estimate of transition radiation or of the radiative recombination rate in Cirlex. 

At SuperCDMS SNOLAB, 
there will be eight HV detectors with a Ge target, 
each with a mass of 1.39~kg, 
and four HV detectors with a Si target,
each with a mass of 0.61~kg.  
These targets are placed in two separate towers, with each tower surrounded by its own copper housing, and with each tower consisting of four Ge and two Si HV detectors.  
Each HV detector inside this tower is held by 12 clamps (six on the top, six on the bottom) made of Cirlex.  
Each clamp has a mass of 0.17~g.  
In addition, four straight and four curved detector interface boards made of Cirlex sit on the edge of the copper housing (near the top and bottom of each HV detector) and have a direct view to the HV detector; the mass of each such detector interface board is $\sim$0.21~g. 
The total mass of Cirlex that sits inside or along the side of the copper housing for each HV detector is therefore about $\sim$3.72~g, or about $\sim$22.3~g for one tower consisting of six HV detectors. 

Cirlex is a non-adhesive laminate made from 100\% Kapton polyimide film~\cite{Cirlex}.  
Cirlex has the same chemical, physical, thermal, and electrical properties as Kapton polyimide.  Since we are unable to find in the literature the measured optical properties of Cirlex specifically, we assume in our estimate below that its optical properties are the same as for Kapton polyimide~\cite{Arakawa:1981,Ahmed:2012}.   
Kapton polyimide has a density of 1.43~g/cm$^3$ and a dielectric constant of $\epsilon\sim 3$~\cite{Arakawa:1981} (which varies $\lesssim$10\% for energies in the relevant $\mathcal{O}$(eV) range, see below).  
Cherenkov photons can thus be generated from electrons that have a kinetic energy $\gtrsim$115~keV.  

The dominant source of high-energy electrons will come from beta decays of the radioactive impurities $^{238}$U, $^{232}$Th, and $^{40}$K, which are present in the Cirlex~\cite{Agnese:2016cpb}.
These radioactive impurities are found to have the following concentrations: $\sim$14~mBq/kg for $^{238}$U, $\sim$4.5~mBq/kg 
for $^{232}$Th, and $\lesssim$5.3~mBq/kg for $^{40}$K~\cite{BenLoer-discussion,SuperCDMS-background-paper-to-appear}.  
Based on this information, we estimate that in the 22.3~g of Cirlex present in one tower of six detectors, the rate for producing energetic electron tracks from radioactivity is $\sim$200~events/day or $\sim$75,000~events/year.   
A significant fraction of these tracks will have energies above the $115\, \textrm{keV}$ threshold for the production of Cherenkov photons in Cirlex, namely 
\begin{equation}
N_\textrm{events}^\textrm{Cirlex} \sim 130~\textrm{events/day} \sim 50,000~\textrm{events/year}\ .  
\end{equation}
In addition, the Cirlex may have some luminescence (from, \textit{e.g.}, recombination of \eh-pairs), 
and we are unaware of measurements in the literature that constrain this possibility.  
We focus our remaining discussion in this subsection on Cherenkov radiation, although some of our comments will also be applicable to luminescence. 

To understand the possible signals generated in the HV detectors by these beta decays, we need to know how many Cherenkov photons are produced and how many of these are detected in the HV detectors.  
Cirlex (assuming it has the same optical properties as Kapton polyimide) has a direct bandgap of $\sim$1.5~eV and an indirect bandgap of~$\sim$1.8~eV~\cite{Ahmed:2012}, so that photons with energies $\gtrsim$1.5~eV are likely absorbed quickly once produced.  
However, photons with energy $\lesssim$1.5~eV could leave the Cirlex and be absorbed in the HV detector.  
As discussed in Sec.~\ref{subsec:cherenkov-theory} (\textit{c.f.}~Fig.~\ref{fig:spectrum})
the number of Cherenkov photons emitted by an electron depends on the track energy, which sets the track length.
The largest number of Cherenkov photons that can be generated by the above radioactive contaminants comes from the highest-energy beta 
decay, which is 3.27~MeV at the endpoint of the $^{214}$Bi decay, which appears in the $^{238}$U decay chain. 
The electron from this beta decay could travel as much as $\sim$12.7~mm inside the Cirlex and produce a maximum of $\sim$106 ($\sim$236)~Cherenkov photons, with energies uniformly distributed in the range 1.16~eV to 1.5~eV (0.74~eV to 1.5~eV), where the lower bound on the energy was chosen to correspond to the Si (Ge) bandgap~\cite{VARSHNI1967149} and the upper bound corresponds to the direct energy gap of the Cirlex.  
In contrast, an electron with an energy of 200~keV would on average produce only $\sim$1.3 ($\sim$3.0) Cherenkov photons in the relevant energy range for Si (Ge).  

A careful simulation is needed to determine how many Cherenkov radiation events make it to the detector and remain after several obvious 
analysis cuts. 
First, a gamma-ray or x-ray may accompany the beta-decay and be absorbed in an HV detector, allowing the event to be vetoed. 
Second, events that produce many Cherenkov photons can be vetoed, 
since they will produce a coincident signal in two or more HV detectors; however, events that produce only a few photons may be difficult to remove with an anti-coincidence veto.  
 Another useful discriminant could be to veto events that do not penetrate deeply into the cylindrical side walls of the HV detectors.  
This allows one to veto some events that produce Cherenkov photons that all hit a single detector, since such events would often contain multiple photons, some of which would be well above the Si or Ge bandgap and hence would be absorbed on the surfaces. 
However, even in this case, some beta decays could produce a few photons all of which are within $\sim$0.1--0.2~eV of the Si or Ge bandgaps, 
so they would have a large absorption length and penetrate into the bulk. 
 
In~\cite{Agnese:2016cpb}, the expected number of events (before any analysis cuts) with energy $\lesssim$50 (100)~eV in the HV detectors, from the HV detector-bulk contamination and ambient background incident on the HV detectors, is $\sim$9 (18)~events/year in one Si HV detector and $\sim$2 (4)~events/year in one Ge HV detector.  Based on our discussion above, it is conceivable that the Cherenkov background dominates in this energy range over the backgrounds that were considered previously.  This would affect the projected SuperCDMS SNOLAB sensitivity on various dark matter models.  
However, a careful study of the optical properties of Cirlex and a detailed simulation of the Cherenkov-photon background is needed to determine these implications precisely and whether this background is a real concern. 
In addition, measurements that determine the luminescence of the Cirlex would also be important.  
Appendix~\ref{app:SuperCDMS-SNOLAB} contains additional details relevant to our discussion above, as well as brief comments on the possibility of operating some, or all, of the six HV detectors with a zero bias voltage. 

\subsection{Future Skipper-CCD Experiments}

We will discuss in~\cite{SENSEI-radiative-to-appear} the implications of radiative backgrounds for the future Skipper-CCD searches SENSEI at SNOLAB (100~g), DAMIC-M at Modane (1~kg)~\cite{Castello-Mor:2020jhd}, and Oscura (10~kg)~\cite{BRN-announcement2}.  We note, however, that these upcoming and proposed experiments will have a  drastically improved shielding, 
which will reduce radiative backgrounds coming from high-energy tracks.  
For example, SENSEI at SNOLAB is aiming for a background rate that is about three orders of magnitude better ($\sim$5~DRU at $\mathcal{O}$(keV) energies) than the SENSEI data taken at MINOS ($\sim$3000~DRU at $\mathcal{O}$(keV) energies~\cite{Barak:2020fql}), while DAMIC-M and Oscura are aiming for a background rate of about 0.1~DRU and 0.01~DRU, respectively.  
This would lead to a very low single-electron event background rate from track-induced radiation, but a careful analysis of 
these backgrounds is needed.  

\subsection{Future Dark Matter Detectors Searching for Photons with Scintillators and Molecules}\label{subsec:future-photons} 

Radiative backgrounds could mimic a dark matter signal at proposed detectors that search for one or more photons from dark matter interactions in, \textit{e.g.}, solid-state scintillators~\cite{Derenzo:2016fse}, molecular targets~\cite{Arvanitaki:2017nhi,Essig:2019kfe}, and organic aromatic materials~\cite{Blanco:2019lrf}.  Radiative backgrounds could either excite the target material, which subsequently produces a photon, or directly hit the photodetector.  

As a concrete example, we consider the proposed SPICE detector, which aims to detect 
the scattering or absorption of light-dark matter on a GaAs target  
by measuring a scintillating signals obtained from the event~\cite{SPICE,Derenzo:2016fse} (SPICE also aims to detect phonons from dark matter events, see Sec.~\ref{subsec:future-collective}). 
The proposed detector setup consists of multiple targets with a volume of order $\textrm{cm}^3$ each,
with independent sensing devices attached to each target.
Cherenkov or recombination radiation from high energy tracks passing through non-instrumented dielectric materials, 
such as the detector holders, or transition radiation from high-energy tracks passing through two or more layers 
with different refractive indices, can produce photons that could make it into their detector target.
In the GaAs target, photons with energy $\gtrsim$1.5~eV can be directly absorbed by the photodetector or they can be absorbed by the GaAs target that subsequently scintillates and emits a secondary photon; in both cases such photons would mimic a dark matter signal. 

\subsection{Future Dark Matter Detectors Searching for Collective Excitations}
\label{subsec:future-collective}

The search for collective modes in solid-state detectors, such as phonons and magnons, 
have been suggested as a way to probe sub-MeV dark matter scattering off, and sub-eV dark matter being absorbed by, 
a target material~\cite{Knapen:2017ekk,Trickle:2019ovy,Trickle:2019nya}.   
The background processes discussed here can also mimic dark matter signals for these searches. 

As a concrete example, we again consider the SPICE detector (see also Sec.~\ref{subsec:future-photons}).  
One of the goals with SPICE is to detect the scattering or absorption of matter on a sapphire target by 
measuring sub-eV phonon signals obtained from the event~\cite{SPICE}.
While achieving sub-eV energy resolutions is currently still challenging,
the proposal aims to develop novel transition-edge sensors to obtain the required sensitivities.

The authors of the proposal identify two possible backgrounds affecting future dark matter searches in their detectors, 
neutrinos and radiogenic backgrounds, 
and indicate that after discriminating high-energy events from their low-energy signals, 
zero backgrounds are to be expected~\cite{Knapen:2017ekk}. 
Another potential background comes from the low-energy coherent scattering of a photon off nuclei~\cite{Robinson:2016imi}, but 
these can likely be mitigated using an active veto. 
However, backgrounds such as Cherenkov, transition, or recombination radiation, as well as non-radiative recombination have not been discussed previously in the literature.
 
Regarding Cherenkov or recombination radiation, high energy tracks passing through non-instrumented dielectric materials, 
such as the detector holders, lead to photons with a variety of energies. 
Such photons may escape the non-instrumented materials, 
and make it into their detector target.
In particular, Cherenkov photons can have energies that match the lattice modes of the sapphire detector.
In this case, they will convert into optical phonons in the target, and thus constitute a background.
A plausibility argument for this background can be made as follows.
Take the radioactivity of the teflon holders at EDELWEISS as an example (10~mBq/kg), 
a total mass of 1~g in holders made of some insulating material, 
a detector with 1~kg of fiducial target, 
and an effective exposure of 10 kg-year as indicated in~\cite{SPICE}. 
In this case, 
$\mathcal{O}(3000)$ charged tracks are expected in the holders. 
According to the estimate in Eq.~\eqref{eq:naiveestimate}, each track will lead to $\mathcal{O}(1)$ Cherenkov photons with energies below 100~meV, which may then convert to phonons in the sapphire detector.
It is likely that most of these events will be vetoed, 
as tracks will also radiate higher energy Cherenkov photons, which will lead to large energy depositions in the detector that are not consistent with the dark matter signal region.
However, there will also be low-energy charged tracks;  
in this case, Cherenkov photons are radiated only 
when the dielectric constant of the insulating holders is extremely large, which happens only near lattice modes of the holders, 
as discussed in Sec.~\ref{subsec:cherenkov-theory}. 
Thus, such tracks would \textit{only} emit Cherenkov photons with energies of the order of phonon modes ($\sim$100~meV and below).
These events, although rare, would likely not be vetoed, and could pass as a dark matter signal. 
A concrete example of this possibility is shown in Fig.~\ref{fig:spectrum}, 
where we see that a 20~keV electron track in $\alpha$-quartz only leads to Cherenkov radiation near lattice modes.
It may be desirable therefore for any material near the detector to consist of a non-polar material, or of a polar material with lattice modes 
that do not match the lattice modes of the target, 
or a material that does not produce Cherenkov radiation altogether like a conductor.  
However, even in this case photons may be generated by transition radiation, which needs to be studied carefully.  

Another potential background for a phonon search is non-radiative recombination. 
A track passing through uninstrumented materials will lead to a large number of \eh-pairs. 
As discussed in Sec.~\ref{sec:recombination}, 
these pairs may relax and recombine emitting phonons, 
or go towards band minima via the emission of phonons. 
If an un-instrumented material is in contact with the detector, 
a few such phonons may be transferred into the detector, 
mimicking the dark matter signal. 

In all cases, a precise characterization of the properties of all material surrounding the detector is crucial to avoid 
these low-energy backgrounds. 


\section{Radiative Backgrounds at Neutrino Experiments and Superconducting Qubits}\label{subsec:other}
Up to this point we have focused on sub-GeV dark matter searches, but it is worth noting that the backgrounds that we have identified could also be a  concern for other research areas.  For example, these backgrounds would also appear in searches for coherent scattering between neutrinos and nuclei with Skipper-CCDs such as CONNIE~\cite{Aguilar-Arevalo:2019jlr} and $\nu$IOLETA~\cite{Fernandez-Moroni:2020yyl} .

Moreover, the radiative backgrounds we pointed out in this paper have could have a significant impact on quantum computing as they could limit the coherence time of superconducting qubits. It has been shown in several references~\cite{Vepsalainen:2020trd,Cardani_2021,Wilen:2020lgg,Mcewen:2021ood} that high rates of charged particles from radioactivity and cosmic rays correlate with high quasiparticle (broken Cooper pairs) densities in superconducting qubits and reduced coherence times. 
The current explanation is that phonons generated by these charged tracks in a dielectric substrate interact with the qubits and create quasiparticles. 
However, the impact on qubit coherence times of low energy photons from Cherenkov radiation, transition radiation, and luminescence from recombination have not been discussed in the literature. As mentioned in previous sections, the Cherenkov process and luminescence from recombination in semiconductors typically lead to photons with energies $\mathcal{O}(\eV)$ or below. Superconducting qubits made of Al efficiently absorb photons at these energies. 
Once absorbed, these low energy photons can break Cooper pairs and create quasiparticles. 
Therefore, low energy photons have similar effects on qubits as phonons and they must be carefully investigated. 
While a detailed analysis is beyond the scope of this paper, we now present a simple estimate that demonstrates the relevance of Cherenkov and recombination photons produced by tracks passing through the dielectric substrate. 

We focus on photons near or below the bandgap since after being produced, they can travel sufficiently far in the substrate to eventually reach the qubits and create quasiparticles.
Let us first consider Cherenkov photons.
A single track passing through the substrate leads to $\sim \alpha\times E_g \times \Delta x$ sub-gap Cherenkov photons, 
 where $\Delta x$ is the track length (see Eq.(\ref{eq:cherenkov})). 
Thus, the total energy in sub-gap Cherenkov photons created by a single track is $\sim \alpha\times E_g^2 \times \Delta x$.
For example, a sapphire (Al$_2$O$_3$) substrate has bandgap of 
 $\sim$$7\ev$. 
 Thus, a  $300\um$ track passing through such substrate can release $\sim$$200\ev$ in the form of long-lived, sub-gap Cherenkov photons, which can create a large number of quasiparticles as they are absorbed by the qubits.
Note that a substrate with larger bandgap will produce more energy that can be absorbed by the qubit. 

Moreover, substrates with high radiative recombination efficiency 
will produce a large number of luminescence photons as tracks pass through.
For instance, 
the luminescent yield of sapphire is $\sim1\%$~\cite{Coron:2004iy}, which means that a $\sim 100 \kev$ track can release more than 1~keV in the form of $\mathcal{O}(\eV)$ photons. Since the sapphire substrate is basically transparent to these low energy photons, almost all such photons will be absorbed in the qubits if there are no other absorptive materials. 

These consideration suggest that materials with low bandgaps and low radiative recombination efficiency, such as Si and Ge, could dramatically reduce radiative backgrounds for superconducting qubits, and may thus constitute optimal substrates to mitigate them.

 
 \section{Characterization and Mitigation of Radiative Low-Energy Backgrounds}\label{sec:mitigation}
 
The radiative backgrounds discussed in this paper have significant implications for future searches of low-mass dark matter, both in designing  future detectors and in calculating their low-energy backgrounds.  Moreover, these backgrounds affect searches for dark-matter-induced electromagnetic excitations (electrons or photons) and excitations of collective modes (such as one or more phonons or magnons).  
Fortunately, there are several mitigation and characterization strategies.  Several of these strategies are already employed to reduce radiogenic backgrounds and are not specific to mitigating the backgrounds discussed in this paper, but others are not universally used.  
\begin{itemize}[leftmargin=0.1cm,itemindent=.5cm,labelwidth=\itemindent,labelsep=0cm,align=left]\addtolength{\itemsep}{-0.6\baselineskip}
\item \textbf{Increase passive shielding.}  Passive shielding with lead, copper, and other materials will reduce the number of photons that can interact in the detector or the material that surrounds it, thereby reducing the number of backgrounds from Cherenkov radiation, transition radiation, recombination, and bremsstrahlung.  
\item \textbf{Increase active shielding.}  In detectors with excellent timing resolution, an active veto can detect a high-energy event in coincidence with a low-energy event,
in which case the low-energy event can be vetoed.
  For example, the CRESST-III detector partially employs an active shield, for which which the materials inside the copper housing will scintillate when interacting with a high-energy event.  
\item \textbf{Minimize non-conductive materials near sensors.}  
Reducing non-conductive materials and, when feasible, either replacing them with conducting materials or covering them with a conductive surface, will reduce Cherenkov radiation and recombination-induced backgrounds.  
It may also be possible to add coatings on the non-conductive materials to absorb low-energy photons, although it is important to check that the coating itself does not lead to additional radiation that can mimic low-energy events in the sensor. 
We note that transition radiation can be generated in all materials.
\item \textbf{Radiopure materials.}  In addition to high-energy events from ambient background radiation, a second, often dominant, source of high-energy events can arise from radioactive contaminants inside the detector materials.  
These can, for example, beta-decay to produce electrons that create Cherenkov radiation.  
The careful selection and screening of materials will be crucial to reduce these sources of high-energy events that can create Cherenkov radiation, transition radiation, and recombination-induced backgrounds.
\item \textbf{Multiple sensors.}  Having multiple nearby sensors (inside a common copper shield) can drastically reduce low-energy events that mimic dark matter by allowing events to be vetoed in which two or more sensors detect events in the same time window.  For example, Cherenkov radiation often produces multiple low-energy photons, which can be vetoed if the photons get absorbed in two or more sensors.  However, Cherenkov radiation in which only a one or a few photons are produced that are all absorbed by one sensor would still constitute a background. It may also be challenging to veto transition radiation using multiple sensors, since few photons are typically produced.

\item \textbf{Precise measurements of properties of all materials.}  
Our work makes it clear that the low-energy backgrounds at different experiments depend very sensitively on the detailed properties of detector materials. 
Estimating the low-energy background rates requires precise knowledge of each material's relative permittivity and extinction coefficient, 
recombination coefficients, 
electron and hole mobilities, etc, 
all of which must be specified at the detector's operating temperature. 
These properties are often not known precisely in the literature for the materials used in different dark-matter detectors, 
so should be measured in each case. 

\item \textbf{Thinning of backside Skipper-CCDs.} 
The backside of CCDs are routinely thinned for astronomical applications, but it naively seemed not to be necessary for dark matter searches.  However, we have shown that a careful study of the backside thickness and doping levels is needed.  A ``thick'' layer heavily doped with phosphorus will help absorb near-bandgap photons from Cherenkov and radiative \eh\ recombination, preventing such photons to travel far away from the high-energy charged particle track.  However, such a thick layer will also lead to many photons from radiative \eh\  recombination.  We will discuss this further in~\cite{SENSEI-radiative-to-appear}. 

\item \textbf{Thin layer of absorptive material on surfaces.}  
It may be useful to cover all surfaces with a thin layer of material that absorbs low-energy photons.  Such layers may create more radiative backgrounds, but could be beneficial if they absorb light that can create events in the target. 

\item \textbf{Substrate with low bandgap and low luminescence for superconducting qubits.}  
As mentioned in Sec.~\ref{subsec:other}, low energy photons generated from the substrate can create quasiparticles in qubits and reduce the coherence time. 
Using substrates with low bandgaps and small luminescence rates, such as Si or Ge, could alleviate these backgrounds. 

\end{itemize} 

The mitigation strategies above quite plausibly reduce these backgrounds to manageable levels.  However, a detailed investigation of these backgrounds is necessary for for each experiment.

 \section{Summary and Conclusions}\label{sec:conclusions}
 

In this paper, we discussed four processes that can produce low-energy events in low-threshold direct-detection experiments and which have not been previously considered as important backgrounds: Cherenkov radiation, transition radiation, and luminescence or phonons from recombination.  These processes can generate low-energy ($\sim$meV to few-eV) photons or phonons that can be absorbed by the target or sensor and, depending on the specific target properties, create one or a few \eh\-pairs, phonons, or photons that can mimic the signal expected from a wide range of sub-GeV dark matter interactions.  
Cherenkov radiation is generated when high-energy charged particles (electrons and muons, from cosmic-rays or radioactivity) above some threshold energy interact with the detector target material or any non-conductive material surrounding the target.  
Transition radiation is generated when charged particles cross the boundary between two materials with different relative permittivities.  
Recombination can occur between electron and holes generated after the passage of a high-energy charged track.  

We discussed these backgrounds in the context of several existing and proposed experiments: 
\begin{itemize}[leftmargin=0.1cm,itemindent=.5cm,labelwidth=\itemindent,labelsep=0cm,align=left]\addtolength{\itemsep}{-0.6\baselineskip}
\item 
A sizable fraction of the $\sim$450/g-day of single-electron events seen in SENSEI data from a Skipper-CCD at a shallow underground site arise from Cherenkov radiation and radiative recombination.  Cherenkov photons close to the Si bandgap are generated by charged particles interacting (mostly) in the Si Skipper-CCD, which can travel ``far'' away from the high-energy track before being absorbed.  
In addition, photons close to the bandgap are also generated by the radiative recombination of \eh-pairs created by charged tracks in a $\mathcal{O}$($\mu$m) layer of Si heavily doped with phosphorus on the backside of the Skipper-CCD.  
Transition radiation provides a negligible contribution to the observed rate.  
We defer to a companion paper~\cite{SENSEI-radiative-to-appear} a precise calculation of these backgrounds, which is challenging and must include a careful modeling of the Skipper-CCD structure near and on its surfaces.  
The above backgrounds needs to be evaluated carefully for future searches using Skipper-CCDs, including SENSEI at SNOLAB, DAMIC-M, and Oscura, although all of them will be greatly aided by the planned excellent passive shielding. 
\item 
We find that Cherenkov and transition radiation provide a plausible origin of the events containing 2 to 6 electrons in the SuperCDMS HVeV surface data.  Cherenkov radiation, which likely dominates over transition radiation, is produced in the printed-circuit-boards and plastic connectors located inside the copper housing near the Si HVeV detector, while transition radiation is produced when charged tracks cross into the copper housing or between two materials.  A detailed simulation and the detailed optical properties (currently unavailable) of the non-conductive materials in the detector housing are required to evaluate these backgrounds precisely. 
\item 
Even experiments that are well-shielded could be affected by these backgrounds.  For example, we showed that Cherenkov radiation from radioactive contaminants in the $\sim$0.35~g of teflon and sapphire holders may produce a raw rate of $\mathcal{O}(1)$~potential low-energy-events/day in the EDELWEISS Ge detector.  This is too small to explain their currently observed low-energy event rate, but is sufficiently large to warrant a careful investigation for future searches.  
\item We find that our backgrounds are unlikely to explain the excess events observed at CRESST-III~\cite{Abdelhameed:2019hmk,Abdelhameed:2019mac}, since the materials inside the copper vessel scintillate when interacting with a high-energy particle and Cherenkov events would thus be vetoed. 
\item We find that Cherenkov radiation or luminescence would not be able to explain more than about 
10\% of the large observed low-energy event rate at the SuperCDMS CPD detector~\cite{Alkhatib:2020slm}. 
\item 
We pointed out that beta-decays of radioactive contaminants ($^{238}$U, $^{232}$Th, and $^{40}$K) in the $\sim$22.3~g of Cirlex located inside or on the copper side walls of a tower of six SuperCDMS SNOLAB detectors will produce a raw rate of $\sim$75,000~potential events/year containing one to several Cherenkov photons.  While not all will make it to the detector and while most can be easily vetoed, a careful analysis is needed to determine the surviving background. 
\item We discussed how the radiative backgrounds could also mimic a dark matter signal in proposed searches in which the signal consists of one or more photons.  Typical targets consist of, \textit{e.g.}, solid-state scintillators (as for the SPICE detector), molecular gases, or organic aromatic materials. 
\item 
We also showed that the radiative and non-radiative backgrounds can mimic a dark matter signal in proposed searches in which the signal consists of one or more phonons  (such as SPICE) .  In particular, sub-gap photons with energy near the phonon modes of the target material can be produced in materials surrounding the target and be absorbed by the target to produce a phonon.
Phonons from \eh recombination generated in detector clamps or holders may also constitute backgrounds.
\item Finally, we pointed out that the backgrounds that we have identified could also be a concern for other, non-dark-matter experiments.  Examples include searches for coherent scattering between neutrinos and nuclei with Skipper-CCDs (such as CONNIE) and the decoherence of superconducting qubits.  
\end{itemize}

Fortunately, having now identified these unexplored radiative and non-radiative backgrounds, we can point out several mitigation and characterization strategies.  Mitigation strategies include using a large passive shield, using an active shield, using radiopure materials, minimizing non-conductive materials near the target material and the sensors, using multiple nearby sensors to veto coincident low-energy events, and (for detectors using Skipper-CCDs) investigating if thinning the backside of CCD detectors can reduce backgrounds. 
The first three strategies are already employed to reduce radiogenic backgrounds, but the latter are not universally used.  

In addition, a careful calculation of the backgrounds studied here is of utmost importance, 
as it could enable the different experimental collaborations to perform background subtraction and enhance their reach to detect dark matter.
Calculating these backgrounds requires detailed and detector-specific simulations, which could be carried out with GEANT4 \cite{Agostinelli:2002hh}, using the input from theorists, the different experimental collaborations and GEANT4 developers.
Note that these simulations also require precise knowledge of the optical properties of all materials in the detector. 
While these are not necessarily difficult to measure, they are not currently available to sufficient precision for many materials commonly found in dark matter detectors.

Finally, we note that there are other radiative backgrounds that we did not consider in detail, such as bremsstrahlung and diffraction radiation.  
The bremsstrahlung rate is $\mathcal O(\alpha^3)$ but enhanced at low photon-emission energies and at low charged-particle energies.  However, a simple estimate suggests, for example, that its contribution to the single-electron event rate in the SENSEI MINOS data is about two orders of magnitude less than the observed rate.  
Diffraction radiation occurs when a charged particle travels close to the interface between two materials. 
We leave a detailed study of these and other backgrounds to future work. 

 \acknowledgments

We are especially grateful to Donald Groom, Stephen Holland, Noah Kurinsky, Ben Loer, Javier Tiffenberg, and Sho Uemura for several extensive discussions and correspondence.  
We have also benefitted greatly from numerous insightful discussions and/or correspondence with Quentin Arnaud, Julien Billard, Mariano Cababie, Alvaro Chavarria, Cyrus Dreyer, Juan Estrada, Marivi Fern\'andez-Serra, Enectali Figueroa-Feliciano, Yonatan Ben Gal, Jules Gascon, Daniel Gift, Alexandre Juillard, Ben Loer, Guillermo Fernandez Moroni, Sravan Munagavalasa, Federica Petricca, Franz Pr\"obst, Matt Pyle, Florian Reindl, Alan Robinson, Jochen Schieck, Aman Singal, Christian Strandhagen, Raimund Strauss, Sho Uemura, Jerry Va'vra, and Tomer Volansky.  
RE acknowledges support from DoE Grant DE-SC0009854, Simons Investigator in Physics Award~623940, the Heising-Simons Foundation Grant No.~79921, and the US-Israel Binational Science Foundation Grant No.~	2016153.  
PD and MS are supported in part by Simons Investigator in Physics Award~623940.  
PD is also supported in part by NSF award PHY-1915093. 
DEU is supported by Perimeter Institute for Theoretical Physics. Research at Perimeter Institute is supported
in part by the Government of Canada through the Department of Innovation, Science and Economic Development Canada and by the Province of Ontario through
the Ministry of Economic Development, Job Creation
and Trade.


\appendix


\section{Carrier concentrations in semiconductors}
\label{app:semiconductors}
The background (equilibrium) number densities of electron and holes in a semiconductor,
$\bar{n}_{e,h}$,
 are determined by the intrinsic concentration and the density of dopants. 
The intrinsic carrier concentration corresponds to the density of electrons and holes in the absence of dopants, purely due to the thermal excitation of electrons into the conduction band. 
Approximating the band structure as parabolic and assuming a non-degenerate semiconductor, 
the intrinsic carrier concentration is given by~\cite{Hu2009ModernSD} 
\begin{eqnarray}
n_i&=&\sqrt{N_C N_V} e^{-E_g/2T} ~ \\
\bar{n}_e&=&\bar{n}_h=n_i~~~~~~\textrm{(intrinsic semiconductor)} , \nonumber
\label{eq:intrinsic}
\end{eqnarray} 
where $E_g$ is the bandgap and $N_C$ ($N_V$) is the effective density of states in the conduction (valence) band. 
The effective densities of state are \cite{Hu2009ModernSD}
\begin{eqnarray}
\nonumber 
N_C&=&2 \bigg[ \frac{m_e^* T}{2\pi} \bigg]^{3/2}  \quad ,\\
N_V&=&2 \bigg[ \frac{m_h^* T}{2\pi} \bigg]^{3/2}  \quad ,
\label{eq:NCNV}
\end{eqnarray}
where $m^*_{e,h}$ are the electron and hole effective masses. 
We present the bandgaps and effective masses in Table~\ref{tab:semicondprop2} for Si, Ge, and GaAs. 
\begin{table}[t]
 \begin{tabular}{|c|c|c|c|}
 \hline
 & Si &  Ge   & GaAs  \\
 \hline
 $E_g$ [eV] \cite{VARSHNI1967149} & 1.11 & 0.66 & 1.43 \\
$m^*_e$  \cite{Hu2009ModernSD} &  0.26 $m_e$ & 0.12 $m_e$ & 0.068 $m_e$ \\
$m^*_h$  \cite{Hu2009ModernSD}  &  0.39 $m_e$ & 0.3 $m_e$ & 0.5 $m_e$ \\
 \hline
\end{tabular}
\caption{The room-temperature bandgap, effective electron mass, and effective hole mass for Si, Ge, and GaAs. 
}
\label{tab:semicondprop2}
\end{table}

When considering doped materials,
the background value of the electron and hole densities $\bar{n}_{e,h}$ differs from the intrinsic carrier concentration of Eq.~\eqref{eq:intrinsic}. 
In this case, the number density of n-type (p-type) dopants $n_d$ sets the concentration of electrons (holes).
The concentration of the opposite carrier can then be obtained by using the equilibrium condition $\bar{n}_e\bar{n}_h=n_i^2$, which holds true in presence of doping ~\cite{Hu2009ModernSD}.
Thus, for doped materials, the carrier concentrations are
\begin{eqnarray}
\nonumber \bar{n}_e&=&n_d \quad ,\quad \bar{n}_h=n_i^2/n_d \quad \quad \textrm{(n-doped)} \\
\bar{n}_e&=&n_i^2/n_d \quad ,\quad \bar{n}_h=n_d \quad \quad \textrm{(p-doped)} \quad .
\label{eq:doped}
\end{eqnarray}
The carrier concentrations in doped materials usually significantly exceed the room-temperature intrinsic carrier concentrations.

The background equilibrium density of electrons and holes determines the Fermi level of a material.
For non-degenerate semiconductors and using the parabolic band structure approximation, 
the Fermi level can be obtained by using Eqns.~\eqref{eq:intrinsic} or \eqref{eq:doped} and the expressions
\begin{eqnarray}
\nonumber
\bar{n}_e &=& N_C  \exp\big[-(E_C-E_F)/T\big]  \quad ,\\
\bar{n}_h &=& N_V  \exp\big[-(E_F-E_V)/T\big]  \quad  \quad .
\label{eq:elandholes}
\end{eqnarray}
Here $E_C$ ($E_V$) is the energy of the conduction (valence) band and $E_F$ is the Fermi energy. 
For instance, in the case of an intrinsic semiconductor, 
using the intrinsic electron or hole density Eq.~\eqref{eq:intrinsic} and Eq.~\eqref{eq:elandholes} we obtain
\begin{equation}
E_F^i=\frac{1}{2}(E_C+E_V)+\frac{1}{2}T \log(N_V/N_C) \quad. 
\end{equation}
Note that in an intrinsic semiconductor, the Fermi level lies approximately midways inbetween the valence and conduction bands.
In doped semiconductors,  
the Fermi level can be easily found by using~\eqref{eq:doped} in~\eqref{eq:elandholes}, 
with the result that in n-doped (p-doped) semiconductors the Fermi level lies close to the conduction (valence) band.


\section{Details on Cherenkov, transition radiation, and recombination in SENSEI}~\label{app:SENSEI} 

In this appendix, we provide a few additional details for how we estimated Cherenkov radiation, recombination, and transition radiation rates in Sec.~\ref{subsec:SENSEI}. 
A detailed simulation will be presented in~\cite{SENSEI-radiative-to-appear}. 

\begin{center} \textbf{Cherenkov Radiation} \end{center}
For Cherenkov radiation we consider here only those tracks that pass through the CCD, 
and neglect radiation generated in the pitch adapter and epoxy. 
The charged tracks passing through the detector can be either electrons or muons, and SENSEI presented the aggregated data in~\cite{Barak:2020fql}.
Comparing with the muon-only tracks in~\cite{Bogdanova:2006ex},
we find that muons contribute less than $10\%$ of the number of electron tracks,
so in what follows we neglect them, and assume that all track data comes from electrons.

Assuming that the electron events are isotropic, 
most electrons traverse the Skipper-CCD along its shortest dimension, 
$675~\mu$m.
However, electrons can be stopped in the Skipper-CCD before exiting, 
depending on their mean range $\ell_e$ shown in Fig.~\ref{fig:meanrange}, which depends on the track energy. 
Thus, the typical length of an electron track is 
\begin{equation}
L=\textrm{min}(675\mum, \ell_e)\,.
\label{eq:typicalelectrontrack}
\end{equation}

Once the photons are emitted by the charged tracks,
the distance they travel away from the originating track is determined by the photon absorption length, which depends on the photon energy. 
In particular, and as discussed in Sec.~\ref{subsec:cherenkov-theory}, 
photons with energies close to the Si bandgap may travel for macroscopic distances before converting into \eh-pairs.
From Fig.~\ref{fig:photonabs}, we see that photons with energies in the range $1.1\,\eV \lesssim \omega \lesssim 1.2 \, \eV$ are absorbed 60~pixels or more away from the track. 
Using our electron track distribution data, the typical track length \eqref{eq:typicalelectrontrack}, 
the Si dielectric function in Fig.~\ref{fig:dielectric}, and integrating the Cherenkov differential rate Eq.~\eqref{eq:cherenkov} over photon energies in the range  $1.1\,\eV \leq \omega \leq 1.2 \, \eV$, 
we obtain our estimate for the number of single-electron events expected at SENSEI from Cherenkov backgrounds,
\begin{equation}
R_{1e^-}^{\textrm{Cherenkov}} \sim 500/\textrm{g-day} \quad .
\end{equation}

 \begin{center} \textbf{Radiative Recombination} \end{center}
 
For radiative recombination, we obtain the number and spectrum of the tracks as done for the Cherenkov estimate above. 
The phosphorus doping profile in the backside of the CCD is taken from \cite{CCDtalk}.
The doping concentration varies from $10^{20}  \cm^{-3}$ close to the backside of the CCD, 
to $10^{17} \cm^{-3}$ at a distance of $\sim$4~$\mu$m from the backside.
For each track, we obtain the number of \eh-pairs created per unit-track length in the doped region, $dN_h/dx$, 
using Eq.~\eqref{eq:NeNh} and the Si ionization energy from~\cite{Rodrigues:2020xpt}. 
A fraction of these holes recombine radiatively with the electrons donated by the dopant.
This fraction is given by  (see Sec.~\ref{sec:recombestimates}),
\begin{equation}
\frac{dN_\gamma}{dx} = \frac{dN_h}{dx} \frac{\tau}{\tau^{\textrm{direct}}} \quad ,
\end{equation}
where $\tau$ is the hole lifetime and $\tau^{\textrm{direct}}$ the radiative recombination timescale in Eq.~\eqref{eq:timedirect}. 
Both these lifetimes depend on the electron-carrier concentration, which is set by the doping density discussed above.
The radiative recombination coefficient at the SENSEI operating temperature, required to calculate $\tau^{\textrm{direct}}$, is obtained from~\cite{doi:10.1002/pssa.2210210140}.
Regarding the hole lifetime, we take it be the minimum between three timescales: (i) the trap-assisted recombination time, which we take to be $10^{-6}\,\s$ (ii) the Auger recombination time, 
calculated in Eq.~\eqref{eq:timescales2}, and (iii) the diffusion time required for the \eh-pairs to diffuse away from the doped regions, 
\begin{equation}
\tau = \textrm{Min}(10^{-6}\,\s, \tau^{\textrm{Auger}},\tau^{\textrm{diff}}) \quad .
\end{equation}
We approximate the diffusion time by $\tau^{\textrm{diff}}\sim r^2/D_h$ as in Eq. \eqref{eq:timescales2}, where $D_h$ is the hole diffusion constant, which we take from \cite{dorkel1981carrier},
and $r$ is the length over which the holes need to diffuse.
We take $r=2~\mu$m, which is the scale that roughly characterizes layers with the same order-of-magnitude doping, 
according to \cite{CCDtalk}. 
Then, we integrate $dN_\gamma/dx$ along the length of the track passing through the doped region to obtain the number of 
photons per track assuming that the track goes perpendicular to the CCD. 
We finally sum over tracks to obtain the total amount of radiation.

To obtain the radiative recombination spectrum,
we use the radiative absorption coefficient from~\cite{RAJKANAN1979793} in Eq.~\eqref{eq:B}. 
Doping affects the Si bandgap, which in turn affects the absorption coefficient formulas in~\cite{RAJKANAN1979793}, 
and we include this effect by correspondingly shifting the bandgap using~\cite{van1987heavy}.

Given the total number of recombination photons and their spectrum, 
we obtain the fraction of them that escape the doped region into the active CCD area by using the  photon absorption coefficient above, but now including also absorption by free-carriers in doped Si, described in \cite{green1995silicon,tsai2019interband}.
As for the Cherenkov photons, we only consider the recombination photons in the energy range $1.1\,\eV \leq \omega \leq 1.2 \, \eV$, 
which pass the halo-mask cut. 
In this way, the corresponding rate of recombination photons leading to events at SENSEI is estimated to be 
\begin{equation}
R_{1e^-}^{\textrm{recombination}} \sim 500/\textrm{g-day} \quad .
\end{equation}

\begin{center} \textbf{Transition Radiation } \end{center}
 We now estimate the contribution to single-electron events at SENSEI from transition radiation generated from charged particles passing through material surfaces. We will only consider the inner walls of the copper housing, since these constitute the largest surface area compared to the other surfaces in the detector.\footnote{We neglect, \textit{e.g.}, the CCD-epoxy, CCD-vacuum, pitch-adapter-vacuum, pitch-adapter-epoxy interfaces (see SENSEI detector schematic in Fig.~\ref{fig:schematics}), as well as various layers on the front and backside of the CCD.}

In order to be registered as an event, the photons from transition radiation need to penetrate into the CCD bulk.  Most of the CCD frontside is covered by the pitch adapter, and even the part that is not covered has a 0.6~$\mu$m layer of polysilicon.  Photons entering the backside need to penetrate at least $\sim$6~$\mu$m to be registered as an event.  These considerations immediately limit the energies of the photons that can penetrate into the bulk.  Since only photons with energy $\lesssim$1.8~eV can travel more than $\sim$6~$\mu$m in silicon, we only consider photons with energies between $1.1$~eV and $1.8$~eV.
We take the dielectric functions for copper from~\cite{PhysRevB.11.1315} at 78~K.  Since charged particles can pass the surfaces traveling in either direction, we consider both forward and backward emission at the copper-vacuum interface.
 
The next ingredient is the spectrum of charged particles that pass through the copper-vacuum surface. We use the cosmic muon energy and angular spectra from~\cite{Bogdanova:2006ex}.  For determining the electron spectrum, we will make some simple assumptions, rather than doing a full background simulation.  The measured spectrum at the Skipper-CCD consists of electrons generated inside the CCD (e.g., from $\gamma$ rays) and, to a lesser extent, of electrons coming directly from the outside. We then assume the ambient photon spectrum inside the copper housing is the same as the measured electron spectrum, while the electrons relevant for transition radiation are only generated from photons interacting with the copper. We also assume the angular distributions are uniform. Under this assumption, the high-energy photon flux at the inner surface of the copper housing will be a factor of $A_{\rm housing}/A_{\rm CCD}$ larger than the flux at the CCD, where $A_{\rm housing/CCD}$ is the area of inner surface of the copper housing/CCD.  To estimate the flux of electrons generated inside the copper that can pass through the copper-vacuum interface, we will use the (energy-dependent) mean range of electrons inside copper, $\ell_e^{\rm Cu}$. Assuming each photon converts all its energy to one primary electron once it is absorbed, the number of electrons escaping from the copper housing into the vacuum towards the detector is determined by the number of photons being absorbed within $\ell_e^{\rm Cu}$ from the copper-vacuum interface.   Since the photon flux drops over a distance $L$ inside copper as $\propto e^{-L/\ell^{\rm Cu}_\gamma}$, where $\ell^{\rm Cu}_\gamma$ is the photon absorption length inside copper, the flux of electrons passing the surface can be estimated as $R^{\rm housing}_{e}(E)\sim  R^{\rm CCD}_{\gamma}  (e^{\ell^{\rm Cu}_e/\ell^{\rm Cu}_\gamma}-1) A_{\rm housing}/A_{\rm CCD}$, where  $R^{\rm CCD}_{\gamma}$ is the high-energy photon flux at the CCD.

With the above simple assumptions, the rate of single-electron events in SENSEI from transition radiation can be expressed as
\begin{equation}\label{eq:TR_SENSEI_1e}
R_{1e^-}^{\textrm{TR}} \sim \sum_{i=e,\mu}\int dE\, R^{\rm housing}_{i}(E)\int^{1.8\, \rm eV} _{1.1\, \rm eV}  d\omega \frac{ d N^{\rm TR}_\gamma(E)}{d\omega},
\end{equation}
where $R^{\rm housing}_{e,\mu}(E)$ is the spectrum of electrons (muons) passing through the inner surface of the copper housing and 
$d N^{\rm TR}_\gamma/d\omega$ denotes the rate of transition radiation (see Eq.~(\ref{eq:TR_general})), which includes both forward and backward contributions. 
We estimate 
\begin{eqnarray}
 R_{1e^-}^{\textrm{TR}} &\sim& 0.2/\textrm{g-day}.
\end{eqnarray}
These estimates for transition radiation are clearly negligible compared to those from Cherenkov (Eq.~(\ref{eq:senseicherenkov})) and recombination (Eq.~(\ref{eq:senseirecomb})).  We do not expect our conclusions to change qualitatively once we include the other interfaces mentioned above.

\section{Details on Cherenkov and transition radiation in SuperCDMS HVeV}~\label{app:SuperCDMS-HVeV} 

In this appendix, we provide a few additional details for how we estimated Cherenkov and transition radiation for SuperCDMS-HVeV~\cite{Amaral:2020ryn} in Sec.~\ref{subsec:SuperCDMSHVeV}. 

 \begin{center} \textbf{Cherenkov Radiation} \end{center}\label{app:HVeV_Ch}
 
The HVeV detector target is a $0.93 \textrm{ g}$ Si crystal with dimensions $1\, \cm \times 1\, \cm \times 0.4\, \cm$. 
It likely has some layers of composite elements (such as SiO$_2$) on its surface, but the precise composition of the surfaces is unknown.  The detector is not perfectly black and will thus at least partially reflect some optical photons. 
Several nonconductive materials are located inside the copper enclosure that constitutes the SuperCDMS HVeV detector module (see Fig.~\ref{fig:SuperCDMS_setup}).  
This includes the two PCB boards, the plastic connectors, and the flex cable.  In our estimate, we only include the PCB boards, which we expect to dominate.   

We now estimate the rate of Cherenkov events produced in the two PCBs and the resulting rate and spectrum of \eh-pairs observed with the HVeV detector.  We make several simplifying assumptions. 
First, we need to know the high-energy background rate and spectrum.  The backgrounds consist of cosmic muons and electrons from radioactivity.  The muon flux is well measured, and for our estimate, we take a muon flux at sea level of $\sim 9\times10^{-3}/(\textrm{cm}^2\,\textrm{s\,sr})$, with a zenith angular distribution of $\cos^2\theta$~\cite{Bogdanova:2006ex}. As discussed in the main text, for the high-energy electron background, we  simply assume it is the same as that measured by SENSEI~\cite{Barak:2020fql}, but scaled higher by a factor of 60.  

Second, to estimate the Cherenkov radiation generated inside the PCBs, we need to know the PCB's dielectric properties and the typical track length of high energy electrons/muons in the PCB. 
We do not know the precise composition of the PCB used in the HVeV detector, nor have we found in the literature a measurement of its refractive index and extinction coefficient at optical and infrared frequencies.  
Given this limitation, in order to perform our estimate we simply assume that the composition of the PCB is equal to the one of the ``NEMA FR4 plate'' found at~\cite{PCB-FR4}. 
This assumes that the PCB is made of $61\%$ SiO$_2$, $15\%$ epoxy, and $24\%$ of bromine and oxygen.  
The dielectric properties of SiO$_2$ were discussed in Sec.~\ref{subsec:cherenkov-theory},
while the epoxy properties are taken from~\cite{doi:10.1002/app.33287}. 
Since~\cite{doi:10.1002/app.33287} only provides data for epoxy between  $1.8$~eV and $6.5$~eV, we simply assume the absorption length in epoxy below $1.8$~eV (above $6.5$~eV) are constant and the same as that at $1.8$~eV ($6.5$~eV).
We consider only photons with energies above $\sim$1~eV, as these are energetic enough to create an \eh event in the SuperCDMS Si detector.
For photons frequencies $1\,\eV \leq \omega \leq 8\,\eV$, both SiO$_2$ and the epoxy have a refractive index of $\sim$1.5, so we assume that the refractive index of the PCB board is $n_{\rm PCB}=1.5$, ignoring the contributions from the bromine and oxygen.  
We only account for absorption in the $\textrm{SiO}_2$ and epoxy, and neglect possible absorption due to the oxygen and bromine.
With our assumptions, 
the threshold energy for electrons (muons) to produce Cherenkov radiation in the PCB is $E_e^{\rm th}=175$~keV ($E_\mu^{\rm th}=36$~MeV). 

The typical track length, on the other hand, is obtained from the track mean range and the PCB dimensions. 
For each track, we assume that its length is given by the minimum of the electron or muon mean range,  
$\ell_{e,\mu}(E)$ (where $E$ is the kinetic energy of the incoming particle), 
and the size of the PCB. 
Since the thickness of the upper PCB and lower PCB is $0.7$~mm and $1.5$~mm, respectively, and the electrons and muons typically pass the PCB at an angle, we take the maximum track length to be 1~mm (2~mm) for the upper (lower) PCB. 
The track length is therefore given by 
\bea 
\Delta x_{\rm PCB,upper}(E)=\textrm{Min}[1\,\textrm{mm},\ell_{e,\mu}(E)]\nonumber\\
\Delta x_{\rm PCB,lower}(E)=\textrm{Min}[2\,\textrm{mm},\ell_{e,\mu}(E)]\,.
\eea
For electrons, 
the mean range can be obtained according given the composition of FR4 from~\cite{NIST}.
For muons, the mean range  $\ell_\mu$ inside PCB is much longer than the PCB sizes, so the muon path is limited to 1~mm (2~mm) for the upper (lower) PCB. 

We estimate then the total number of Cherenkov photons per high energy charged particle as  
\bea\label{eq:N_ch_PCB}
N^{\rm Ch}(E)&\approx&\alpha\Delta \omega\Delta x_{\rm PCB}(E) \left(1-\frac{1}{v^2 n^2_{\rm PCB}}\right)\\
&\approx& 140 \left(\frac{\Delta \omega}{7\,\textrm{eV}}\right)\left(\frac{\Delta x_{\rm PCB}(E) }{1\,\textrm{mm}}\right)~~~~(v\approx 1).\nonumber
\eea

The PCB parts that can produce Cherenkov photons that can leave the PCB and be absorbed by the HVeV detector are determined by the PCB dimensions and the photon absorption length. Since most of the PCB's surfaces are coated with copper, Cherenkov photons generated in the middle of the PCB far away from any opening will likely bounce around and not be able to escape before being absorbed again. Only those photons close to the edge or the surfaces where there is no copper covering can potentially escape and reach the detector. The relevant PCB parts are the open surface (denoted as ``bare PCB'' in Fig.~\ref{fig:SuperCDMS_setup} with the estimated area being $1.2\times 1.3\, \textrm{cm}^2$) and those parts whose distance to the edge is less than the photon absorption length. 
Since this absorption length is frequency dependent, 
the Cherenkov photons that are able to escape the PCB inherit a characteristic spectrum, shown in Fig.~\ref{fig:HVeV_Ch_number} 
(even if the Cherenkov emission spectrum itself, Eq.~\eqref{eq:cherenkov}, is rather flat for our range 
of frequencies).\footnote{To calculate the number of photons that escape the PCB, we make the simplifying assumption that upon emission, photons take the shortest path towards the edge of the PCB and do not internally reflect. As a consequence, on top of our estimates a geometric penalty factor must be included to account for photons that take longer paths towards the edges of the PCB or that internally reflect (see discussion below Eq.~\eqref{eq:Ch_HVeV_final}).} 
Thus, the number of photons that escape the PCB per track is, on average, 
obtained by direct multiplication of Eq.~\ref{eq:N_ch_PCB} with the spectrum shown in Fig.~\ref{fig:SuperCDMS_setup}.
Note that a sharp drop in the spectrum of photons that can escape the PCB is observed at frequencies $\gtrsim$8~eV, 
where the SiO$_2$ becomes highly absorptive,
so in what follows we drop photons with $\omega \geq 8~\textrm{eV}$.

Having calculated the number and spectrum of photons per track that escape the PCB, 
we must now calculate the number of electrons that each individual track produces in the detector,
which determines into which $n$-electron category, $1\leq n \leq 6$, the track-induced event falls. 
We assume that single photons with energies $1~\textrm{eV} \lesssim \omega\lesssim 6~\textrm{eV}$ create a single \eh-pair inside the HVeV detector, while photons with energies $6~\textrm{eV} \lesssim \omega\lesssim 8~\textrm{eV}$ can create two \eh-pairs~\cite{Ramanathan:2020fwm}; 
we refer to these two energy ranges as ``low'' and ``high.''
Under these assumptions, a track leads to two-electron events either by emitting one high-energy photon, or by emitting two low-energy photons.
The $n$-electron events with $n> 2$ are obtained by multiple emission of low or high-energy photons.
If a track leads to $N^{\rm{Ch}}_{\textrm{ low} E}$ low-energy and $N^{\rm{Ch}}_{\textrm{ high} E}$ photons that escape the PCB, 
and assuming that each one of these photons has the same probability $f$ of reaching and being absorbed in the detector, 
the probability $P_n$ that a track will lead to an $n$-electron event can be obtained from a binomial distribution
that accounts for all possible combinations of photons of different energy that can create exactly $n$ \eh-pairs. 
For example, $P_4$ is given by
\bea
P_4(E)&=&\left(\begin{array}{c}
N^{\rm{Ch}}_{\textrm{ low} E}\\
4
\end{array}
\right)f^{4}(1-f)^{N^{\rm Ch}_{\textrm{low} E}+N^{\rm Ch}_{\textrm{high} E}-4}\nonumber\\
&+&
\left(\begin{array}{c}
N^{\rm{Ch}}_{\textrm{ low} E}\\
2
\end{array}
\right)\,
\left(\begin{array}{c}
N^{\rm{Ch}}_{\textrm{high} E}\\
1
\end{array}
\right)
f^{3}(1-f)^{N^{\rm Ch}_{\textrm{low} E}+N^{\rm Ch}_{\textrm{high} E}-3}\nonumber\\
&+&
\left(\begin{array}{c}
N^{\rm{Ch}}_{\textrm{ high} E}\\
2
\end{array}
\right)
f^{2}(1-f)^{N^{\rm Ch}_{\textrm{low} E}+N^{\rm Ch}_{\textrm{high} E}-2} \quad .
\eea
Note that $P_n(E)$ is a function of the track energy, 
as the number of high and low energy photons emitted by each track $N^{\rm Ch}_{\textrm{low,high} E}$ is fixed by the track energy via Eq.~(\ref{eq:N_ch_PCB}),
in conjunction with the average spectrum of photons leaving the PCB per track shown in Fig.~\ref{fig:HVeV_Ch_number}.
Thus, the $n$-\eh-event rate, $R_n$, 
can be obtained by multiplying the number of tracks of a given energy passing through the areas of the PCB previously discussed, with the probability that such tracks lead to an $n$-electron event, 
 \bea \label{eq:Ch_HVeV_final}
R_{n}= \int dE R_{e,\mu}(E) P_n(E),
\eea

 The probability $f$ is likely energy-dependent and hard to quantify without a detailed simulation. For a qualitative estimate, we simply note that the predicted spectrum of $R_n$ for $f\approx0.0016$ agrees well with the SuperCDMS HVeV data~\cite{Amaral:2020ryn} for different bias voltages. 
The small value of the probability, $f\ll 1$, for each photon immediately tells us that a very large number of Cherenkov photons are created in the PCBs. 
Moreover, a sub-percent probability is possible, because it is the combination of the geometric penalty factor for escaping the PCB, the probability of reaching the detector, and the detector efficiency.  We note that the geometric penalty factor should really be included in the values of $N^{\rm Ch}_{\textrm{low,high} E}$, but we checked that it can be absorbed into $f$ without significantly affecting the predicted values of $R_n$. 

To summarize, we have shown that Cherenkov radiation plausibly explains the observed events in SuperCDMS HVeV.  However, we have made several simplifying assumptions with unknown uncertainties.  A dedicated simulation and more precise knowledge of the optical properties of the PCB are needed for us to draw a definitive conclusion on whether the observed events in SuperCDMS HVeV are mostly from Cherenkov. 

\begin{center} \textbf{Transition Radiation } \end{center}

We now discuss the contribution of transition radiation in the SuperCDMS HVeV detector. As mentioned in Sec.~\ref{subsec:SuperCDMSHVeV}, transition radiation can be generated from charged particles passing through inner surfaces of copper vessel, surfaces of PCBs and any other material in the copper vessel. As long as those charged particles do not pass the detector directly, the subsequent transition radiation will likely be registered as an event if they reach the detector. 
For our estimate, we consider only transition radiation at the inner surface of the copper vessel, while neglecting other surfaces due to their  smaller areas. For the dielectric function for copper, we use the data in~\cite{PhysRevB.11.1315} at 78~K for photon energies below 6.6~eV, and the data in~\cite{Cudata} at room temperature for photon energies above 6.6~eV.

\begin{figure}[t]
\centering
\includegraphics[width=8cm]{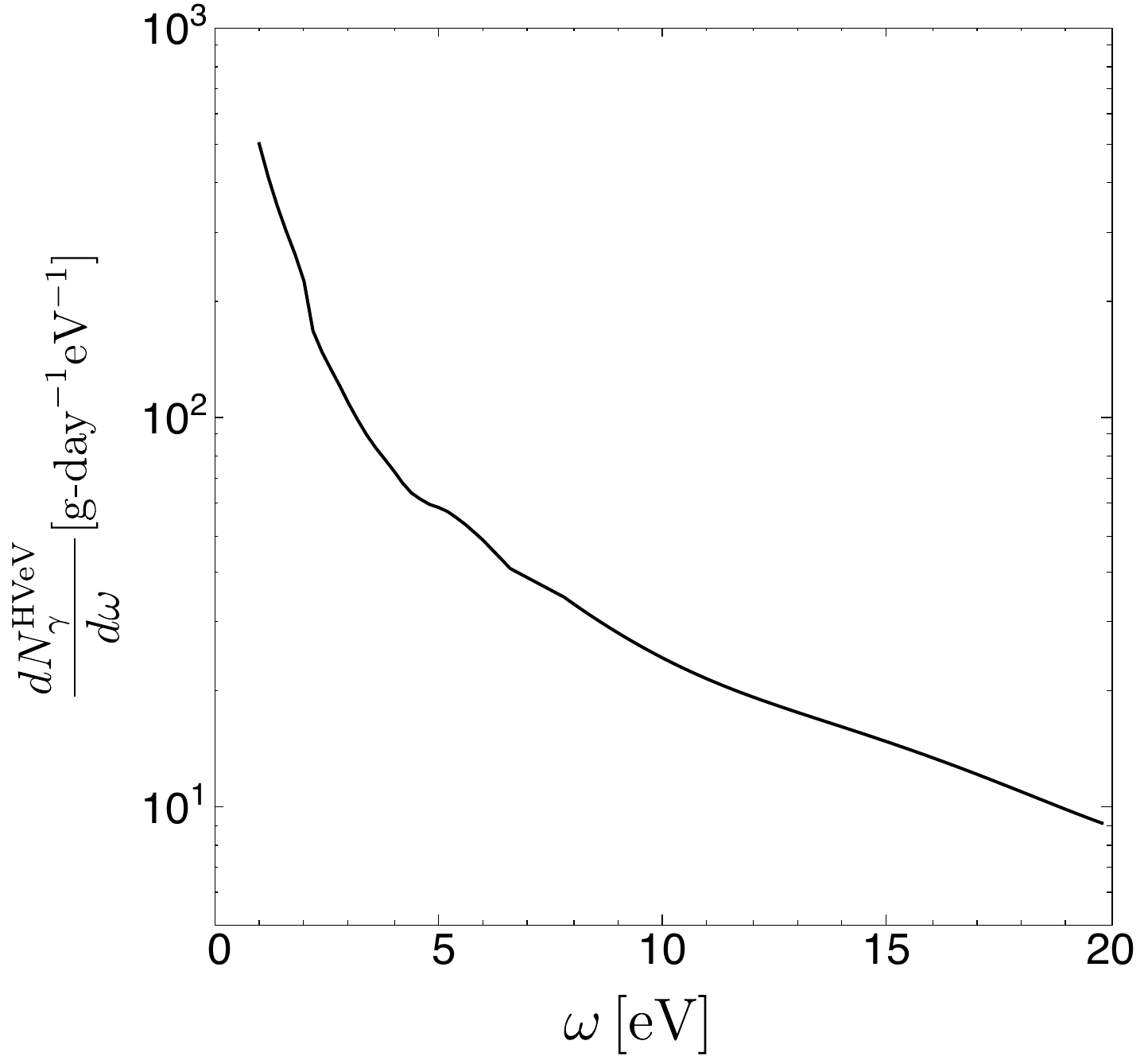}\\
\caption{The estimated number of photons (per g-day exposure) created in transition radiation from charged particles passing the interface between the copper vessel and vacuum at the SuperCDMS HVeV detector, $dN^{\rm HVeV}_\gamma/d\omega$, as a function of the emitted photon energy $\omega$ . 
 \label{fig:TR_HVeV_number}
 }
 \centering
 \end{figure}

We make some simple assumption to determine the high-energy charged particle background.  First, we take the muon spectrum from~\cite{Bogdanova:2006ex}.  Next, we assume that the ambient high-energy photon spectrum through the copper vessel looks like the SENSEI electron spectrum but scaled up by a factor of 60, which brings the background rate in line with the measured background rate at SuperCDMS CPD~\cite{Alkhatib:2020slm,NoahKurinsky-discussion}. Similar to our estimate of transition radiation at SENSEI (see discussion around Eq.~(\ref{eq:TR_SENSEI_1e})), 
the number of low-energy photons created in transition radiation at SuperCDMS HVeV detector is given as 
\begin{equation}\label{eq:TR_HVeV_spectrum}
\frac{ d N^{\rm HVeV}_\gamma}{d\omega}=\sum_{i=e,\mu}\int dE\, R^{\rm vessel}_{i}(E)  \frac{ d N^{\rm TR}_\gamma(E)}{d\omega}\, ,
\end{equation}
where $R^{\rm vessel}_{e(\mu)}$ is the spectrum of electrons (muons) passing the inner surface of the copper vessel. In particular, $R^{\rm vessel}_{e}= R^{\rm detector}_{\gamma}A_{\rm vessel}/{A_{\rm detector}} (e^{\ell^{\rm Cu}_e/\ell^{\rm Cu}_\gamma}-1)$, where $R^{\rm detector}_{\gamma}$ is the high energy photon spectrum at the detector, $A_{\rm vessel( detector)}$ is the surface area of the copper vessel (detector), and $\ell^{\rm Cu}_{e(\gamma)}$ is the mean range of an electron (the absorption length of a photon) inside copper.  The factor $d N^{\rm TR}_\gamma/d\omega$ denotes the rate of transition radiation (see Eq.~(\ref{eq:TR_general})), which includes both forward and backward contributions from the copper-vacuum interface.
The estimated number of low-energy photons (per g-day exposure) created in transition radiation, $dN^{\rm HVeV}_\gamma/d\omega$, as a function of the emitted photon energy $\omega$ is shown Fig.~\ref{fig:TR_HVeV_number}.

To estimate the contribution to the observed $n$-electron events ($R_n$) at SuperCDMS, we need to introduce the detection probability $f$ of each emitted photon as discussed in Sec.~\ref{subsec:SuperCDMSHVeV} and above for Cherenkov radiation. Unlike the case for Cherenkov radiation, where typically tens or hundreds of Cherenkov photons are generated at (almost) the same time, typical only one photon is  produced within the time window of the detector resolution. This means only one power of $f$ is needed to obtain the rate. Moreover, since transition radiation has a spectrum of emitted photon that extends to $O(10)$~eV, they can create $n\geq 1$ \eh-pairs at the Si detector with the probability defined as $\eta_n$ . Therefore, $R_n$ is given as 
\begin{equation}\label{eq:TR_HVeV_R_n}
R_n=\int d\omega \frac{ d N^{\rm HVeV}_\gamma}{d\omega} \eta_n(\omega) f\, ,
\end{equation}
where $\eta_n(\omega)$ is take from~\cite{Ramanathan:2020fwm} with $\sum_n\eta_n(\omega)=1$ for any $\omega$.

\begin{figure}[t!]
\centering
\includegraphics[width=8cm]{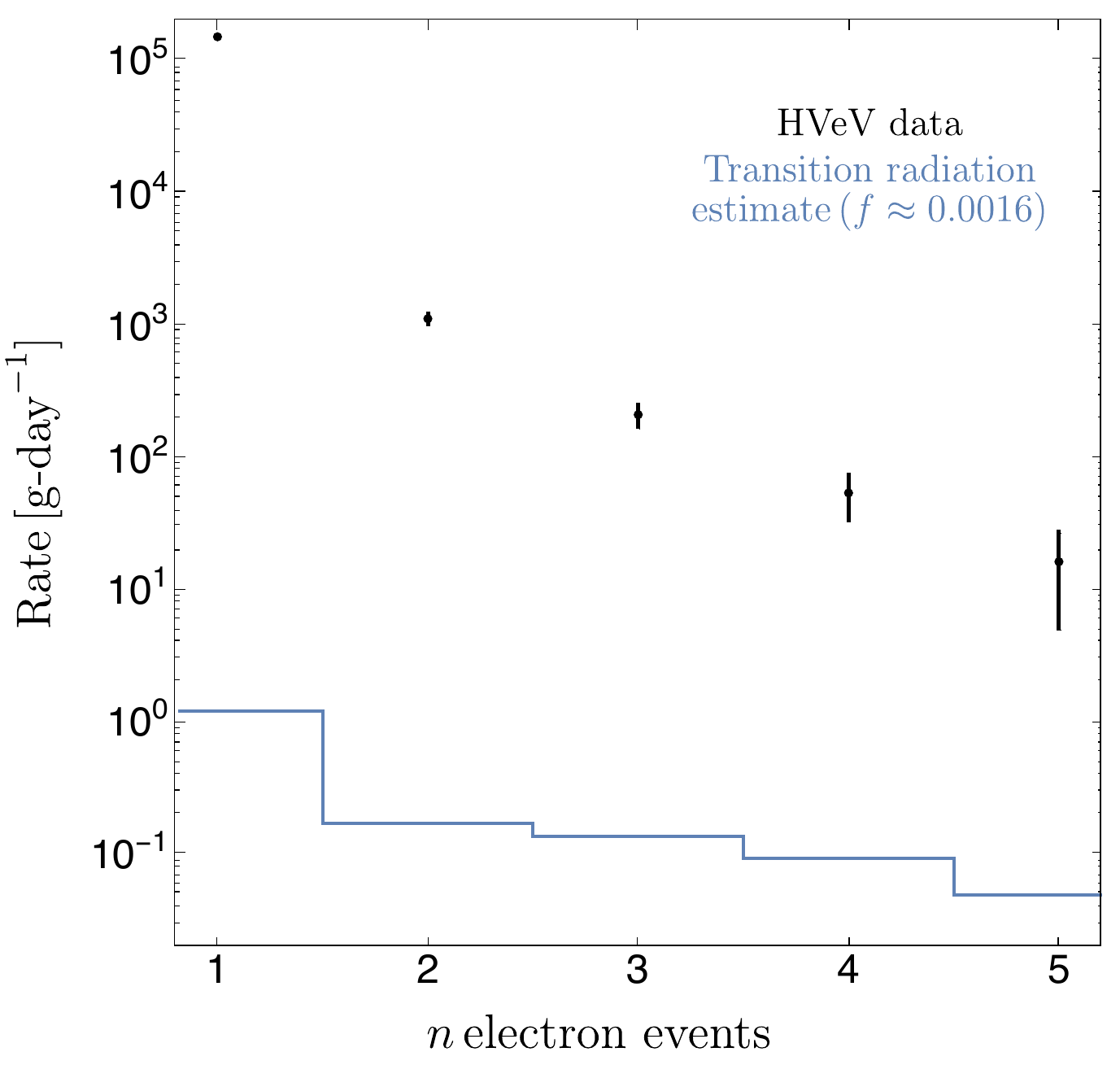}\\
\caption{The estimated $n$-electron event rate from transition radiation at SuperCDMS HVeV (blue) based on a simple model with several assumptions and a photon detection probability $f \approx 1.6\times 10^{-3}$ (see text for details). The experimental data with $100$~V bias voltage is shown in black with the error bars corresponding to the $3\sigma$ statistical uncertainty~\cite{Amaral:2020ryn}.   
 \label{fig:TR_HVeV_estimation}
 }
 \centering
 \end{figure}

Since it is hard to estimate $f$ without a simulation, for illustration we simply use the inferred detection probability from our estimate of Cherenkov photons, $f\approx 0.0016$ (see above discussion for Cherenkov radiation). The contribution to $n$-electron events at SuperCDMS HVeV detector is shown in Fig.~\ref{fig:TR_HVeV_estimation}. This estimate suggests that transition radiation is not the dominant source of the observed events.  However, we have made several simplifying assumptions with unknown uncertainties, so that we hesitate to draw a definitive conclusion on the importance of transition radiation.  A detailed simulation is needed to quantify the contribution from transition radiation at the SuperCDMS HVeV detector.

\section{Cherenkov Radiation for SuperCDMS SNOLAB Detector: further details}~\label{app:SuperCDMS-SNOLAB}

We provide a few additional details to our discussion in Sec.~\ref{subsec:SuperCDMS-SNOLAB} on the Cherenkov background at the SuperCDMS-SNOLAB experiment. 
We focus on the HV detectors, which are expected to be more sensitive than the iZip detectors to the sub-keV energy depositions expected from sub-GeV dark matter. 

Each HV detector inside the tower is held by 12 sensor clamps consisting of Cirlex.  Each clamp has a mass of 0.17~g; it has an irregular shape, but approximating it as a cuboid, its dimensions are $\sim$6~mm$\times$14~mm$\times$1.5~mm.  In addition, four straight and four curved detector interface boards consisting of Cirlex (of mass $\sim$0.21~g) sit on the edge of the copper housing near each HV detector and have a direct view to the HV detector.  The shape of the detector interface boards are also irregular, but approximating them as cuboids, their dimensions are $\sim$3.8~mm$\times$25.5~mm$\times$1.5~mm. 
We neglect the Cherenkov contribution from the cables connecting the HV sensors to the outside. 
The total mass of Cirlex that sits inside or along the side of the copper housing for each HV detector is therefore about 3.72~g, or about 22.3~g for one tower consisting of six HV detectors. 

In order to estimate the Cherenkov generated inside the copper housing by the Cirlex, we need to know the rate of such high-energy events that interact with the Cirlex.  The SuperCDMS Collaboration has published the expected background spectrum only up to $\sim$100~keV for the HV detectors~\cite{Agnese:2016cpb}, finding that detector-bulk contamination with $^{32}$Si (in the Si HV detectors) and $^3$H (in the Ge HV detectors) dominate the background spectra at low energies.  These contaminants will not be present inside the Cirlex, but other radioactive impurities will be present in it.  
These radioactive impurities are listed in Table~IV of~\cite{Agnese:2016cpb}, and have  recently been assayed more precisely~\cite{BenLoer-discussion,SuperCDMS-background-paper-to-appear}.  Beta-decays of these impurities can produce electron events with energy above 115~keV, which can create Cherenkov radiation.  There will also be an 
``ambient'' environmental background that includes (i) gamma-rays that Compton-scatter off electrons or pair-convert to electron-positron pairs inside the Cirlex and (ii) electrons produced in other materials that interact with the Cirlex.  
A simulation is required to calculate this ambient background rate and spectrum above 115~keV, although it is likely subdominant to the contribution from the radioactive impurities inside the Cirlex.  In any case, our estimate of the Cherenkov event rate based on the radioactive impurities alone should be viewed as conservative.  

Three radioactive impurities in Cirlex, $^{238}$U, $^{232}$Th, and $^{40}$K, listed in Table IV of~\cite{Agnese:2016cpb} will dominate the production of high-energy electrons inside the Cirlex:
\begin{itemize}[leftmargin=0.1cm,itemindent=.5cm,labelwidth=\itemindent,labelsep=0cm,align=left]\addtolength{\itemsep}{-0.6\baselineskip}
\item $^{238}$U has a concentration of $\sim$14~mBq/kg~\cite{BenLoer-discussion,SuperCDMS-background-paper-to-appear}.\footnote{All quoted concentrations assume that the entire decay chain is in secular equilibrium~\cite{BenLoer-discussion}.} 
The $^{238}$U-decay chain contains six beta decays with various energy releases $\Delta E$, namely 
$^{234}$Th$\to$$^{234{\rm m}}$Pa ($\Delta E 	\simeq 0.27$~MeV), 
$^{234{\rm m}}$Pa$\to$$^{234}$U ($\Delta E \simeq 2.27$~MeV), 
$^{214}$Pb$\to$$^{214}$Bi ($\Delta E \simeq 1.02$~MeV), 
$^{214}$Bi$\to$$^{214}$Po ($\Delta E \simeq 3.27$~MeV), 
$^{210}$Pb$\to$$^{210}$Bi ($\Delta E \simeq 0.06$~MeV), and 
$^{210}$Bi$\to$$^{210}$Po ($\Delta E \simeq 1.16$~MeV).  
Using Geant4, we expect about 3.8~electrons above the Cherenkov threshold of 115~keV from the $^{238}$U decay-chain~\cite{BenLoer-discussion}. 
\item $^{232}$Th has a concentration of $\sim$4.5~mBq/kg~\cite{BenLoer-discussion,SuperCDMS-background-paper-to-appear}.  
The $^{232}$Th-decay chain contains the following beta decays with various energy releases $\Delta E$, namely 
$^{228}$Ra$\to$$^{228}$Ac ($\Delta E \simeq 0.05$~MeV), 
$^{228}$Ac$\to$$^{228}$Th ($\Delta E \simeq 2.12$~MeV), 
$^{212}$Pb$\to$$^{212}$Bi ($\Delta E \simeq 0.57$~MeV), 
$^{212}$Bi$\to$$^{212}$Po ($\Delta E \simeq 2.25$~MeV, 64\%), and 
$^{208}$Tl$\to$$^{208}$Pb ($\Delta E \simeq 1.80$~MeV, 36\%). 
Using Geant4, we expect about 2.6~electrons above the Cherenkov threshold of 115~keV from the $^{232}$Th decay-chain~\cite{BenLoer-discussion}. 
\item $^{40}$K was assayed to have an upper limit on its concentration of $\lesssim 5.3$~mBq/kg~\cite{BenLoer-discussion,SuperCDMS-background-paper-to-appear}.  We will take the upper limit in our estimate of the Cherenkov event rate below.  About 89\% of $^{40}$K undergo a beta-decay to $^{40}$Ca, emitting a total energy of 1.31~MeV.  About 90\% of the electrons released will have an energy above 115~keV~\cite{K40-spectrum} and produce Cherenkov photons.  
\end{itemize}
Based on this information, we estimate that in the 22.3~g of Cirlex present in one tower of six detectors, the total rate of beta decays 
is $\sim$200~events/day or $\sim$75,000~events/year.  
The rate for producing electron events above the Cherenkov threshold of 115~keV is $\sim$130~events/day or $\sim$50,000~events/year.  

Many of these events can be vetoed easily, although it is plausible that several Cherenkov events will remain in the final dark matter search data set:  
\begin{itemize}[leftmargin=0.1cm,itemindent=.5cm,labelwidth=\itemindent,labelsep=0cm,align=left]\addtolength{\itemsep}{-0.6\baselineskip}
\item Beta decays will typically leave the daughter nucleus in an excited state, and can lead to a gamma-ray or x-ray being emitted that gets subsequently absorbed in the detector.  This could veto many events that produce Cherenkov radiation~\cite{BenLoer-discussion}, but of course not all high-energy photons will be observed in the detector.  
\item A fraction of high-energy electrons produced in the beta-decays will leave the Cirlex and could interact in the HV detector.  Nevertheless, the range of an electron with an energy of 1~MeV in Cirlex is only about 3.3~mm, so that many beta decays will produce electrons that do not leave the Cirlex.  Moreover, many electrons that leave the Cirlex would simply interact with the surrounding copper without causing a signal in the HV detector.  Finally, the small increase in heat from a beta-decay in the Cirlex clamp is likely not detectable by the HV detector.  It seems therefore challenging to veto many Cherenkov events using the accompanying energy deposition of the high-energy electron.  
\item Beta decays that produce electrons with an energy well above the Cherenkov threshold of $\sim$115~keV will produce a large number of photons.  Such events are easy to veto, since they will have a high probability of producing a signal in more than one detector.  However, beta decays that produce electrons with an energy close to 115 keV, or that produce electrons that leave the Cirlex before coming to a full stop (without subsequently interacting with an HV detector) are more problematic.  These would produce only a few photons, which part of the time could interact only in one detector.  
\item Some of the photons, especially those produced in the detector interface boards along the side wall of the copper housing, could leave the copper housing.  
This may allow at least some beta decays that produce electrons with energy well above the Cherenkov threshold to produce only a few photons that are actually observed in a single HV detector. 
\item  It is possible that events produced close to the surface on the cylindrical side walls of the HV detectors can be distinguished from those that penetrate some distance into the bulk from the side walls.  
This would allow many events to be vetoed that produce Cherenkov photons that all hit a single detector, since they would produce multiple photons, several of which would be well above the Si or Ge bandgap and hence be absorbed on the side walls.  
However, photons that hit the top or bottom surfaces of the detector are likely harder to distinguish from those that penetrate into the bulk~\cite{MattPyle-discussion}.  
Even in this case some beta decays could produce a few photons all of which are within $\sim$0.1--0.2~eV of the Si or (with lower probability) the Ge bandgaps and have a large absorption length that allows them to penetrate into the bulk. 
\end{itemize}

Finally, we assumed in our discussion that the HV detectors will be operated with a non-zero bias voltage.  It is also possible that they will operate without a bias voltage. (This would not dramatically affect their sensitivity to nuclear recoils from dark matter, although would largely remove any sensitivity to electron recoils.)  Since the design goal of the phonon energy resolution for the Si (Ge) HV detectors are about 5~eV (10~eV) (see Table I of~\cite{Agnese:2016cpb}), the detector ``threshold'', taken to be $7\sigma$ above their resolution, is about 35~eV (70~eV).  
This implies that about 35 (70) Cherenkov photons with an energy of $\mathcal{O}$(eV) are needed in Si (Ge) to create a background event above the noise threshold.  This is possible with the beta decays that produce electrons well above the Cherenkov threshold of 115~keV.  While the Cherenkov photons from such high-energy beta decays will likely be absorbed in multiple HV detectors, it will not be above threshold in all of them.  A detailed simulation is needed also in this case to determine whether the Cherenkov background is a concern.  One interesting option might be to operate some HV detectors with a non-zero bias voltage and use these as a veto for the HV detectors that are being operated with a zero bias voltage.

\bibliographystyle{utphys}
\bibliography{cherenkov}
 
\end{document}